**Shanitamol Sojan Gracy**

# A global analysis of data breaches from 2004-2024 using Microsoft Power BI

**Supervisor: Mr. Martin Warren**

Submitted as part of the requirements for the award of the
MSc in Information Security
at Royal Holloway University of London.

Information Security Group

Royal Holloway, University of London

August 2024

# Anti-Plagiarism Declaration

I affirm that the dissertation produced is my own work and I have properly acknowledged other's work with quotations, citations and references.

I confirm that I have read and understood the Academic Misconduct Guide as directed in the ISG Project handbook and I submit my dissertation in compliance with it.

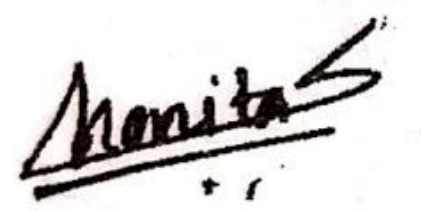

Shanitamol Sojan Gracy

**Date: 25/08/2024**

# Acknowledgements

I am deeply grateful to my supervisor for his expertise and guidance which have greatly contributed to my research experience. I appreciate his willingness to generously give his time and support to shape my dissertation and motivate me to push my boundaries and produce a well-researched work.

Royal Holloway University of London and Library staff have been extremely helpful in getting all the requested materials and resources needed for this research. I wish to extend my thanks to the University for supporting me and helping me complete my dissertation on time.

This academic journey has been extremely overwhelming. I could only navigate through it because of my strong pillars – my parents. I am eternally grateful for their unconditional love, prayers and encouragement.

I could only chase my dreams because I knew my elder sister was strong enough to take care of my family. I am grateful for her reassurance and for constantly challenging me to achieve new heights.

A special thanks to my dear ones for their unfailing support, sacrifice, and love.

Most importantly, thank you God for helping me with my personal challenges and giving me more strength to strive harder and fulfil my goals.

# Table of Contents

# List of Figures

# List of Tables

# List of Abbreviations

**AI:** Artificial Intelligence

**AIMS:** Artificial Intelligence Management System

**APT:** Advanced Persistent Threat

**APWG:** Anti-Phishing Working Group

**AWS:** Amazon Web Services

**BEC:** Business Email Compromise

**BI:** Business Intelligence

**CCPA:** California Consumer Privacy Act

**CSPM:** Cloud Security Posture Management

**CVC:** Card Verification Code

**CVV:** Card Verification Value

**DAST:** Dynamic Application Security Testing

**DLP:** Data Loss Prevention

**DPA:** Data Processing Agreement

**EDR:** Endpoint Detection and Response

**FBI:** Federal Bureau of Investigation

**GDPR:** General Data Protection Regulation

**GPT:** Generative Pre-trained Transformer

**HIPAA:** Health Insurance Portability and Accountability Act

**IDS/IPS:** Intrusion Detection System / Intrusion Prevention System

**IP:** Intellectual Property

**IP header:** Internet Protocol Header

**ITRC:** Identity Theft Resource Centre

**MOD:** Ministry of Defence

**MSA:** Master Service Agreement

**MFA:** Multi-Factor Authentication

**NHS:** National Health Service

**NASA:** National Aeronautics and Space Administration

**NIST:** National Institute of Standards and Technology

**NSA:** National Security Agency

**OSINT:** Open-Source Intelligence

**PHI:** Protected Health Information

**PII:** Personally Identifiable Information

**RaaS:** Ransomware as a Service

**RBAC:** Role-Based Access Control

**SIEM:** Security Information and Event Management

**SSN:** Social Security Number

**SVR:** Russia's Foreign Intelligence Service

**TCP header:** Transmission Control Protocol Header

**TTP:** Tactics, Techniques, and Procedures

**UFO:** Unidentified Flying Object

**WAF:** Web Application Firewall

# Executive summary

IBM's report [1] highlights that the worldwide average cost of a data breach has reached $4.45 million in 2023. These breaches can cause financial, reputation and legal challenges[2][3]. If businesses can recognise how the types of attacks targeting their industry relate to the value of their data, they can focus their data security efforts on the most critical threats that can hamper their business operations and minimise the cost of the breach. Hence, it is important to understand the current threat landscape to strengthen cyber resilience. For this purpose, this research provides a comprehensive analysis of data breach trends across industries, regions, attack methods and data sensitivity from 2004 to 2024 using Microsoft Power BI. The research seeks to benefit from data visualisation techniques to help policymakers and organisational leaders make informed decisions based on data insights. Understanding the correlation between different data breach categories will enable them to prioritise security based on severity and protect their data. In this regard, this research aims to provide insights into dominant, emerging and declining data breach trends and further explore their implications. Finally, the study concludes by contributing to recommendation strategies and future directions for further research.

**Key findings:**

**Industry Impact:** Web services, healthcare, government and financial services are the most targeted industries with web services alone accounting for over 23% of breaches.

**Prevalent Attack Methods:** Hacking is the most common method accounting for 69% of breaches followed by poor security at 14%. While historically lost or stolen devices were significant, recent trends show credential stuffing and insider threats as more prominent with zero-day exploits posing significant operational and regulatory challenges.

**Data Sensitivity:** Personal Identifiable Information (PII) is the most frequently compromised type of data, followed by credit card details and Protected Health Information (PHI).

**Regional Analysis:** While North America had the most breaches over the years, Europe surpassed it in 2023 due to increased cloud adoption and ransomware attacks.

# 1. Introduction

## 1.1. Research problem:

Businesses are increasingly relying on new technologies to optimise their operations and add value [4], which makes data a critical business asset. However, the use of technology, especially IT services, has introduced significant data security and privacy challenges [5][6]. Alfawzan et al. [7] conducted a study on 23 women's health apps and found out that most of them collect personal data without the user's consent and share it with third-party services. Poor data security and privacy practices in technology innovations make sensitive business information susceptible to risks like unauthorised access, data abuse and security breaches [8] which have become a growing concern for business leaders [9][6].

The worldwide average cost of a data breach has escalated to $4.45 million in 2023 [1]. These breaches not only cause direct financial damage but also bring about long-term harm to reputations as well as substantial regulatory and legal repercussions [2][3]. The US alone reported 3,205 breaches in 2023 affecting around 3.5 million people [10]. According to the UK Government's 2023 Cyber Security Breaches Survey [11], nearly 32% of UK firms reported cyber breaches especially prevalent among large to medium business. Although the percentage of reported breaches dropped from 39% in 2022, it was mainly due to underreporting in small businesses. On average, these breaches cost businesses about £1,100 each [11]. With expected increases in cybercrime costs to USD 10.5 trillion by 2025 [12], data breach analysis remains a hot topic for researchers.

While numerous reports track data breach trends, they often lack clear visualisations that break down data into actionable insights [13]. To strengthen security posture, it is important to identify the incident trends and learn from real-world examples to avoid such failures. For example, a study by the Ponemom Institute [14] revealed that organisations that prioritise encryption, two-factor authentication and security awareness training are better at preventing data breaches than those that do not.

This research mainly focuses on providing a thorough analysis of data breaches trends across various sectors, regions, attack methods and data sensitivity to provide actionable recommendations for improving **cybersecurity resilience**, which is the "**ability of organisations to prepare for, respond to and recover from cyber-attacks and security breaches**" [15].

## 1.2. Research motivation:

The author's experience in data analysis using Power BI and expertise in information security compliance has motivated her to select the research theme. The research draws inspiration from the ICO Data Security Incident Dashboard [16] developed in Power BI. It shows trends and patterns in personal data breaches reported to ICO and facilitates a better understanding of security risks to personal data [16]. Recognising the value of data visualisation, the author aims to design a similar Power BI dashboard tailored to this research. By integrating different datasets within a single dashboard, the author seeks to provide a granular view of data breach patterns including the industries affected, the scale of data loss, methods employed by attackers and types of data compromised. Even non-experts can benefit from this self-service tool to understand incident trends with the option to download the data for further analysis in other tools [17].

## 1.3. Significance of the study

Data breach is a worldwide problem and its economic effects are significant. According to the FBI's 2023 Internet Crime report [18], cybercrime led to losses of over $12.5 billion in 2023 which is twice the loss made in 2021. The cost of data breaches goes beyond monetary value to more severe challenges such as "customer confidence, personal safety and social trust [19]." In 2023, the Ponemon Institute [20] conducted a study on the cost of data breaches and found that the nature of these breaches varies from intricate ransomware attacks to extensive unauthorised access events impacting sectors like healthcare, finance and government due to the confidential data involved. If businesses can recognise how the types of attacks targeting their industry relate to the value of their data, they can focus their data security efforts on the most critical threats that can hamper their business operations and minimise the cost of the breach.

The primary goal of this research extends beyond mere data presentation. The research seeks to benefit from data visualisation techniques to help enhance cyber resilience. The Department of Innovation and Technology [15] highlights "**cyber resilience as the key to operational resilience and business continuity**" [15].

## 1.4. Research questions

This report groups the research questions into three sections- data breach breakdown, time series analysis and cross-filter analysis.

**Data breach breakdown**

**RQ 1:** Which industries or sectors are targeted the most?

**RQ 2:** Which attack methods are more prevalent?

**RQ 3:** What type of data or data sensitivity is at risk of compromise?

**RQ 4:** Are there any regional biases in data breaches especially for government and public entities?

**Time series analysis**

**RQ 5:** How have data breaches evolved and are there any specific trends in various industries, methods used and types of data compromised?

**Cross filter analysis**

**RQ 6:** Is there any relation between industries, attack methods and type of data compromised?

## 1.5. Research objectives

The study focuses on analysing data breach trends and patterns over two decades using data visualisation techniques in Power BI to propose better prevention and remediation strategies.

**OBJ 1:** This study aims to provide an overview of affected industries, attack methods, data sensitivity and regions to understand the overall landscape of data breaches.

**OBJ 2:** This study aims to examine data breach categories over time to identify dominant, emerging and declining trends.

**OBJ 3:** This study aims to identify any relation between targeted industries, attack methods and data compromised to understand the prominent risk factors in various categories.

## 1.6. Research methodology

The report uses two data sources, the “World’s Biggest Data Breaches and Hacks” dataset published on ‘informationisbeautiful.com’ [21] and the “List of data breaches” available on Wikipedia [22]. The data is analysed using Microsoft Power BI. The thesis draws references from books, published papers and journal articles made available in recognised publications. The search engines used are the Royal Holloway Library search, Google Scholar and Google Search. The reference generator used is Mendeley Reference Manager to keep track of references, insert citations and bibliography.

The research answers the findings for the research questions in Chapter 4 and its analysis in Chapter 5. Table 1 provides a mapping between research objectives and research questions with reference to the chapters where they are discussed.

| Research Objectives | Research Questions | Findings | Analysis |
|---|---|---|---|
| OBJ 1 | RQ 1 | Section 4.1.1 | Section 5.1.1 |
| | RQ 2 | Section 4.1.2 | Section 5.1.2 |
| | RQ 3 | Section 4.1.3 | Section 5.1.3 |
| | RQ 4 | Section 4.1.4 | Section 5.1.4 |
| OBJ 2 | RQ 5 | Section 4.2 | Section 5.1.5 |
| OBJ 3 | RQ 6 | Section 4.3 | Section 5.1.6 |

*Table 1: Chapter-wise breakdown of research objectives and research questions.*

## 1.7. Challenging aspects of the research

It is challenging to combine two datasets due to the differences in their data structures. Wikipedia's list contains only basic information. At the same time, the collection "The World's Biggest Data Breaches and Hacks" [21] is populated additionally with more detailed information such as data sensitivity, date field and data breach description. This discrepancy particularly affects time series analysis; the Wikipedia entries specify only the "year", while the other source provides complete "date" information. Thus, comparisons can only be made on an annual basis, and more granular insights on a monthly or quarterly basis are hindered. To merge these datasets effectively, it may be necessary to map fields, eliminate unnecessary fields or possibly introduce new fields to accommodate a unified structure.

There may be inconsistencies, typos, or missing values in both datasets, requiring data cleaning and standardisation. For instance, breaches might be listed in both datasets, potentially with slightly different details [13]. One of the challenging tasks is to identify true duplicates, combine information, and at the same time preserve the uniqueness of the data. This process might also need a manual review of data breaches to identify and remove these duplicates.

For regional analysis, one dataset does not contain a region column and the other one just contains region information for government and public entities making it difficult to filter samples and assign regions manually.

Lastly, the data type needs to be accurately mapped to avoid errors while combining the datasets. For instance, Wikipedia's data breaches list contains few records listed as text (e.g. "10 million") while records in the other dataset contain numerical values (e.g. 10000000). It is important to consider what to do with the inconsistency of the entries e.g., leave fields blank, assign default values or drop error entries, making such integrations difficult.

## 1.8. Scope:

The scope of the research is restricted to cover only the three areas of objectives defined in section 1.5 and answer the six research questions identified in section 1.4. The research does not look into regulatory or legal aspects of data breaches. The study only focuses on the data categories available in the dataset.

## 1.9. Structure of report:

The report is further structured into the following chapters:

**Chapter 2: Literature Review**

The chapter lays the foundation for this research by examining various industry reports and academic papers.

**Chapter 3: Methodology: Data analysis in Power BI**

The chapter describes the ethical considerations and the methods used to gather data and how they are modelled in Power BI to create meaningful visuals.

**Chapter 4: Findings: Data breach insights and trends**

The chapter presents the main findings of the data analysis and uses graphs, charts and tables to visually represent the data.

**Chapter 5: Discussion and Implications**

The chapter discusses any patterns or trends observed, compares the results of Chapter 4 with existing literature and discusses wider implications of research for organisations and policy makers.

**Chapter 6: Conclusion**

The chapter includes a summary of the key findings of the data breach analysis, its limitations, data breach prevention and remediation strategy and recommendations for future research.

## 1.10. Conclusion:

Chapter 1 delved into the significant issue of data breaches, highlighting the research problem, motivation, questions and objectives driving the research. The author outlined the research methodology, acknowledged the challenging aspects and defined the research scope. The next chapter delves into the overview of data breaches, examines the threat actors and threat vectors and reviews existing literature related to the field. Additionally, it discusses how the proposed study extends existing work and what novel contributions it brings to the table.

# 2. Theoretical Background

## 2.1. Data Breach Classification

Sutton et al. [23] suggest characterising a data breach by considering whether the act was deliberate or accidental and evaluating the severity of the resulting damage. Another approach to classifying data breaches includes internal and external threats. Cheng et al. [24] classify enterprise data leak threats as shown in Figure 1. Intentional threats include 'Malware', 'Hacking', 'Social Engineering', 'Cyber Espionage' and 'Sabotage'. Out of these malware and hacking can be associated with cyber-dependent crimes where the crime cannot exist without the use of computers or networks and social engineering potentially relating to cyber-enabled crimes that use technology or the internet as a facilitator for cyber offences [25]. Cyber espionage and sabotage could be internal or external. An example of an external sabotage attempt could include APT groups such as Chinese APT 10, also known as Stone Panda famous for stealing data and intellectual property [26]. Conversely, an example of an internal threat is Tesla's sabotage attempt in 2018 where an employee made unauthorised changes to the company's manufacturing operating systems and transferred large volumes of sensitive data to unidentified third parties [27].

On the other side, inadvertent threats indicating unintentional security risks include 'Accidental publishing', 'Configuration error', 'Improper encryption' and 'Lost computers' which all result from human error or system misconfigurations rather than malicious intent. Privilege abuse can be intentional or inadvertent. A classic example of intentional privilege abuse is Edward Snowden's exfiltration of classified data from the National Security Agency (NSA) [28].

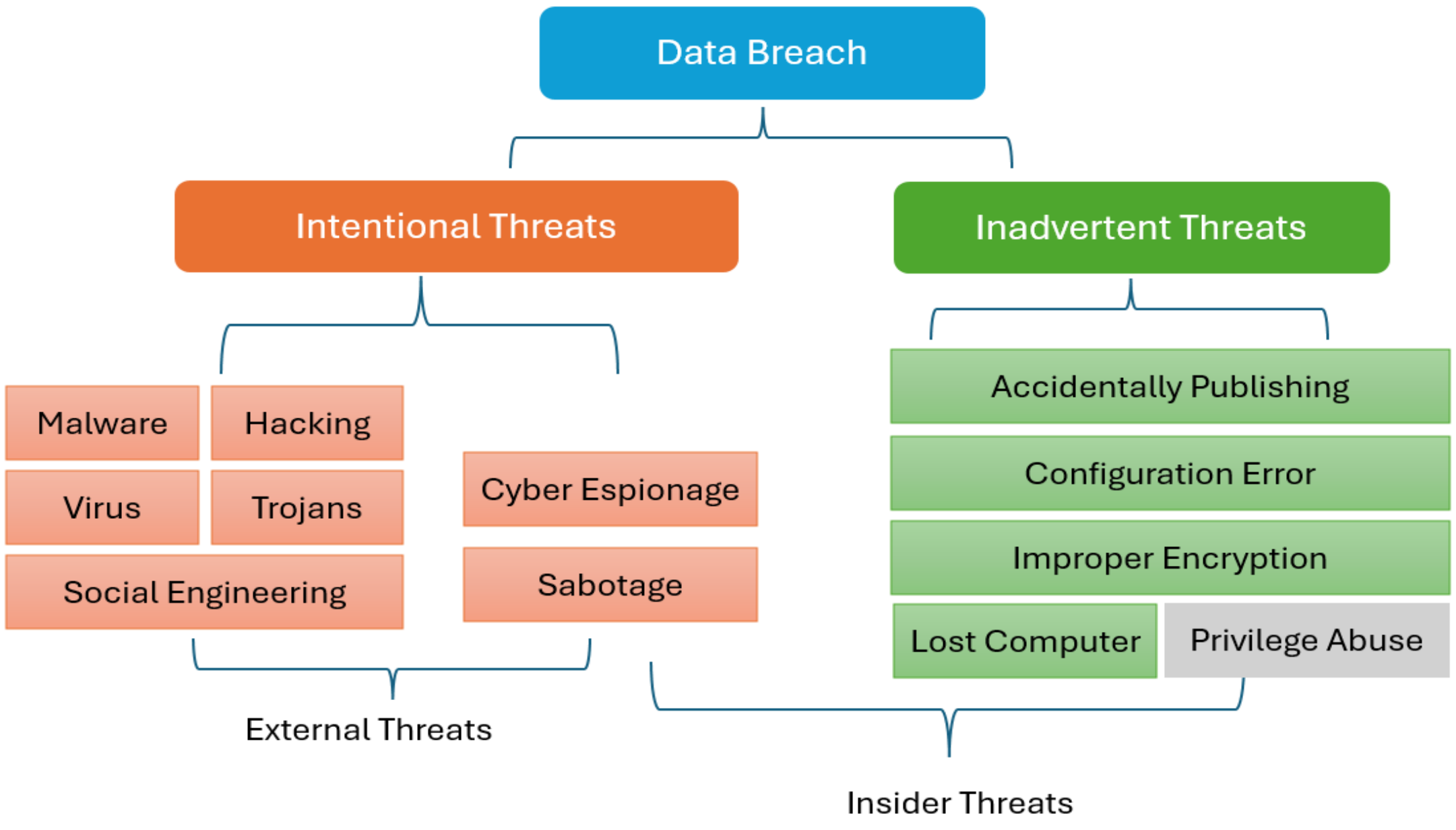

*Figure 1: Data breach classification as proposed by* [24]

## 2.2. Data Sensitivity

According to the National Institute of Standards and Technology (NIST), information that could be compromised in a data breach includes credit card details, personally identifiable information (PII), medical records, client information, corporate secrets and national security data [29]. X. Zhang et al. [30] further organise PII into nine major categories based on a data privacy study conducted by [31], as listed in the Table 2:

| **Type** | **Description** |
|---|---|
| Personal Description | Includes characteristics like gender, date of birth, age, first name, last name, eye and hair colour, tattoos, biometrics and identifiable marks. |
| Documents | Consists of sensitive items such as official documents like passports, driver's licenses and social security numbers; credit cards, receipts and ID cards. |
| Health-related data | Also known as Personal Health Information (PHI) captures a person's current health status such as heart rate and blood pressure, alongside historical medical records like past treatments and hospital visits. |
| Employment information | Relates to a person's job and workplace documents such as salary statements, employee referrals and qualifications. |
| Personal Life | Reflects an individual's religion, sexual orientation, hobbies and education |
| Relationships | Details personal, social and professional connections |
| Whereabouts | Track locations visited and residence details |
| Internet Activity | Includes online identifiers like usernames, email content, cookies and IP addresses. |
| Automobile Data | Includes details like vehicle ownership and license plate numbers. |

*Table 2: Nine categories of PII data* [30]

Similarly, NIST [32, pp. 2–1] proposes a definition of PII that takes into account the linked and linkable nature of personal data. Linkable information includes non-sensitive information such as race, gender, mother's maiden name, religion and so on, that must be used along with a linked identifier such as full name, date of birth, email ID etc to identify an individual uniquely.

Statistics (refer to Figure 2) show that the frequency and impact of data breaches involving PII have intensified [33]. According to IBM's 2023 Data Breach report [1], over half (52%) of data breaches in 2022 involved customer PII marking a notable 5% increase from the previous year. Similarly, breaches compromising employee PII have surged to 40% of all records affected, a notable jump from 26% in 2021. IBM further reports that the financial repercussions of compromised PII are substantial, the cost of a customer's PII is USD 183 and that of an employee's PII is slightly lower at USD 181.

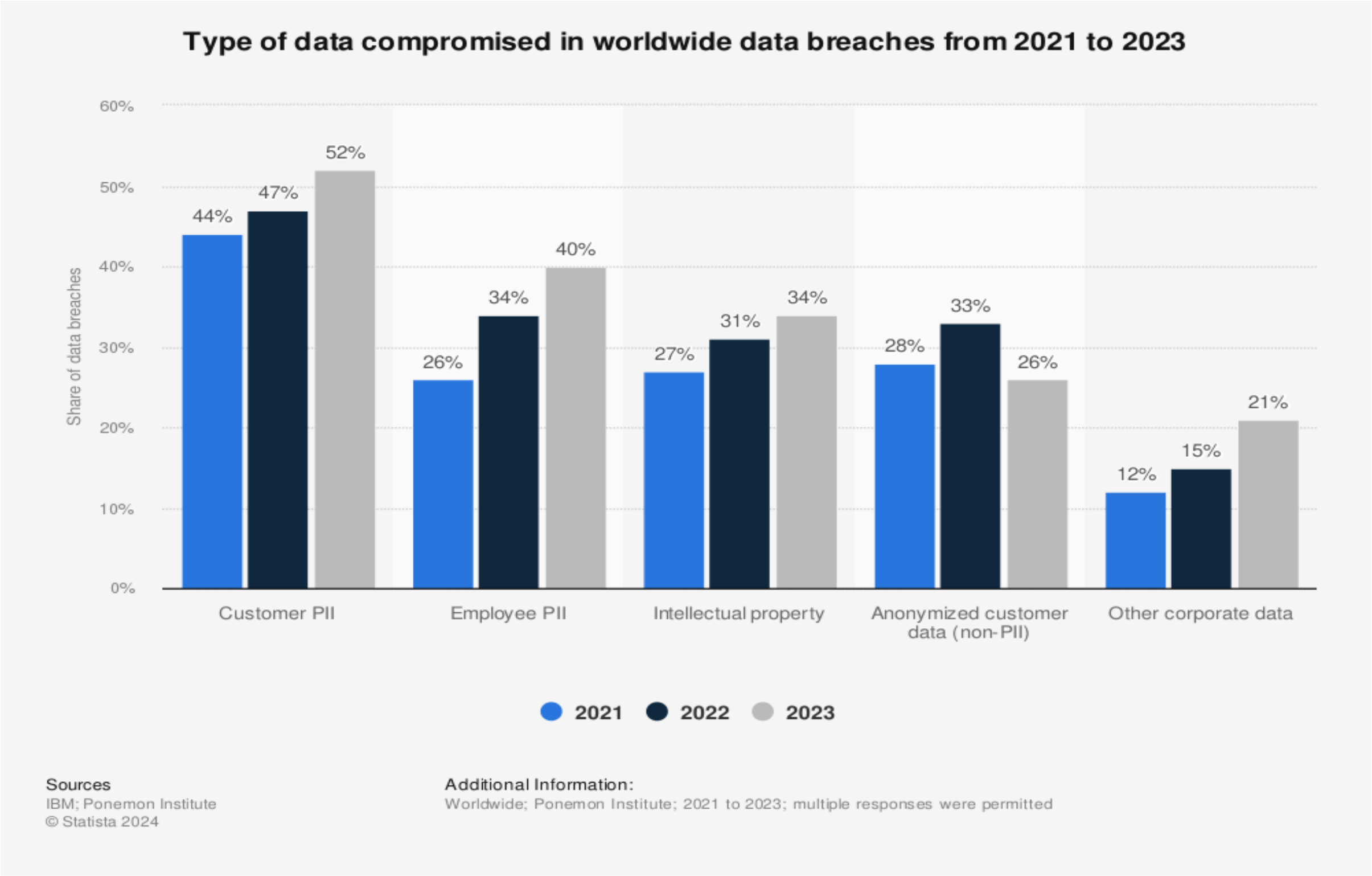

*Figure 2: Type of data compromised in worldwide data breaches from 2021 to 2023* [33]

## 2.3. Threat Actor

Identity Theft Resource Centre (ITRC) defines a threat actor as a "**person or group whose action or negligence can lead to data compromise**" [10, p. 38]. A recent study by Verizon [34] (refer to Figure 3) shows that the threat actor motivation mainly remains financial gain but there has been an increase in espionage from 5% in 2023 to 7% in 2024. The report further reveals (refer to Figure 4) that organised crime is the most seen type of threat actor followed by end-user that involves insider threats or third-party contractors. Attacks caused by state-sponsored groups are nearly 5% of the breaches [34].

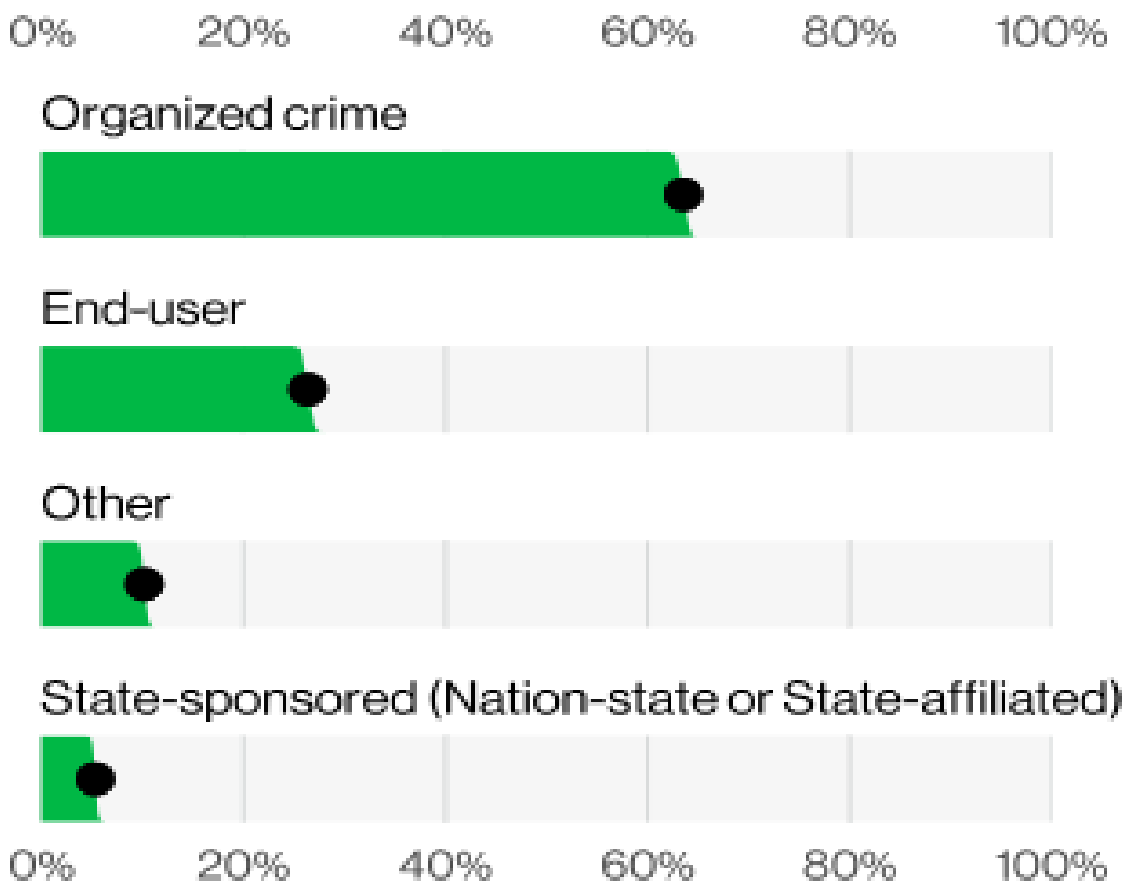

*Figure 3: Threat actor varieties in data breaches* [34]

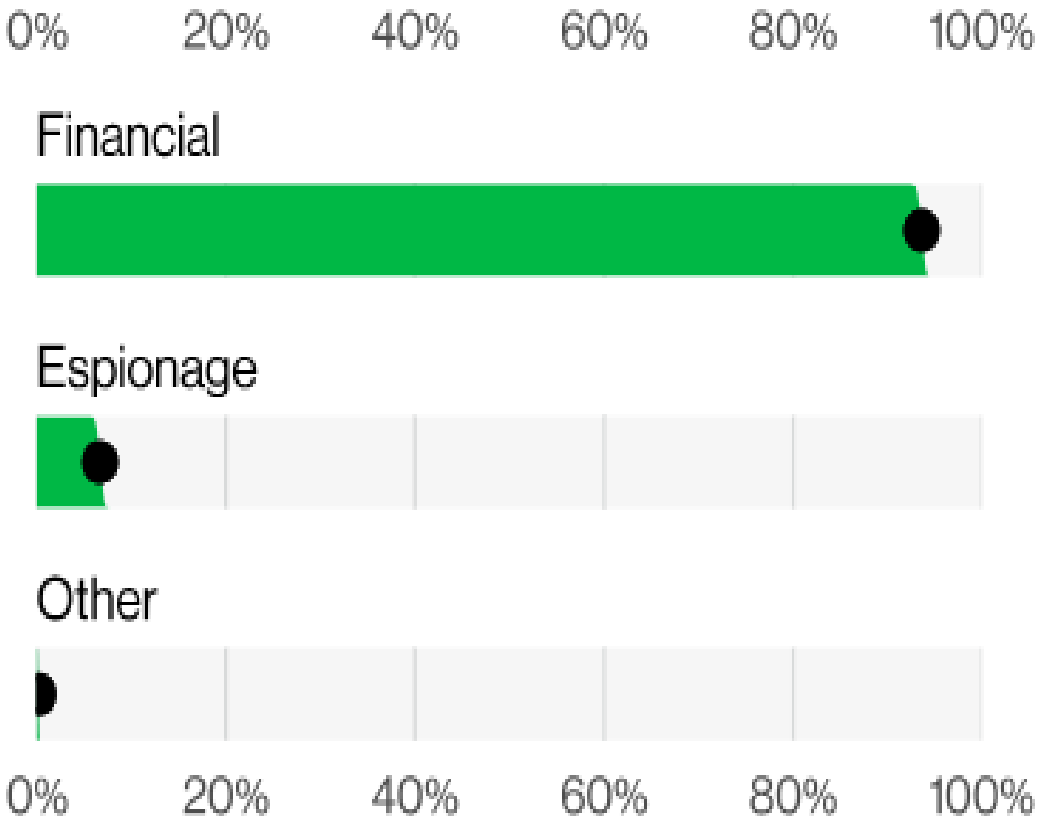

*Figure 4: Threat actor motives in data breaches* [34]

Table 3 summarises threat actors and their motivation based on [35], [36] and provides data breach examples.

| Skill level | Threat actor | Motivation | Data Breach examples |
|---|---|---|---|
| Low | Script kiddies<br>(amateur hackers) | Status/Ego | TalkTalk data breach was caused a 17 year old script kiddie who wanted to show off his computer skills to his friends. He exploited a vulnerability in TalkTalk's website and posted it online. The attackers used this vulnerability to gain access to TalkTalk's network and exfiltrated confidential data of over 15,000 customers. The compromised data consisted for email addresses, phone details and full bank details. The financial fallout was huge. TalkTalk lost around 95,000 customers and incurred a loss of £60 million. ICO also fined £400,00 million for poor security [37]. |
| | Insider employee or third-party contractor | Profit/revenge | In 2019, a dissatisfied employee of Desjardin's group copied data of 4.2 million customers into a USB drive and sold to private lender. The data compromised during this breach involved customer name, social security numbers, email IDs and transaction records. The breach further affected 1.8 million credit card holders and caused significant reputational damage. It took Desjardin around $70 million to $ 108 million to recover from this incident.[38], [39]. |
| | Hacktivists | Ideology<br>(Political or social) | LulzSec attacked Fox.com in 2020 and released personal details of 73,000 X-Factor contestants. This was part of series of attack carried out by Anonymous group on major entities like Fox.com, FBI and Sony Play Station. Their motive was to highlight the insufficient security measures in affected entities. However, for exposing sensitive information online LulzSec members faced legal charges and imprisonment.[40]. |
| | Black hat hackers | Profit/personal motivation<br>(individual) | Gary Mckinnon, a British IT specialist was intrigued by UFOs and believed the US and NASA are hiding secrets from the world. He managed to access 97 computers under the US military and NASA to search for evidence related to UFOs. He scanned for susceptible systems using a dial-up modem and deployed remote access tools to extract documents and data. This resulted in an estimated loss of $700,000. For this, Gary could have faced a possible extradition to the US and a 70 year sentence. However, his extradition was put on hold by the UK Home Secretary in 2012, citing reasons related to human rights and medical conditions [41]. |
| | Organised crimes | Profit<br>(group) | In 2020, Magecart targeted Warner Music Group (WMG). They were able to leak sensitive information such as card numbers, CVC/CVV numbers, customer names, email addresses, telephone numbers, billing addresses and shipping addresses. They used a technique known as web skimming to insert malicious code into checkout pages of |

| Skill level | Threat actor | Motivation | Data Breach examples |
|---|---|---|---|
| | | | WMG's e-commerce website and exfiltrated the customer information during the purchase process [42]. |
| High | Government-sponsored groups (hostile and national intelligence agencies) | Espionage, Data Theft for Geopolitical positioning | The 2020 breach of SolarWinds was caused by APT 29, Russia's Foreign Intelligence Service (SVR). The hackers compromised the build systems of SolarWind's Orion network monitoring software and installed a backdoor allowing access to 18,000 customers including the UK and US governments [43], [44]. |

*Table 3: Data Breach Threat actors and their motivations*

## 2.4. Threat Vector

Identity Theft Resource Centre (ITRC) defines threat vectors, also known as attack vectors as the "**methods used by threat actors to compromise data security**" [10, p. 39]. The selected attack vectors do not constitute an exhaustive list but briefly cover the major ones that are in the scope of this research.

### 2.4.1. Malware and Ransomware

Ransomware is a malware type used to encrypt data and demand ransom in bitcoins to get the decryption key [45]. However, as seen in Figure 5, ransomware has now moved away from locking data to double and triple extortions. In double extortion, first data is exfiltrated before encryption and the victims are threatened to pay ransom to avoid any leak. In triple extortion, Aggarwal [45] states that the attacker also threatens to launch a DDoS attack which can severely disrupt the victim's operations, until the ransom is paid [45]. However, the author argues that triple extortion is not just related to DDoS attacks, it is just an additional tactic used by hackers to build pressure. Triple extortion tactics mainly target customers or partners of the victim organisation to build pressure into paying the ransom or extort more money by launching supply chain attacks.

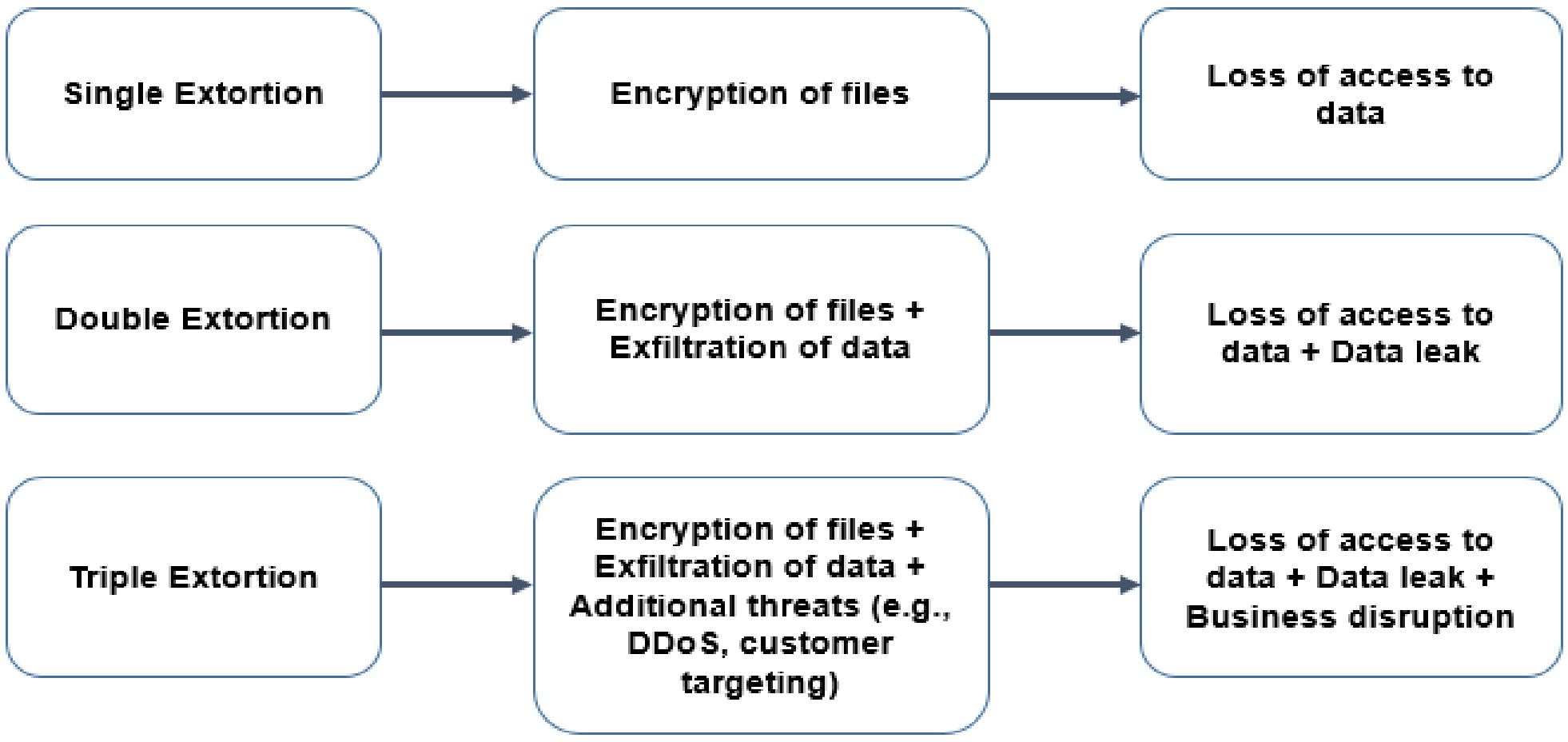

*Figure 5: Ransomware extortion types*

Ransomware as a service (RaaS) business model has grown in scale. For example, in October 2020, one of the ransomware groups – REvil spent 1 million USD for recruiting purposes [45]. IBM X-Force Threat Intelligence Index 2024 report [46, p. 16] reveals that ransomware groups have upgraded their operations and introduced new variants that are difficult to detect. One such is Sphynx developed by BlackCat developers. Even Linux versions of ransomware families have been introduced such as Royal and CLOP. Also, tactics, techniques and procedures (TTP) have been improved. For example, QR codes are used to link to payment instructions or decryption services [46, p. 16]. The top RaaS groups in 2023 were LockBit, CLOP and BlackCat, according to Trend Micro [47].

According to the IBM X-Force Threat Intelligence Index report [46, p. 21], cybercriminals prefer stealing and selling data rather than encrypting them, leading to a rise in infostealer malware by 266% in 2023. Rhadamanthys, LummaC2 and StrelaStealer are some of the info stealers used to steal login credentials and session tokens. Statistics [46, p. 18] show that data theft and leak incidents occurred more than extortion incidents. Cobalt Strike, MimiKatz and Qakbot were the top reported malware groups in 2024, according to Deepwatch ATI [48].

### 2.4.2. Hacking

Hackers use various techniques to gain initial access to the system. Once inside, they try to escalate their privileges and move laterally within the network. Attackers often try to install backdoors that would allow them access in future even if the vulnerability is patched. Once data is exfiltrated, attackers usually try to cover their tracks and remove evidence by tampering with logs and timestamps. Table 4 describes the stages of hacking and the techniques used based on [49][50]. The author further explores Table 4 in Chapter 6 for providing prevention and remediation techniques.

| Reconnaissance | Scanning | Gaining Access | | | Maintaining Access | Data exfiltration | Covering tracks |
|---|---|---|---|---|---|---|---|
| | | Credential Access | Lateral Movement | Privilege Escalation | | | |
| Application Vulnerability | Network Scan | Leaked/Exposed Credentials | Database Links (to connect and traverse databases) | Dynamic SQL (to manipulate Database Queries) | Application Backdoor (by exploiting a piece of code or unpatched vulnerability) | SQL injection | Rootkits (to disable logging and remove existing logs) |
| Exposed Service | Network Sniffing | Brute-force (to guess login details) | Remote Code Execution (to run malicious code) | User Defined Functions (to run commands as root on the system | Create Database User | Dump (transfer large volumes) | Modify Files (to change attributes such as size and date to match original attributes) |
| Third-Party (Supply chain) | | Configuration Files (to extract usernames and passwords) | Database command execution | | Create Operating system User | Slow & Low (transfer small volumes over a long period to avoid detection) | Trojans (Netcat to destroy evidence or modify timestamps) |
| Malicious Insider | | Malware (deploy keyloggers to capture password) | Read/Write files | | Rootkits (to gain repeat access at Kernel Level) | Steganography[1] (to hide information in images and sound files) | |
| Phishing | | | | | Trojans (to gain repeat access at Application level) | Tunnelling (to hide information in unused bits of the TCP and IP headers) | |

*Table 4: Hacking Stages and Techniques*

[1] Image Steganography: Data bits are encoded in every pixel of the image. The output image is visually the same as the original image making it indistinguishable for a human eye [130].

### 2.4.3. Social Engineering and Phishing

Phishing attacks have grown in scale from the famous Nigerian Prince scams in the early 90's to more complex techniques that are difficult to detect [51, p. 31]. According to Verizon's 2023 Data Breach report [51, p. 32], phishing constitutes 44% of social engineering incidents, however, pretexting is more prominent making up to 60%. It is important to note that, phishing still tops the list on the number of confirmed breaches. One example of a pretexting attack is a Business Email Compromise (BEC) attack, where the attackers impersonate someone known to you. It is also an example of spear phishing attack targeted at a group or individual. In this type of attack, the threat actor uses OSINT (Open-Source Intelligence) techniques to gather information about the target that is publicly available or uses existing email threads and context that involves routine tasks such as updating vendor bank accounts. However, the new bank account would be fraudulent making the victim lose their money. BEC attacks are difficult to detect because of the strong groundwork laid before the attack. The threat actor needs to spin up a domain that has a close resemblance to the original domain of the vendor and also update the signature with the fraudulent number for elicitation techniques. As employees fail to identify such subtle changes, BEC attacks have doubled in recent times making almost 50% of the social engineering incidents [51, p. 32]. Among these, the top action seen is persuading the employees to change the bank account for a claimed recipient [51, p. 33].

According to a recent report, one of the main aims of social engineering attacks is to steal credentials. 76% of the data compromised in social engineering attacks included compromised credentials followed by internal data at 28% and personal data at 26% [51, p. 31]. Statistics reveal that phishing sites are created every 11 seconds [52]. APWG (Anti-Phishing Working Group) detected 1.3 million new phishing websites in Q4 of 2022 [53, p. 3]. Among these, Microsoft, Google, Yahoo and Facebook top the spoofing list [54] and PayPal and MasterCard International top the frequently impersonated payment services [55]. Proofpoint's 2024 phishing report [56] reveals that the UK is the most targeted for BEC attacks. The attack volume increased in countries such as Japan and South Korea due to the use of generative AI that allows attackers to overcome cultural or language barriers in crafting more convincing phishing emails.

### 2.4.4. Insider Threat

According to Ponemom Institue, the average cost of preventing insider threat rose to $7.2 million in 2023 from $6.6 million in previous year [57, p. 9]. Another study by DTEX [58] reveals that 12% of employees leave their organisation with sensitive intellectual property (IP) such as customer data, employee data, health records and sales contracts. It is important to note that this 12% does not include non-sensitive IPs like templates and presentations that employees might also take with them [58].

Malicious insiders those with deliberate intent to harm are the most concerning group. They account for around 36% of insider incidents followed by the type “outsmarted insider” who can be easily exploited by phishing or external attacks (33%). Mistaken and Negligent Insiders only contributed to 13-15% of insider incidents [57, p. 32]. These incidents often include insider threats such as accidental disclosure due to a lack of knowledge about data handling policies, transmitting sensitive data without proper encryption, storing sensitive data on unsecured personal devices or loss of devices.

The top two actions performed by malicious insiders include accessing and downloading sensitive information not related to their role or function and sharing sensitive information with outside parties usually in large amounts and outside of usual working hours [57, p. 33]. One such example is Anthem’s data breach in 2017. The employee of a third-party contractor accessed personal health identity information of around 18,500 Anthem customers and sent it to their personal email IDs. This action resulted in both Anthem and the consulting firm facing potential fines and legal consequences [59].

### 2.4.5. Supply chain risks

Third-party risks are identified as the “weakest link” in supply chain attacks [60]. For example, the recent data breach at the UK Ministry of Defence (MOD) on 06 May 2024 involved exploiting the MOD’s payroll system managed by a third-party contractor exposing details of 270,000 service personnel. MOD points at Chinese state-sponsored actors for cyber espionage though there is not much evidence to support this [61]. A similar data breach by state-sponsored Chinese attackers targeted the US Office of Personnel Management (OPM) in 2015. They posed as an employee of a subcontracting company (KeyPoint) and used stolen credentials to establish a foothold and install malware in the OPM network that created a backdoor. They were able to exfiltrate the background verification records of millions including sensitive personal information and fingerprint records. This breach which was carried over two years is expected to have affected 22.1 million US citizens [62], [63].

Tim Bach, Senior Vice President of Security Engineering at AppOmni, said "**If you’ve hardened the system that the attacker is ultimately going after, but a supplier is easier to get into, the supply chain attack is going to look more appealing to the attacker**" [60]. Ryan Sherstobitoff, the Senior Vice President of Threat Research and Intelligence at SecurityScorecard, further added "**The supplier ecosystem is a highly desirable target for ransomware groups**" [64]. This is because victims of third-party breaches often remain unaware until they receive ransom demands, giving attackers ample time to infiltrate multiple companies undetected [64]. Reports from Statista (refer to Figure 6) show that cybercriminals are increasingly targeting supply chains. 2023 witnessed a 52% increase in supply chain attacks from the previous year which was also the highest of all time [65]. Other reports show that the sudden increase was particularly due to a ransom group CLOP (sometimes also written as Cl0p) who exploited a zero-day vulnerability, CVE 2023 34362 (MOVEit), the most cited vulnerability which was later patched on [66, p. 4]. Cl0p contributed to about 64% of the third-party breaches followed by LockBit at 7% [66, p. 13]. According to Black Kite, enterprises with an average revenue of $10 million usually suffer from the aftermath of these ransomware attacks [67, p. 9].

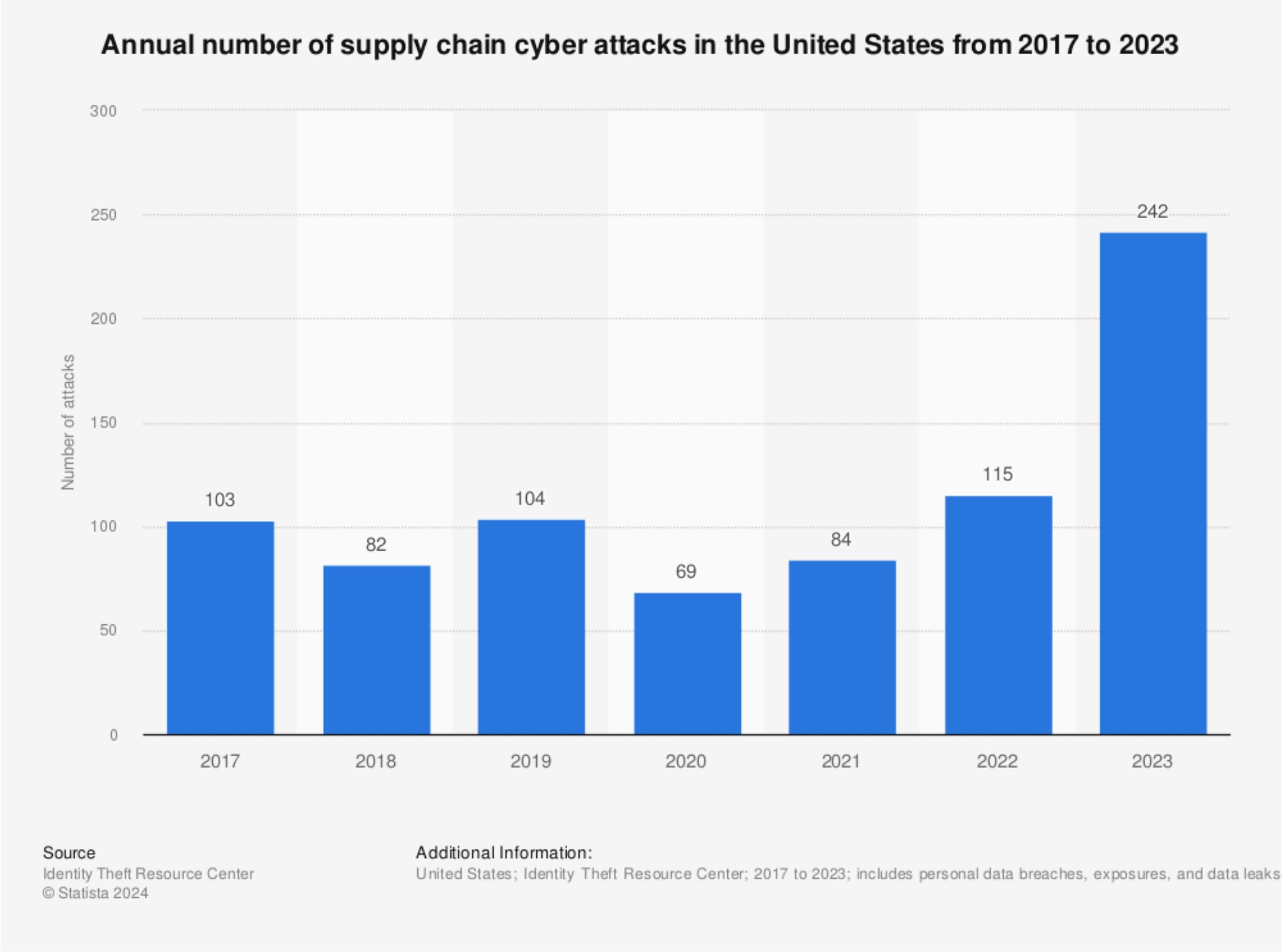

*Figure 6: Annual number of supply chain cyber-attacks in the United States from 2017 to 2023* [65]

### 2.4.6. Security misconfigurations

A study by IBM X-Force states nearly 30% of application weaknesses are due to security misconfigurations. The most common mistake is how concurrent user sessions are handled in applications [46, p. 6]. A study by the Ponemon Institute in 2021 found that cloud security is the third-most common threat vector responsible for 15% of data breaches [68, p. 2]. In most cloud security breaches, cloud storage is not set up correctly. For example, in 2022, Pegasus Airlines exposed around 6.5 terabytes of data due to a misconfigured AWS S3 bucket [69]. As more companies move from on-premises to cloud, such breaches may become more frequent in future. Gartner [70] points out that "**through 2025, 99% of cloud security will be customer's fault**" rather than the cloud service provider [70]. To mitigate these risks, Loureiro emphasises the importance of the "secure by design" principle to ensure proper implementation of the technology stack [71]. Similarly, the Security Intelligence team from IBM highlights the importance of integrating security in DevOps cloud-based workflows as early as possible to prevent cloud risks while maintaining speed and agility [72].

### 2.4.7. Emerging technologies

AI can create phishing emails in 5 minutes, which takes an average of 16 minutes manually [46, p. 9]. Although IBM X-Force has yet to witness confirmed instances of AI-generated campaigns, it is anticipated that cybercriminals will increasingly integrate AI into their activities. Already instances of WormGPT and FraudGPT are found in dark web forums [46, p. 30]. In 2023, X-Force identified over 800,000 references to AI and GPT across illicit markets and dark web forums [46, p. 31].

As the usage of AI is growing, several regulatory requirements are being developed such as the EU AI Act [73]. ISO 42001 for AIMS (Artificial Intelligence Management System) [74] and NIST AI Risk Management Framework [75] have been published to ensure fair, responsible and transparent usage of AI.

A study by McKinsey and Company (refer to Figure 7) highlights AI risk and cloud as key emerging concerns for organisations [76]. La Porte, as quoted by Benjamin David in Infosecurity Magazine [77], warns that misused quantum computers could be dangerous. He compares it to a “ticking time bomb” that can decrypt data. This means that data currently gathered under encryption could be compromised later for malicious purposes. There is concern that cybercriminals and state-sponsored entities might already be accumulating vast amounts of data intending to exploit it once quantum computing becomes available [77].

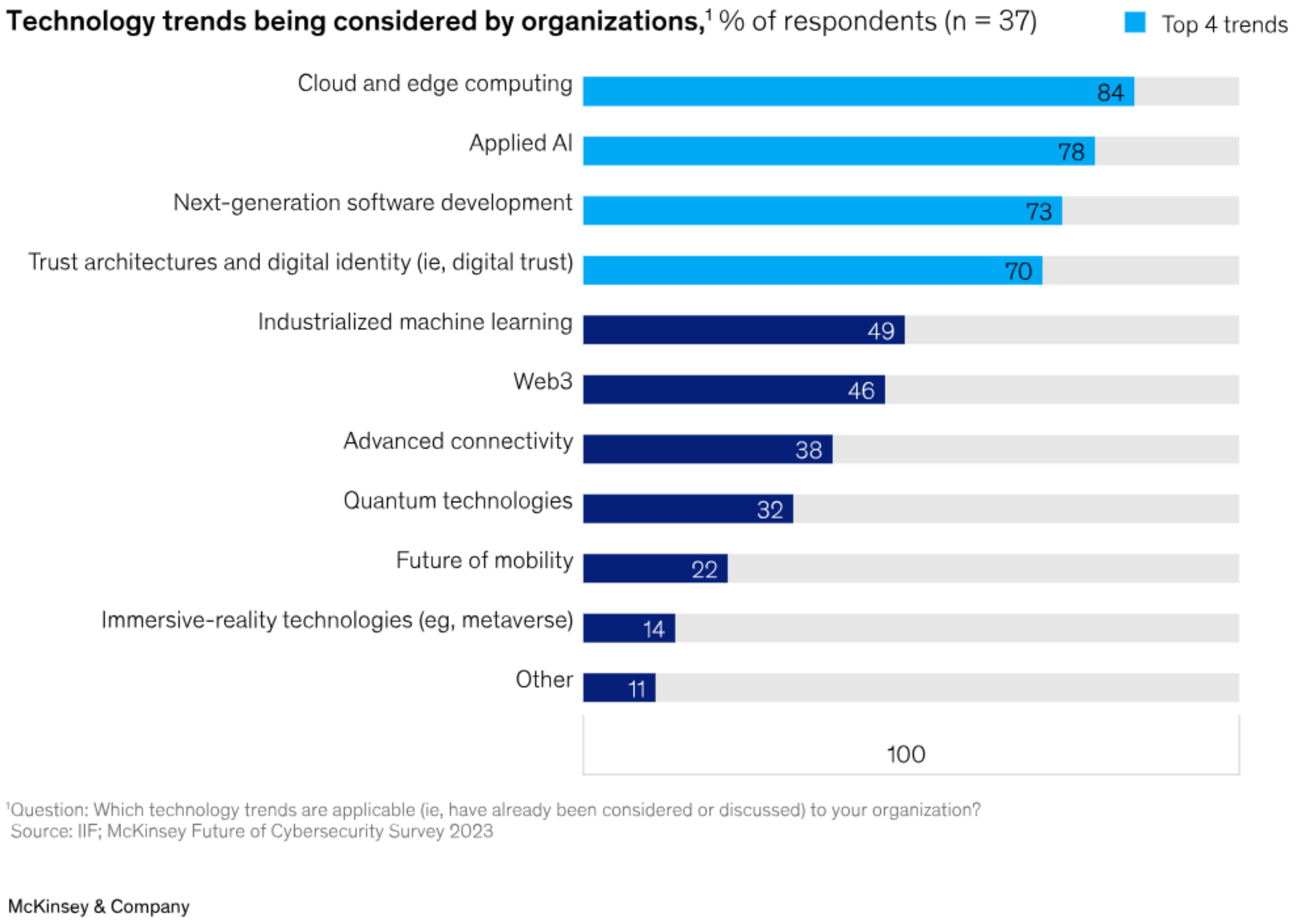

*Figure 7: Emerging Technology Risks* [76]

As attackers have become more sophisticated, it is important to further advance existing work on data breach analysis for better prevention and mitigation strategies. The next section looks at existing literature and identifies gaps to form the foundation for this research.

## 2.5. Related work

Ayyagari [2] studied data breaches that occurred between 2005 and 2011 using content analysis techniques and found that the number of hacking breaches had decreased during this period, while breaches caused by the "human element," such as employee errors or insider threats had increased. In contrast, a study by Hammouchi et al. [78] covering data breaches from 2005 to 2018 revealed that hacking was the most prevalent type of breach. A similar study was conducted by Strupczewski [79] for analysing data breaches in the United States between 2005 and 2006. The study found that over 50% of these breaches affected the healthcare sector in this period and the data breaches caused by unintentional threats were twice as frequent as those caused by malicious threats [79]. In contrast, the author highlights that more than 50% of the cyber attacks are targeted at the web services sector for stealing PII data and data breaches caused by malicious threats are more frequent than unintentional ones.

Liu et al. [19] were among the first researchers to examine the world's largest data breaches from 2004 to 2017 using visualisation techniques. Their studies revealed that the impact of data breaches goes beyond monetary value, and it affects "customer confidence, social trust and personal safety" [19]. Rodrigues et al. [80] performed exploratory data analysis on breaches from 2018 to 2019 to understand factors such as geographical location and economic sector that influence incident severity. Their study revealed that healthcare breaches were severe in countries like India owing to factors such as lack of infrastructure, privacy concerns on storage of data on the cloud, sharing of data without consent, inadequate security policies and weak security controls [80]. While the author agrees with this, it is important to note that the increase in data breaches is also attributed to the rise in ransomware attacks severely affecting countries like the United Kingdom [46, p. 41]. The author argues that attackers target healthcare sectors because disruption can put human lives at risk and affect operations severely. Hence, there are higher chances of ransom being paid. Moreover, attackers can use triple extortion tactics to target patients and build pressure on victim organisations.

Other studies on data breaches cover various aspects such as compliance and policy-based behaviour. Lack of corporate governance, risk assessment, employee training and compliance can influence actions leading to data breaches [81]. Employee behaviour is a significant factor in data breaches [82] [2], with non-compliance with security guidelines being a major cause [83] [2].

With data breaches growing over the years, many researchers have investigated the cost implications of data breaches. Campbell et al. [84] were one of the first to study the economic cost of data breaches on publicly traded US corporations. The study revealed that financial repercussions for data breaches of a confidential nature are much more than non-confidential types of events, also corporations suffering such breaches faced negative effects on market value. Cavusoglu [3] researched on impact of data breaches and found that entities affected by data breach suffered an average loss of 2.1% of market value just in two days of public disclosure. The negative impact of data breaches also includes reputational damage, customer trust and higher expenditure in improving security. It is observed that data disclosures affect the stock market adversely based on factors such as brand name, size and profit rate. For example, Gatzlaff and McCullough's [85] study on the impact of data breaches on shareholder's wealth revealed that investors penalise companies with higher valuations severely, reflecting the negative abnormal returns on the stock price.

In addition to cost implications, threat and risk factors significantly influence the severity of data breaches. Sen and Borle [86] studied theories like the opportunity theory of crime, institutional theory (organisation structure and practices) and institutional anomie theory (social norms and cultures) to identify factors influencing the contextual risk of data breaches [86]. For example, systems are more vulnerable when there is a motivated attacker and insufficient security to guard the system; organisations with dedicated institutional structures and security teams are less likely to face data breaches and organisations that prioritise profitability over security are at risk of data breach.

On the data protection and regulatory aspect, Fan [87] studied the role of data protection laws in reducing the occurrence of data breaches and the effectiveness of data regulation penalties. The study revealed that companies comply with data regulation rules if hefty penalties are imposed. Thus, stricter laws can reduce the occurrence of data breaches [87].

According to a privacy study [88], in response to data breaches on the Facebook platform, users exhibited more cautious behaviour and took protective measures such as adjusting their privacy settings. The study also pointed out the consumer privacy paradox behaviour. For instance, despite being more aware of the risks, users still prioritised the benefits of social media over perceived risks and self-disclosed information. These consumer behaviours are majorly influenced by "trust" [88]. A study by Cheng et al. [24] highlighted that trust in online services and brands significantly declines following a data breach.

Lastly, the impact of data breaches goes beyond financial losses, there is significant emotional and mental stress for individuals whose personal data has been compromised [89]. For example, Palassis et al. [90] studied the psychological impact on victims of hacking and found that violation of security and privacy leads to 1) emotional impact, 2) a sense of vulnerability and 3) a sense of violation. The coping mechanisms of such victims are influenced by stress and the type of data compromised. For example, PII breaches might trigger different coping mechanisms compared to those involving non-personal data [91].

## 2.6. How is the research extending existing work?

This research mainly focuses on descriptive analysis of data breaches using data visualisation techniques. Much of the previous research used the Privacy Rights Clearing House database for analysing data breaches which primarily contained breaches in the United States [2], [78]. This research analyses data breaches to gain a more global perspective. Liu et al. [19] data breaches from 2004 to 2017 using data visualisation, this research extends their work for a more current analysis. The previous study was limited to data breaches with over 30,000 records lost [19]. For a broader analysis, the research uses two public datasets. Moreover, previously used data visualisation techniques were static and lacked interactivity. In contrast, this research utilises an advanced and diverse visualisation tool to reveal deeper insights and provide a better user experience. The research looks into cross-tab analysis to identify the relation between targeted industries, attack methods, impacted regions and data compromised. Furthermore, the research uses time series analysis with a drill down capability to uncover hidden insights.

## 2.7. What is new about the research and its significance?

The purpose of this research is to address gaps identified in previous studies (refer to Chapter 2, Section 2.5) regarding data breaches and refine research questions for a comprehensive analysis. The research uses two datasets for a broader coverage of data breaches and the dashboard operates in real-time, which means whenever data updates on the designated websites, it will automatically reflect in the dashboard. The data refresh frequency can be scheduled in Power BI Service [92] (a cloud-based platform for publishing and sharing dashboards), to automatically load new data without any manual intervention [92]. Upon refresh, all visuals will update on their own to display real-time output eliminating the need to recreate reports and dashboards every time for new inputs. The advantages of this approach to studying data breach patterns are significant. Even if one of the public datasets is not updated regularly, the dashboard will still display distinct data breach records loaded from the other dataset, thus providing a comprehensive view of data breach patterns over the years and in years to come.

## 2.8. Conclusion:

This chapter delved into the theoretical foundations of data breaches, covering definitions, types of compromised information and the motivations of threat actors. It examined various attack vectors, such as ransomware, hacking, phishing, supply chain risks, security misconfigurations and others. Additionally, the author reviewed related literature on exploratory analysis of data breaches to identify gaps in existing literature. The author extends previous research by examining data breaches from 2004 to 2024 using Power BI. The data analysis methodology will be discussed in the next chapter which looks into the data sourcing, preparation and visualisation stages using Power BI.

# 3. Methodology: Data analysis in Power BI

## 3.1. Dataset Description:

The dashboard is compiled from two data sources, Information is Beautiful: World's Biggest Data Breaches & Hacks - Source A and Wikipedia: List of Data Breaches - Source B. The following data fields are used for data analysis: -

**Entity:** The name of the company that experienced a data breach.

**Sector:** The industry or sector to which the organisation belongs.

**Method:** The method of attack used by threat actors to execute the data breach.

**Data sensitivity:** The type of data compromised in the breach.

**Records:** The number of records compromised in the breach.

**Year:** The year in which the data breach was reported.

**Region:** The geographical location where the data breach occurred.

Other data fields such as alternate organisation name, source name URL, and data breach description are not included in the dashboard. Source A and Source B were last updated in June 2024. Any data breaches reported after this date are not included in the dashboard for this research.

## 3.2. Ethical consideration:

The datasets used are publicly available and do not contain any personal data. The research is conducted taking into consideration the ethical standards set by Royal Holloway University of London. The research does not include any research participants for interviews or surveys. The research results and outputs are intended to share insights into data breach patterns and provide recommendations for better prevention and mitigation strategies. Though the dataset captures the company names that were breached, it does not contain any information that discloses confidential information. The data sensitivity field only reveals the type of information compromised. It does not contain any personally identifiable data that could identify the affected individuals. The software used for analysis is Power BI, which is a visualisation and reporting tool. It does not have any harmful impact if used outside of a controlled environment. The tool only loads data in Power BI Desktop in a read-only mode; any manipulation or data cleaning done does not affect the actual data source.

## 3.3. Data analysis tool:

Microsoft Power BI tool is used to create the data breach dashboard. Figure 8 shows Power BI architecture. Power BI collects data from various sources and loads it into the Power BI desktop for data transformation and report creation. Once the report is created, it can be published to Power BI service, a SaaS platform for sharing and collaborating with users.

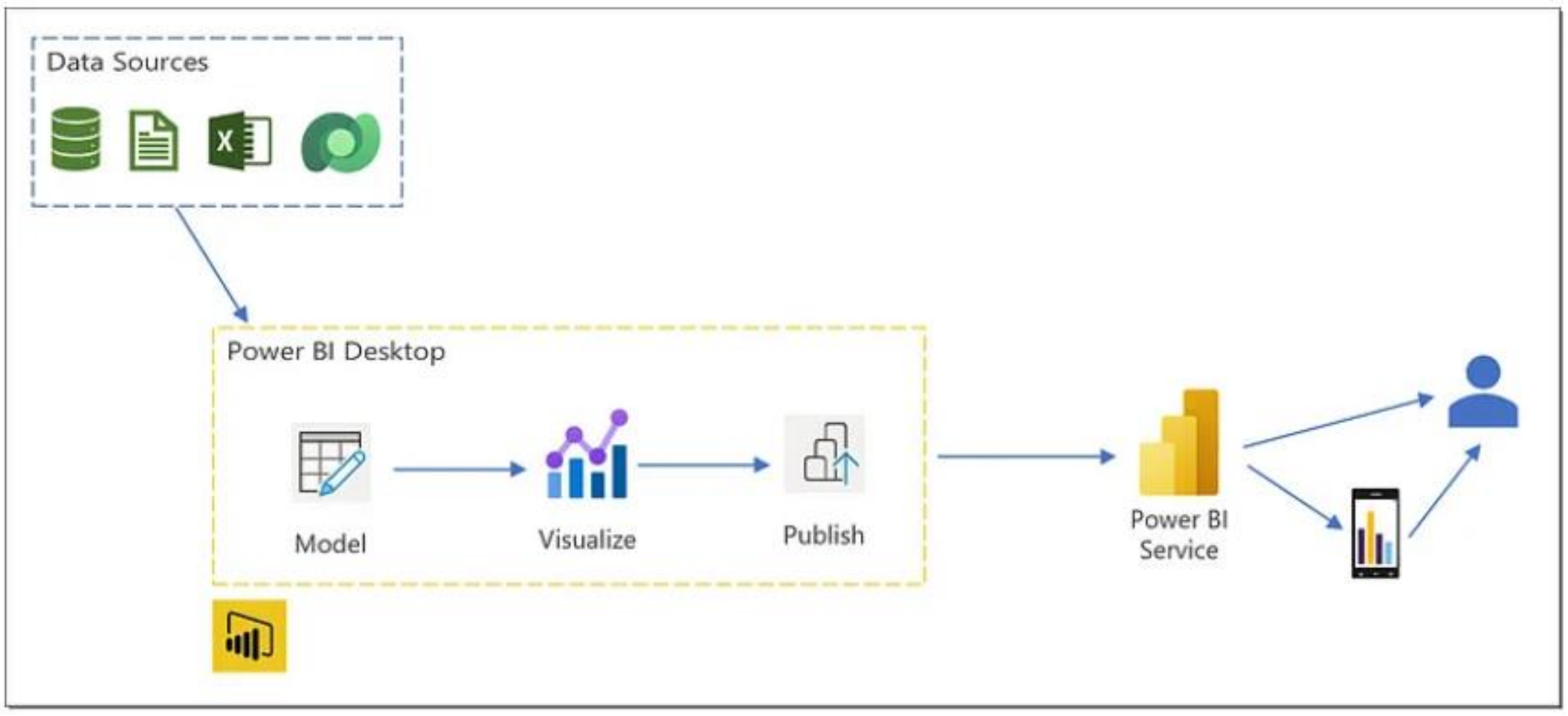

*Figure 8: Power BI architecture* [93]

Previous researchers have used Python [80] and Tableau [94] to create visuals which have certain shortcomings such as coding requirements, performance issues with larger datasets and integration with other platforms. The author chose Power BI because it is simpler for general users or experts to interact with complex data without much knowledge or understanding of data analysis. Interactive visuals, charts and graphs make patterns and relationships easy to understand. Users can explore data insights on their own without having to write complex queries. Moreover, users can easily change visuals from one form to another with simple clicks. Also, it is easier to embed Power BI visualisations and dashboards into other applications or websites.

## 3.4. Data Analysis

The research methodology employed is a descriptive analysis using data visualisation techniques. Figure 9 shows the stages of data analysis in Power BI, aligning them with the research methodology. Manipulation checks were performed to check the validity of the data sources by randomly selecting the media sources linked to the data breaches and verifying their authenticity. In all such cases, the data breach details were found to be accurate, making these two data sources suitable for conducting data breach studies.

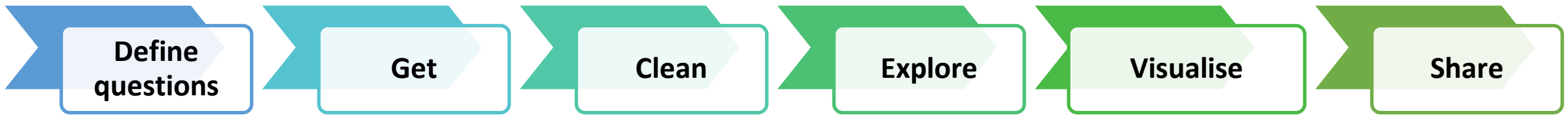

*Figure 9: Data Analysis Stages in Power BI*

### 3.4.1. Data sourcing

Power BI offers to connect with a variety of data sources such as Google Worksheet, Excel, online websites, SharePoint, SQL, MySQL databases and others [95]. As shown in Figure 10, Source A is connected via Google Worksheet and Source B is connected using a website connection.

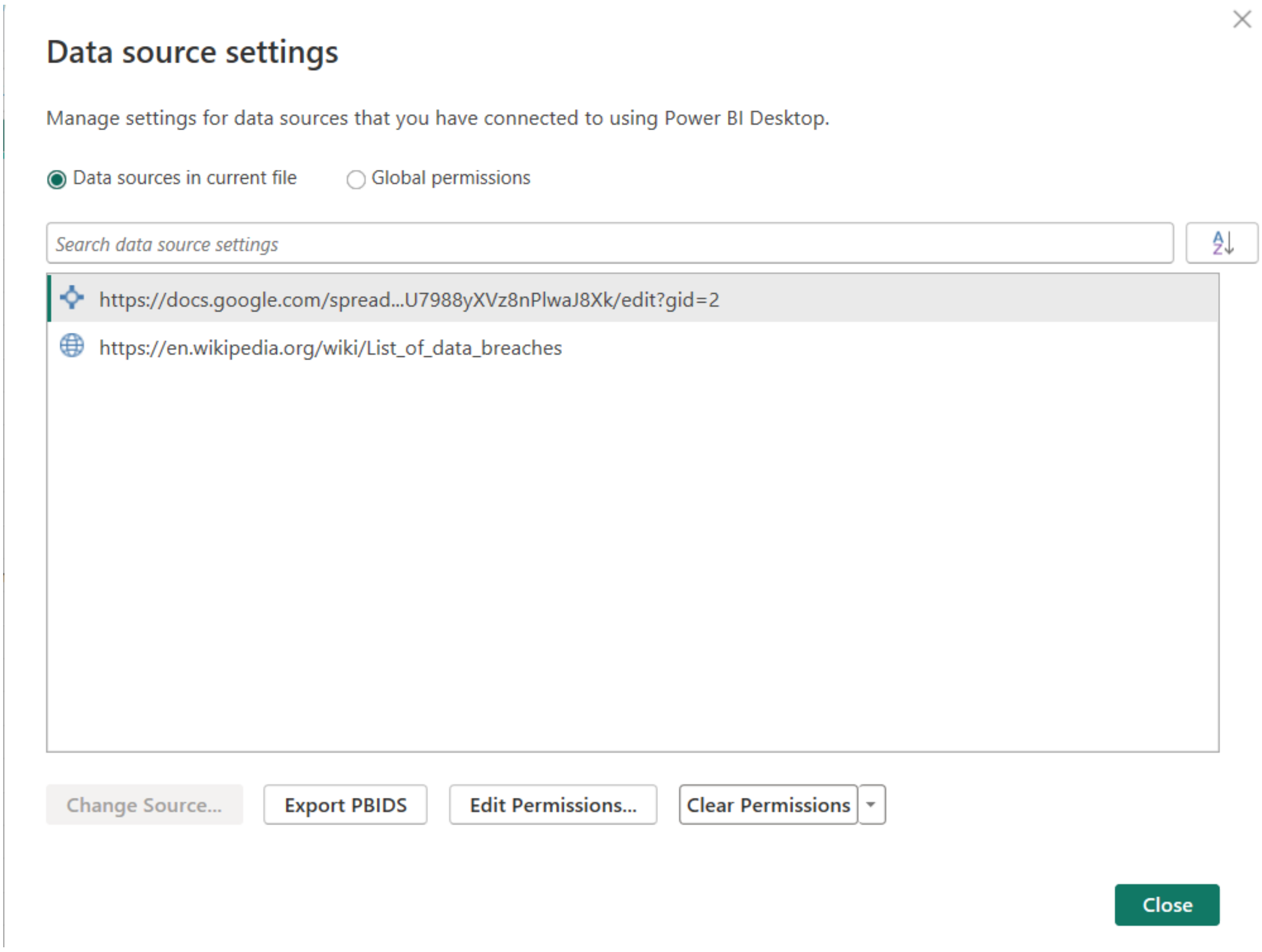

*Figure 10: Data Source Setting in Power BI*

### 3.4.2. Data cleaning

Categories for industries and attack methods are assigned as best fit. All typo errors or redundant groupings are corrected. For instance, industry categories such as finance, financial and QR code payment are mapped to "Financial service". All duplicate values are removed from both data sets based on entity, year and records fields. Entries with errors or missing values are filtered out from the visuals for better data quality and reliability. For regional analysis, country and continent are assigned referring to [96].

### 3.4.3. Data exploration and modelling

"Append Queries" and "Merge Queries" are used to manipulate data. "Append Queries" vertically combines multiple datasets into one complete dataset [97], whereas "Merge Queries" horizontally populates values based on a matching column [98]. As shown in Figure 11, source A and two other data tables from Source B (data breaches involving companies and data breaches involving government and public entities) are combined using Append Queries to create a resulting dataset named "Complete Dataset". The "Attack Method" and "Industry" columns in the "Complete Dataset" table are populated from the "Method Classification" and "Industry Classification" tables by removing redundancy and manually assigning values as best fit. After pre-processing steps, that is combining the three data tables, the resulting table returned 600 unique records which are used for the analysis of data breaches.

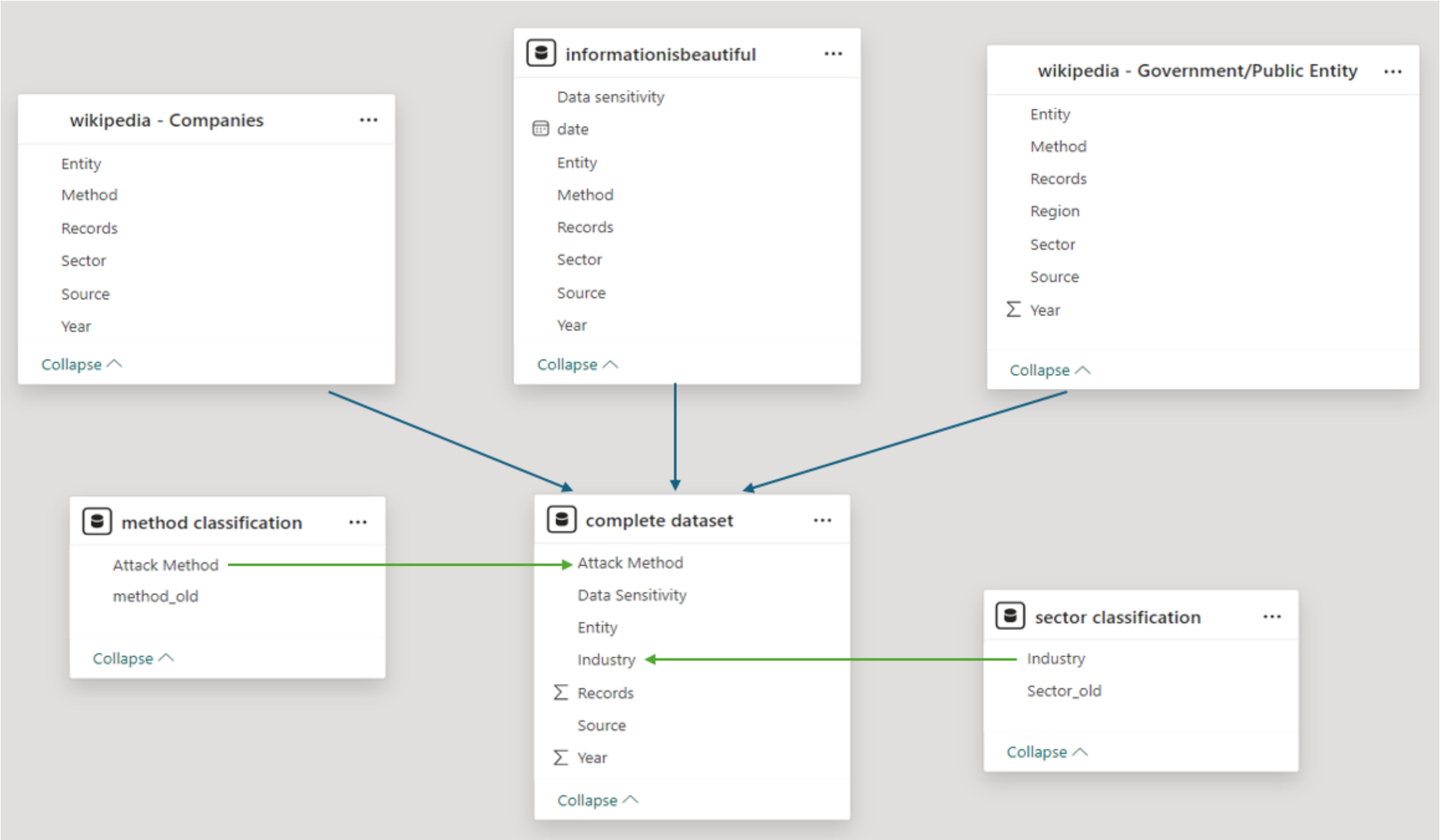

*Figure 11: Data modelling diagram for the complete dataset*

For the data sensitivity report, source A acts as the main data source. As shown in Figure 12, government and public entities are selected from Region A and are combined with Region B to produce 120 unique records. The continent and country fields are manually populated from the "Region Classification" table based on the region column for Region B and the entity column for Region A.

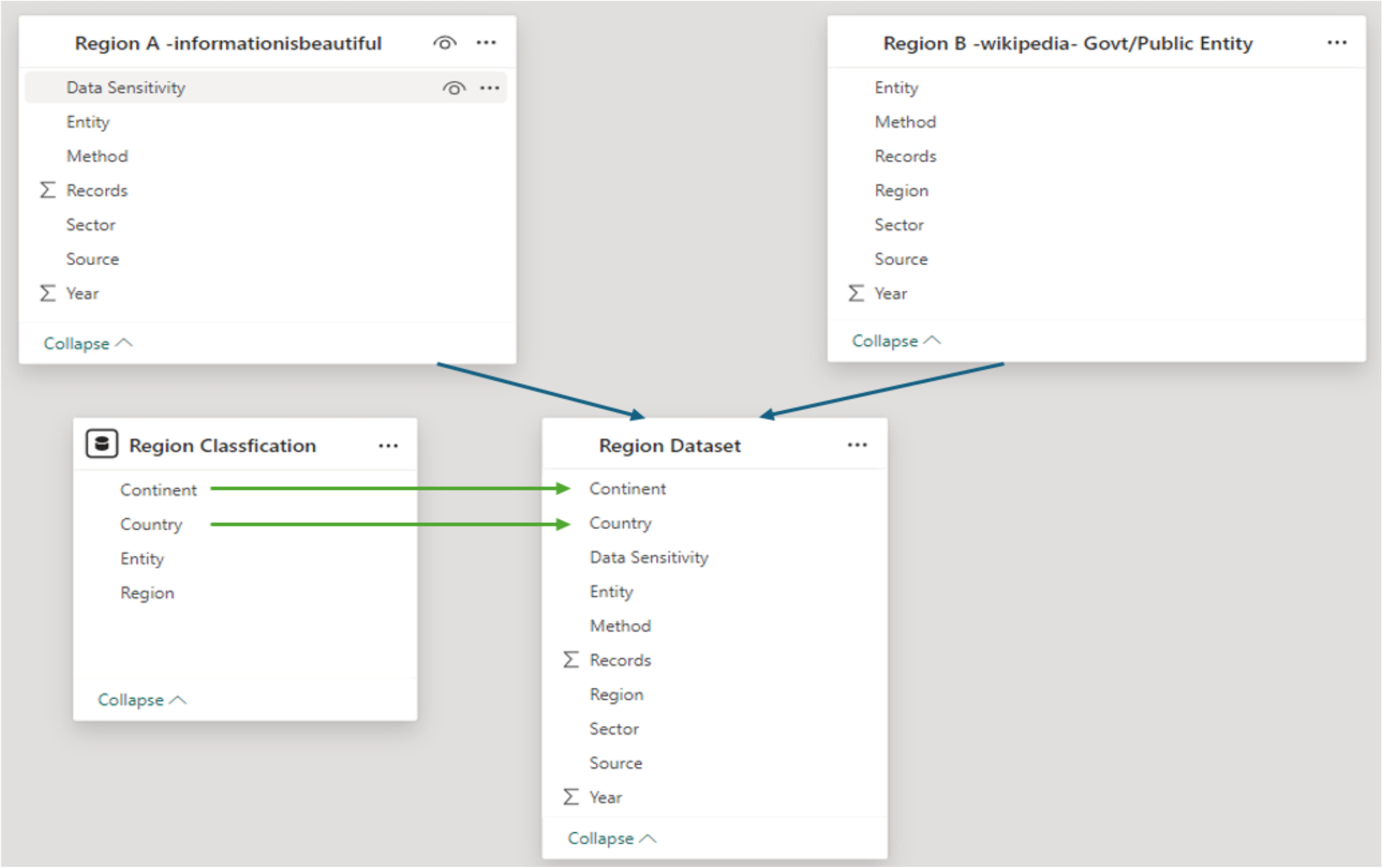

*Figure 12: Data modelling diagram for regional analysis*

Table 5 summarises the records obtained after the pre-processing steps and the datasets that have been used to answer the research questions.

| **Data Sources** | **Source A (World's Biggest Breaches and Hacks - Information is Beautiful)** | **Source B (Wikipedia - List of Data Breaches)** | | **Power BI -Result Dataset** |
|---|---|---|---|---|
| **Industries, Attack Method, Time Series and Cross-Tab Analysis (RQ 1, RQ 2, RQ 5, RQ 6)** | | | | |
| **Data Tables** | informationisbeautiful | Wikipedia-Companies | Wikipedia-Government /Public Entity | complete dataset |
| **Entities Compromised** | 485 | 347 | 80 | **600** |
| **Data Sensitivity Analysis (RQ 3)** | | | | |
| **Data Tables** | informationisbeautiful | - | - | - |
| **Entities Compromised** | **485** | - | - | - |
| **Regional Analysis for Government/ Public Entity (RQ 4)** | | | | |
| **Data Tables** | informationisbeautiful (Region A) | - | Wikipedia-Government /Public Entity (Region B) | Region dataset |
| **Entities Compromised** | 78 | - | 80 | **120** |

*Table 5: Mapping Data tables with the research questions*

### 3.4.4. Data Visualisation

The focus of data visualisation methodology is to tell a story by focusing on information that is important. McCandless [99], a famous data journalist and designer of the popular blog "Information is Beautiful" says "**While it is often said in the industry that "Data is the new oil", he prefers to say, "Data is the new soil". It is a fertile creative medium and data visualisations or infographics as the flowers blooming from this medium**"[99]. Data visualisations combine the language of the brain, which understands numbers and facts, with the language of the eyes, which perceive colours, to reveal interesting patterns that are easily understandable [99]. Inspired by this approach, the author created Power BI visuals with a consistent colour scheme and theme for easy pattern recognition. The author chose visual types based on best practices recommended for data visualisation by [100]. For example, the following data graphs are used for consistent reporting:

**Bar graphs** illustrate the distribution of data breaches across different categories.

**Line graphs** track data breach trends over time.

**Bubble charts** cluster data categories by the size of records compromised.

**Table Matrix** presents detailed data in a tabular format to see granular details.

**Map view** provides a region-based analysis of data breaches.

**Pie charts** show the proportion of different data breach categories.

**Heat maps** visualise the intensity of data breaches across different categories using colour gradients.

**Stacked charts** illustrate the composition and distribution of data breaches over multiple categories helping to understand both individual contributions and the relationship between these categories.

## 3.5. Components of the dashboard

The dashboard includes the following pages

**Home:** Overview of dashboard components and its limitations.

**Industry:** Report shows distribution of data breaches by Industry highlighting sector specific risks.

**Attack Method:** Report shows the distribution of data breaches by attack method, highlighting the prevalent ones.

**Data Sensitivity:** Report shows the distribution of data breaches by data Sensitivity, shedding light on the types of data at the risk of compromise.

**Time Series:** Report shows data Breach trends over time.

**Cross Tabs:** Report shows the relation between data breach categories.

**Bubble Chart:** Report showing records lost by Entity and data breach categories for visualising entities based on the size of the records compromised.

**Region:** Report showing distribution of data breaches by Country and Continent for regional analysis.

Screenshots of the dashboard components are provided in Appendix A. Appendix B includes screenshots of the dataset and pre-processing steps.

## 3.6. Conclusion

"**Dataset changes your mindset; if it can do that, it can change your behaviour. It is a ubiquitous resource that can shape innovation and new insights**" - David McCandless [99].

This chapter broadly looked at the research methodology for analysing data breaches and the steps involved in transforming the data for creating meaningful visuals. The chapter further provided insights into visual analytics and outlined best practices for creating visuals. The next chapter summarises the findings derived from the visuals and highlights key observations for a detailed analysis of data breaches.

# 4. Findings: Data breach insights and trends

## 4.1. Objective 1 - Data breach overview

This section presents the findings related to the first objective of this study i.e. to provide an overview of affected industries, attack methods, data sensitivity and regions to understand the overall landscape of data breaches..

### 4.1.1. Industry-Specific Data Breach Trends

Figure 13 shows the number of data breaches by industry and Figure 14 shows the proportional representation of these breaches. The visuals are filtered to show the top 10 industries. Web services, healthcare and government are the top three industries affected by data breaches. Of this Web services sector is the most impacted accounting for over 23% of all breaches. Other prominent sectors include healthcare (14%), government (13%), financial services (12%) and retail (11%). Comparatively fewer breaches are observed in academic and gaming Industries.

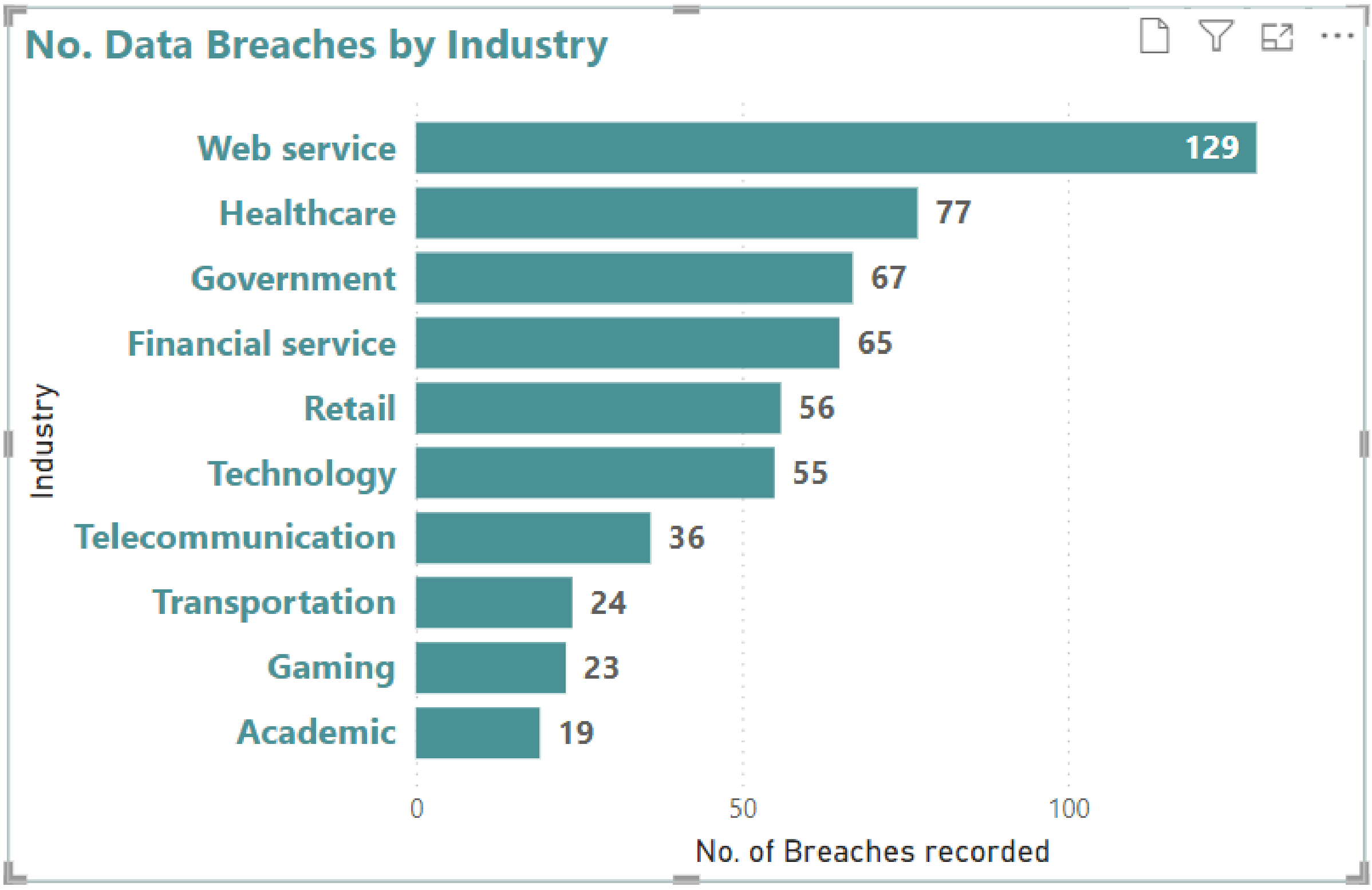

*Figure 13: No. of Data Breaches by Industry (Top 10)*

Other prominent industries contributing to less than 2% of all data breaches included social media, entertainment, hospitality, non-profit organisations, online marketing and energy sectors.

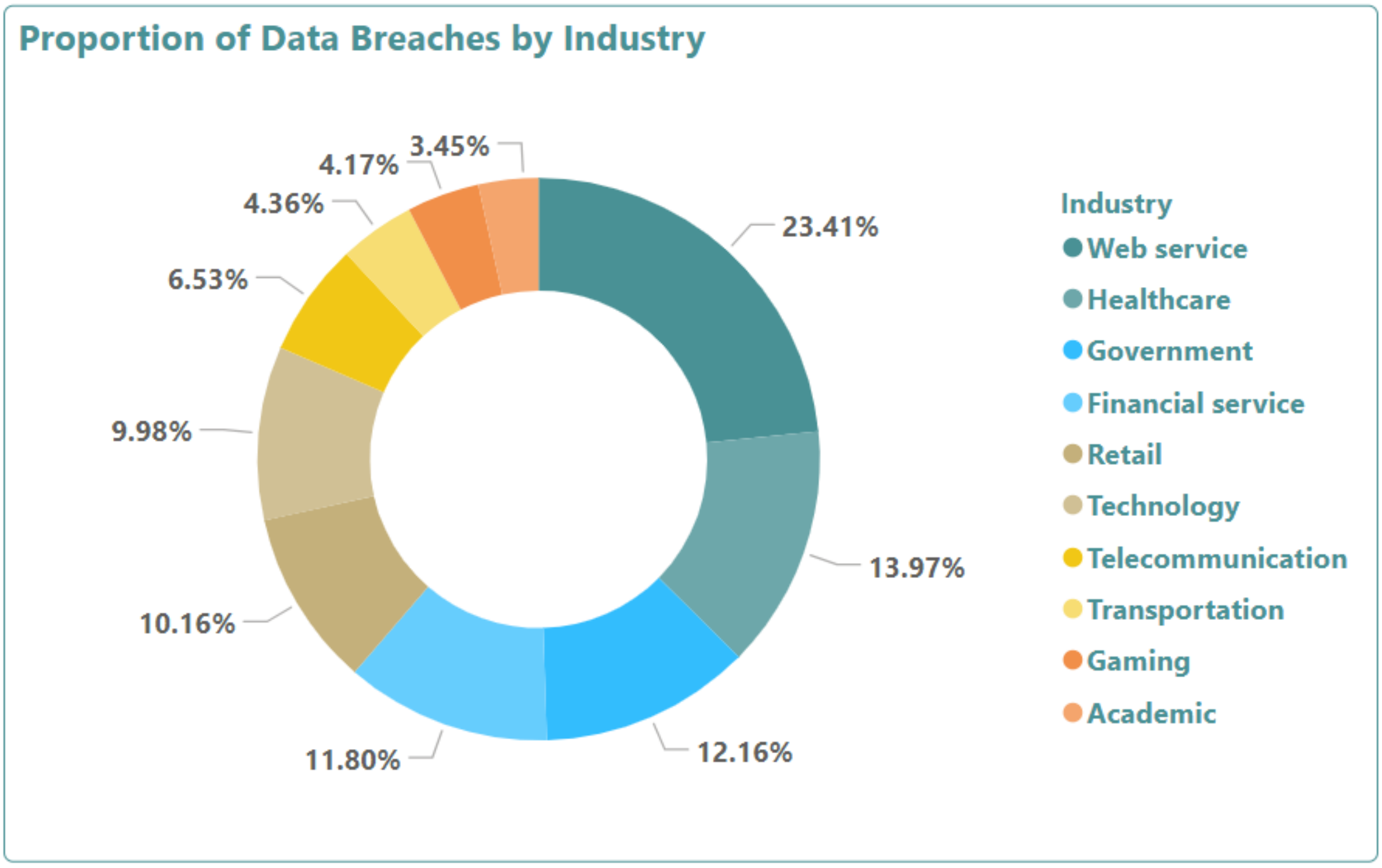

*Figure 14: Proportion of Data Breaches by Industry (Top 10)*

The total number of records lost amounts to nearly 30 billion. Web services accounted for almost 53% of the highest records compromised followed by technology, financial services and telecommunication, each contributing nearly 13% to the breaches (refer to Figure 15). One interesting point to note here is that although online marketing had fewer breaches, it still ranks among the top 5 industries with 2 billion compromised records. This was mainly attributed to the data breach of Verifications.io (a total of two leaks) as seen in Figure 16.

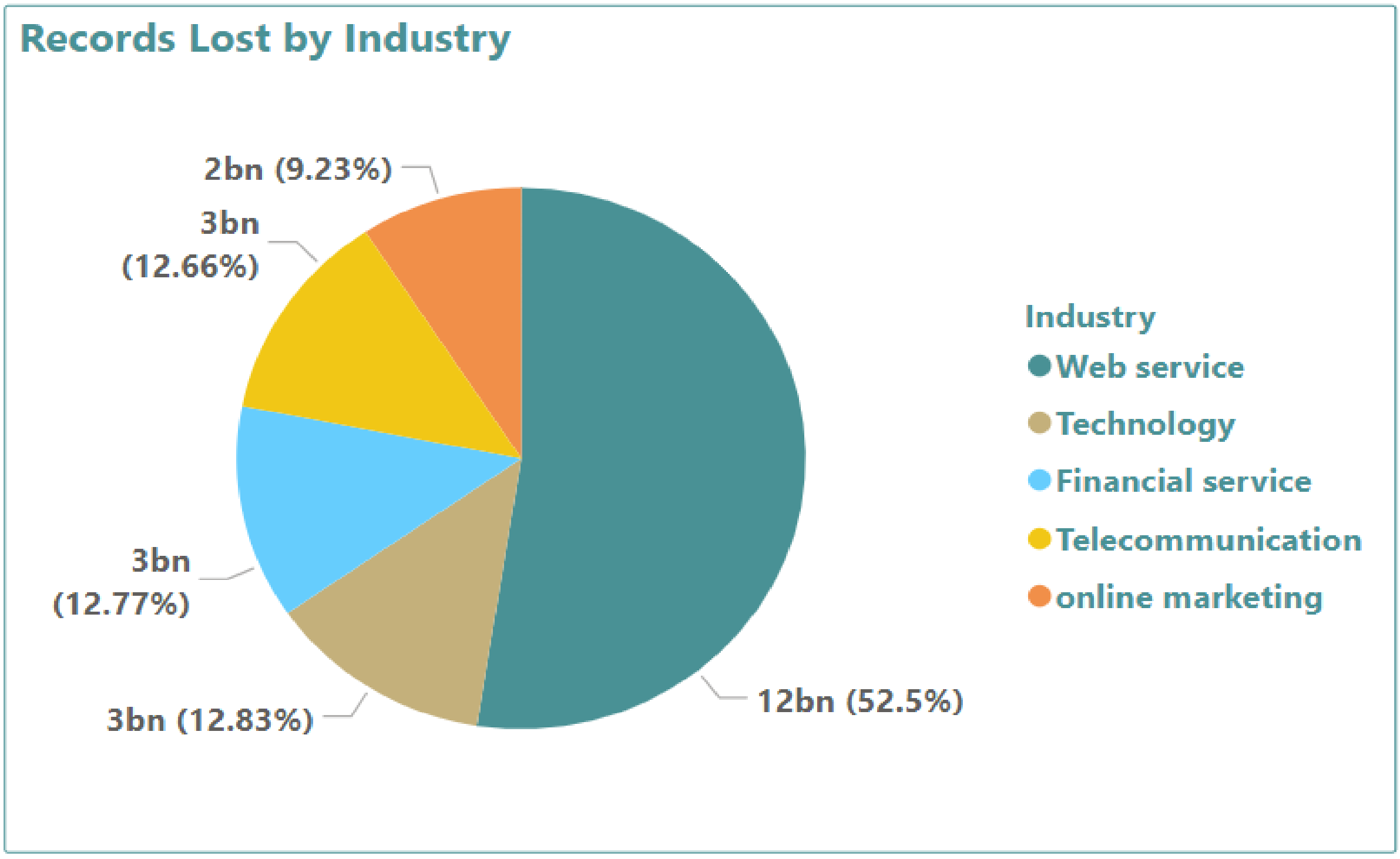

*Figure 15: Records Lost by Industry (Top 5)*

Web services breaches were more devastating and some of the biggest companies affected were Yahoo, LinkedIn, Friend Finder Network and Spambot. As seen in Figure 16, the technology sector had the second biggest share of records compromised, contributed by companies such as Facebook, OxyData and Twitter. This was followed by telecommunication giants such as Indonesian SIM cards, Syniverse, Truecaller and Airtel. Under the government sector, the Aadhar card breach of Indian citizens tops the list. Another significant entity seen is Marriot International, representing the retail sector.

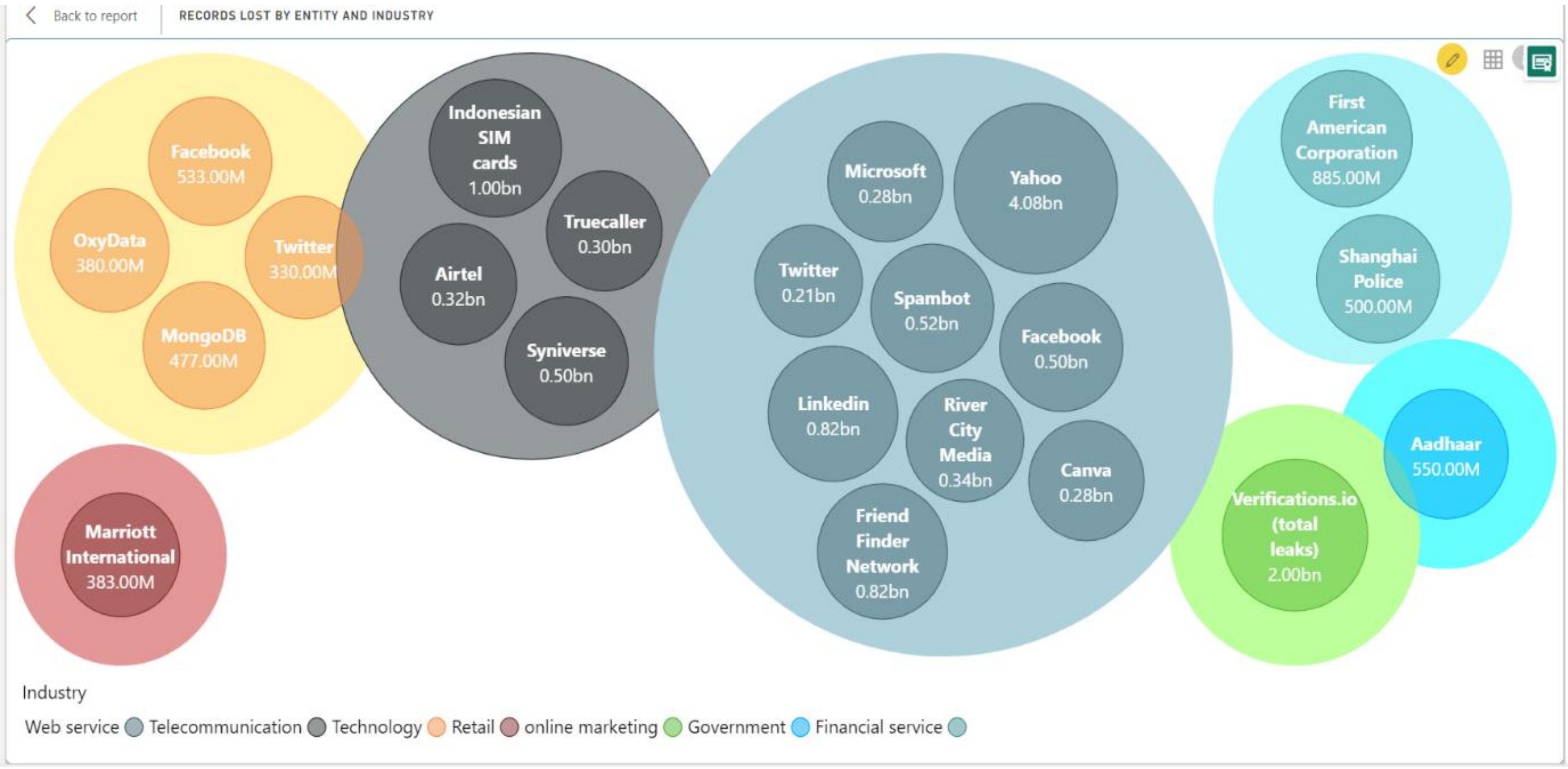

*Figure 16: Records Lost by Entity and Industry*

A recent analysis of data breaches in 2024 (refer to Figure 17) indicates that the technology sector is a primary target with notable breaches such as Dell (49 million records) and Trello (15.11 million records). Following closely are the government and media/entertainment industries. Major incidents contributing to this trend include Ticketmaster (560 million records) and AT&T (73 million records). Additionally, third-party breaches are increasingly common, as evidenced by data breaches at the UK Ministry of Defence and Santander.

Of all the data breaches over two decades, Yahoo still holds the record for the largest breach in history, with 4.08 billion records lost (see Figure 18). This breach is more than twice the size of some other major incidents, such as Verification.io (2 billion records) and Facebook (1.03 billion records).

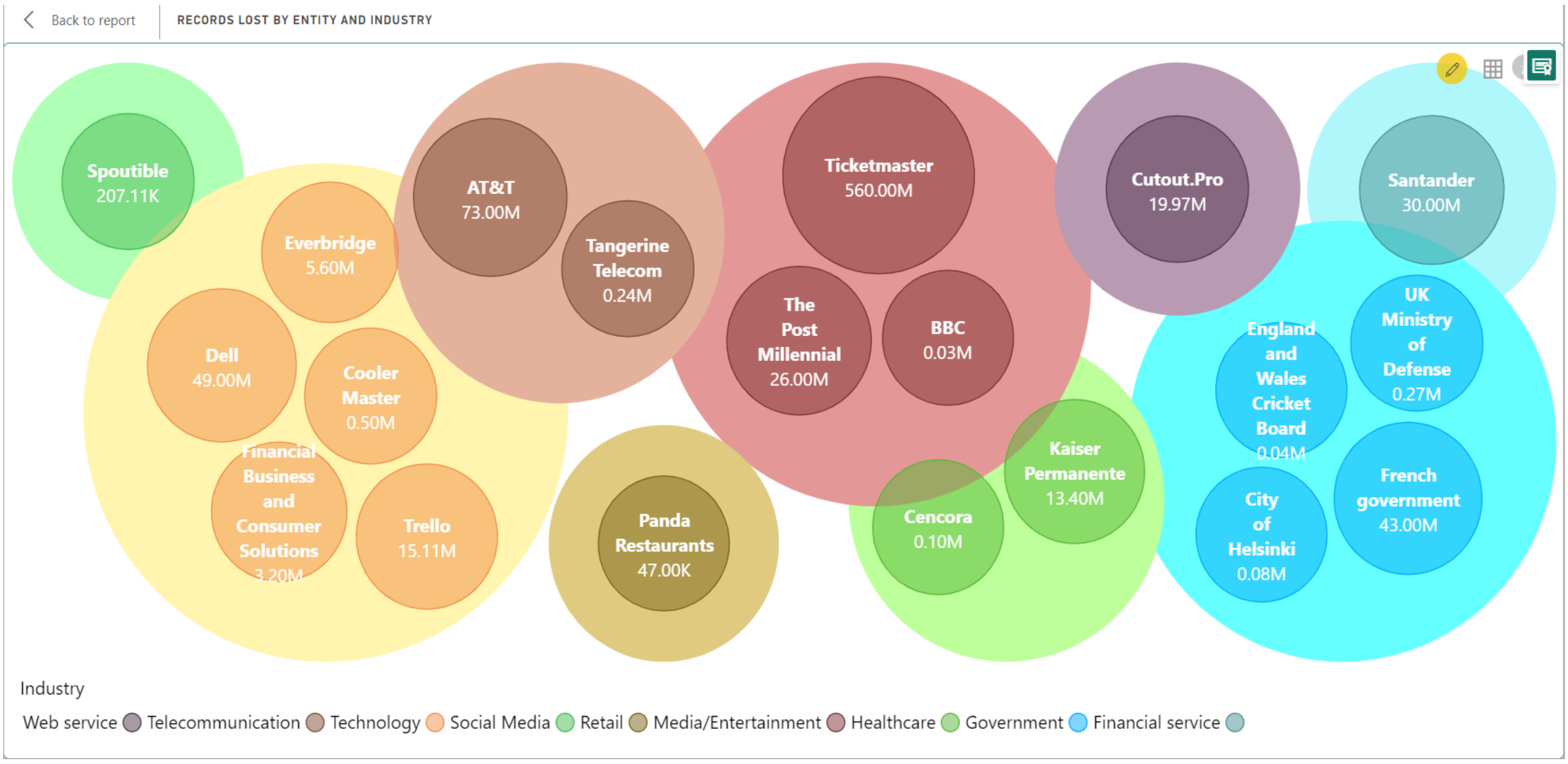

*Figure 17: Record Lost by Entity and Industry in 2024 (Top 20)*

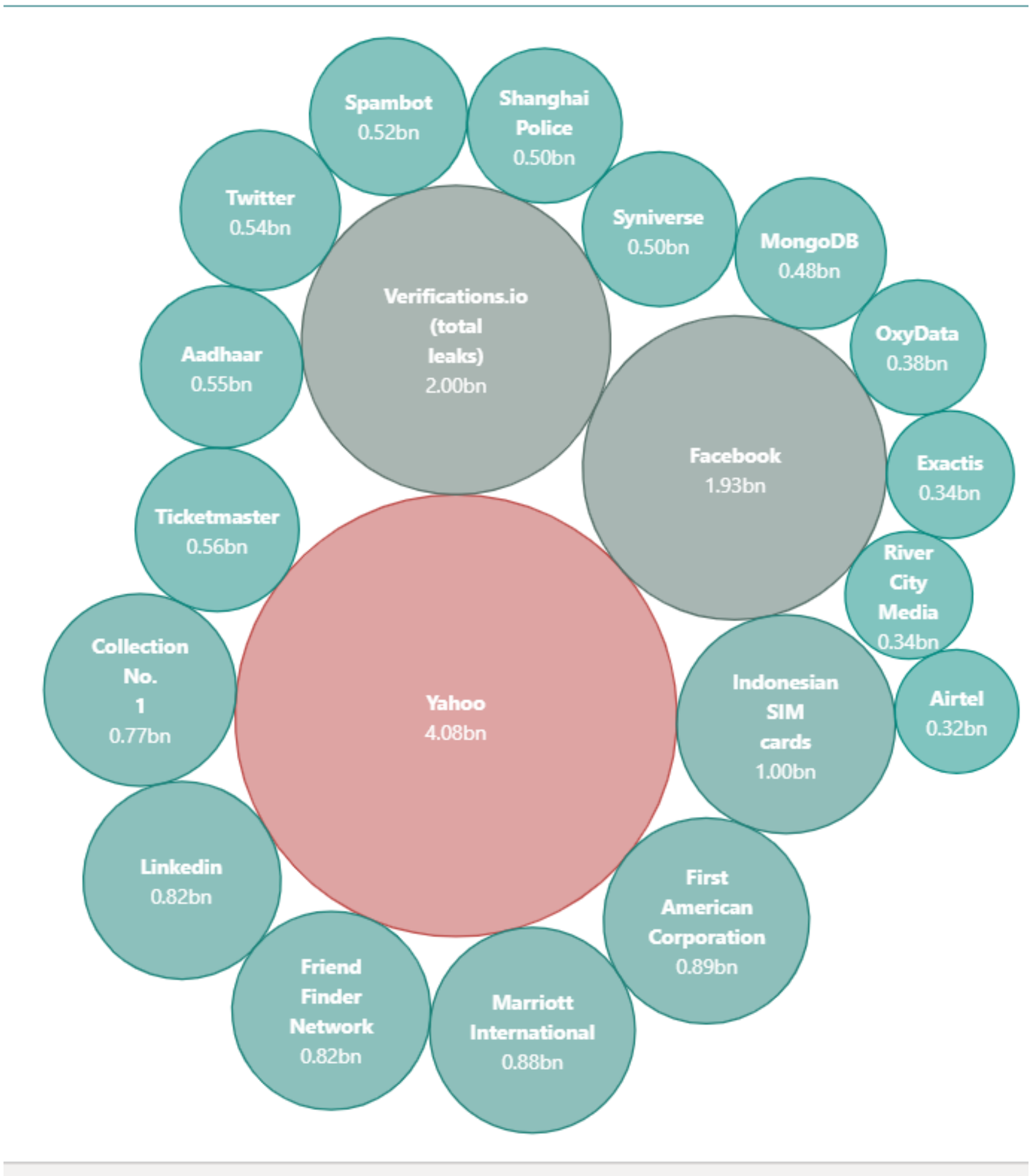

*Figure 18: Records Lost by Entity (Top 20)*

### 4.1.2. Prevalent Attack Methods

Figure 19 shows the distribution of data breaches by attack method. Hacking accounts for most data breaches at 69%, followed by poor security and lost/stolen devices (refer to Figure 20). Other methods like insider threats, misconfiguration and accidental exposure contribute to a much smaller share.

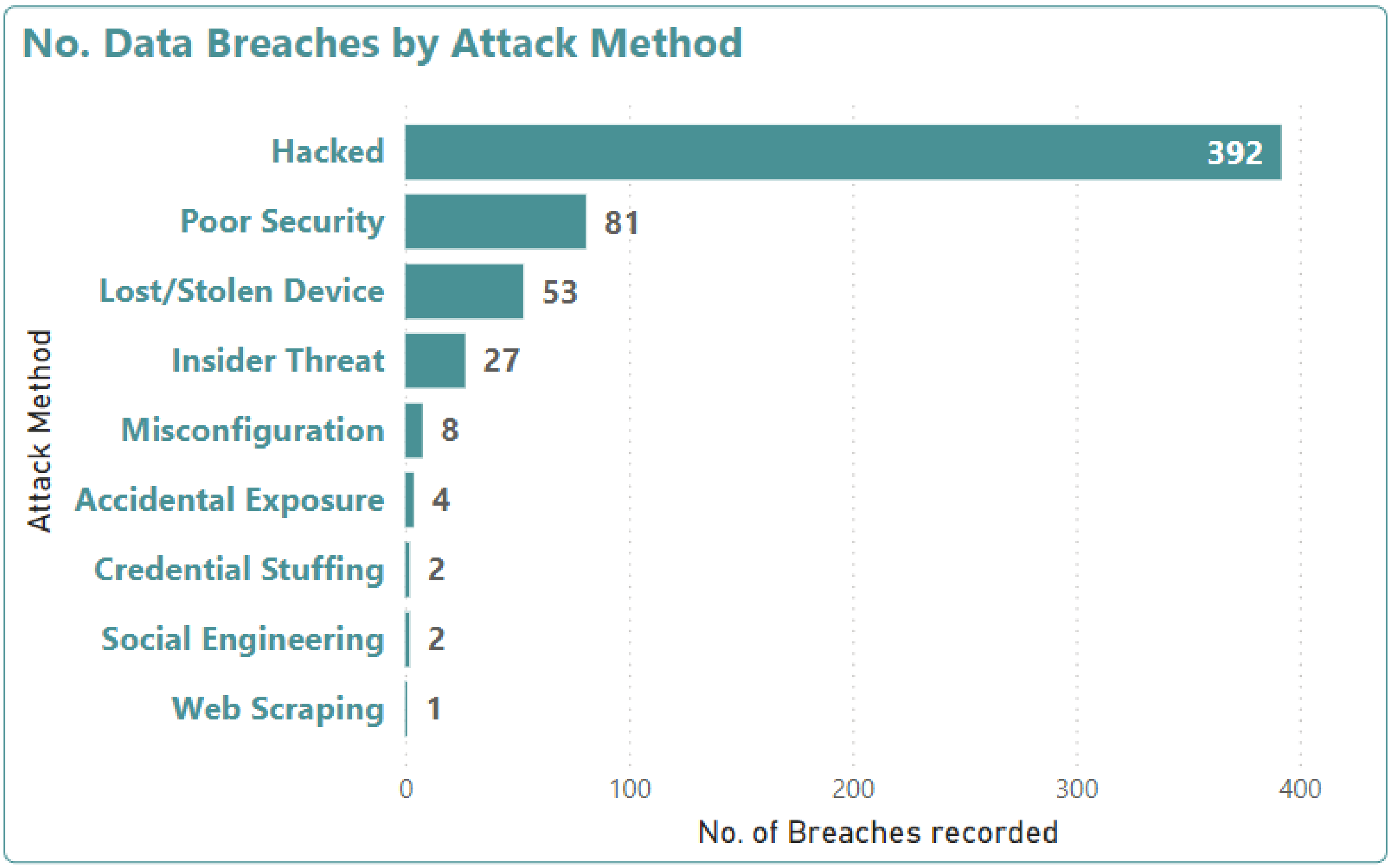

*Figure 19: No. of Data Breaches by Attack Method*

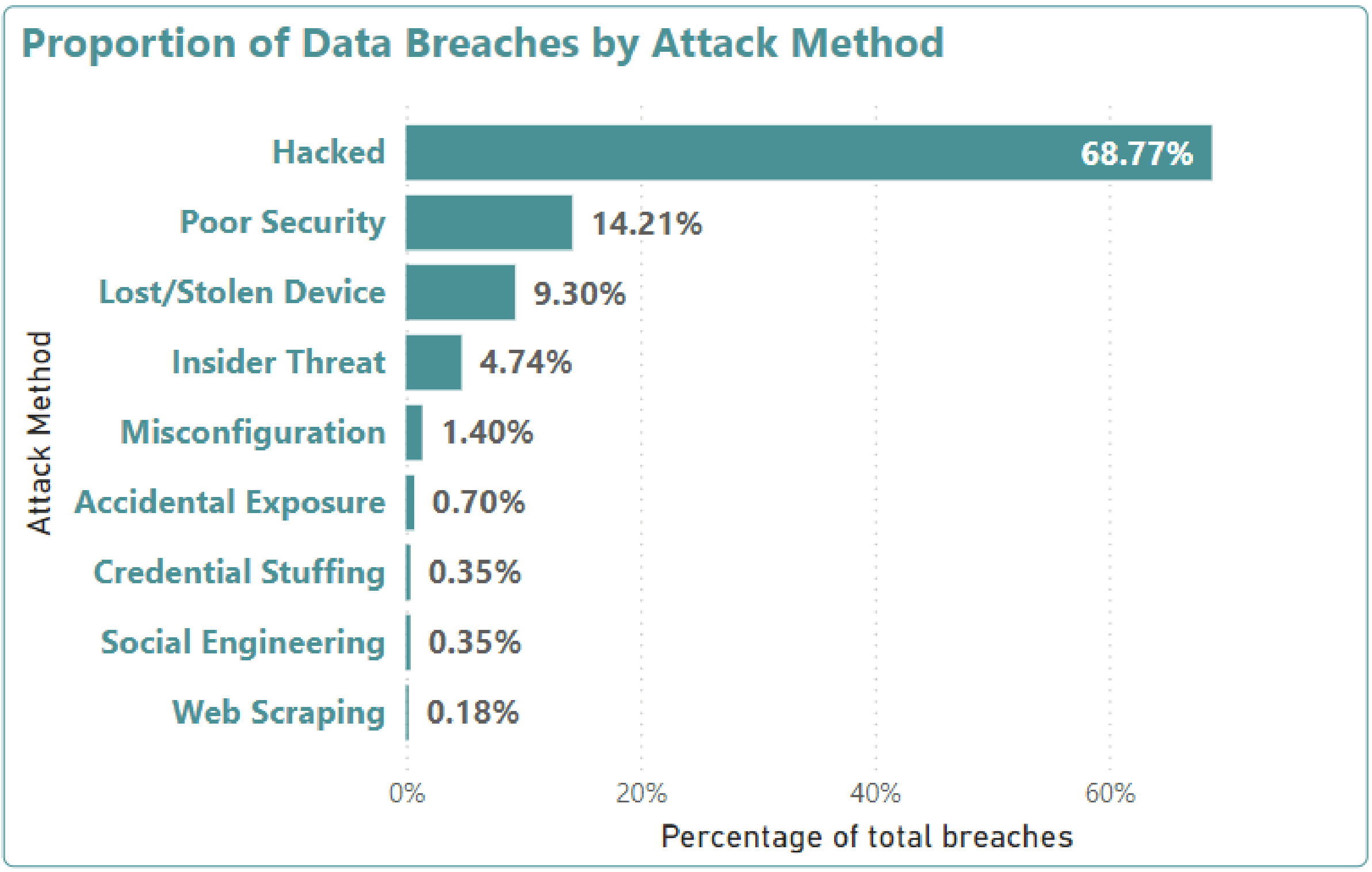

*Figure 20: Proportion of Data Breaches by Attack Method*

Hacking breaches have the highest impact totalling 17.5 billion records, followed by poor security breaches which account for 10.4 billion records. Although lost or stolen devices were historically the third most common cause of data breaches, it is now observed that credential stuffing and insider threats compromise more records compared to lost or stolen devices (refer to Figure 21).

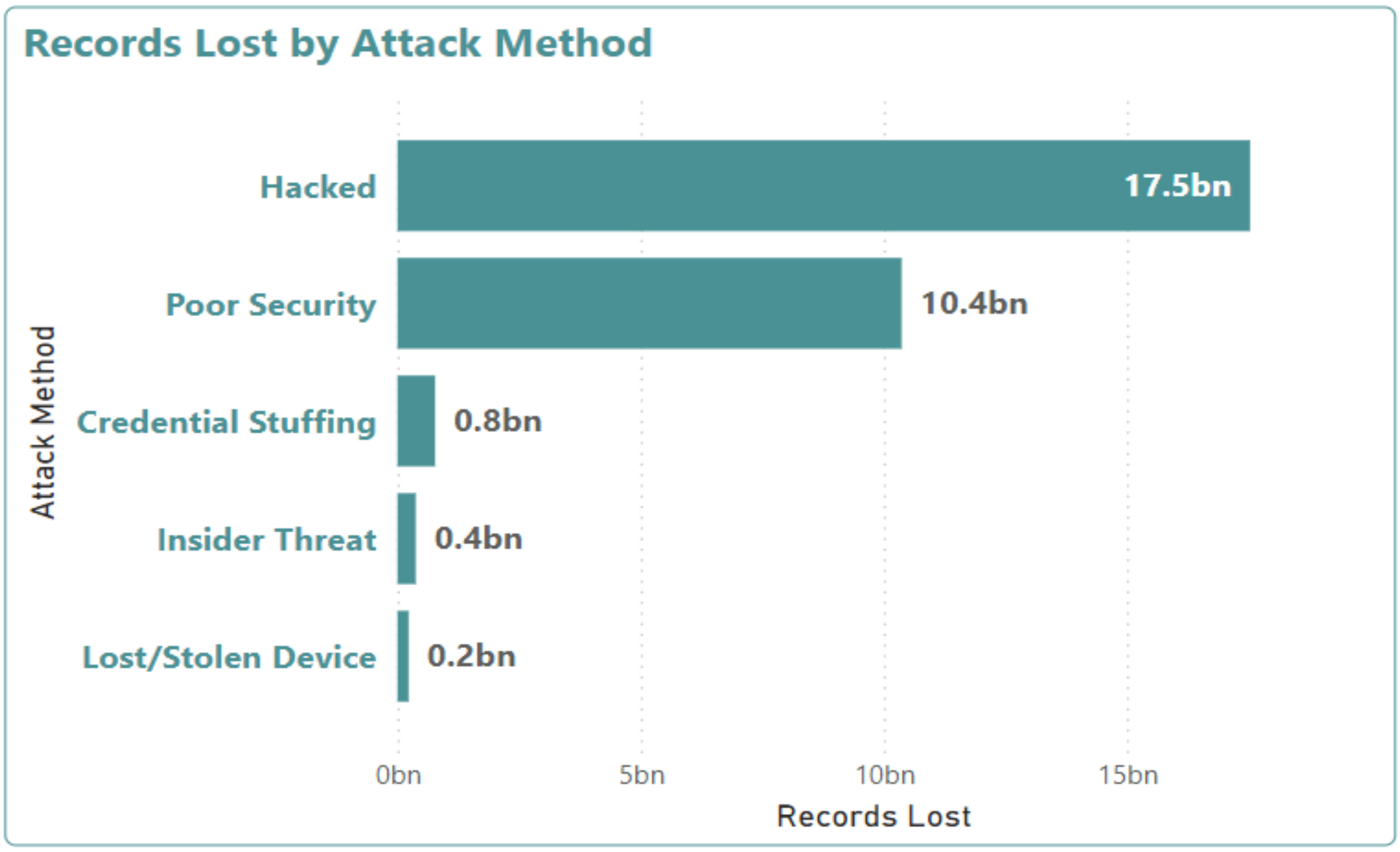

*Figure 21: Records Lost by Attack Method (Top 5)*

Hacked entities like Yahoo and Indonesian SIM cards suffered the largest breaches with records lost in the billions. Poor security breaches significantly affected entities like Facebook, Verifications.io, First American Corporation and Aadhaar, while credential stuffing and accidental exposure also resulted in substantial data losses (refer to Figure 22).

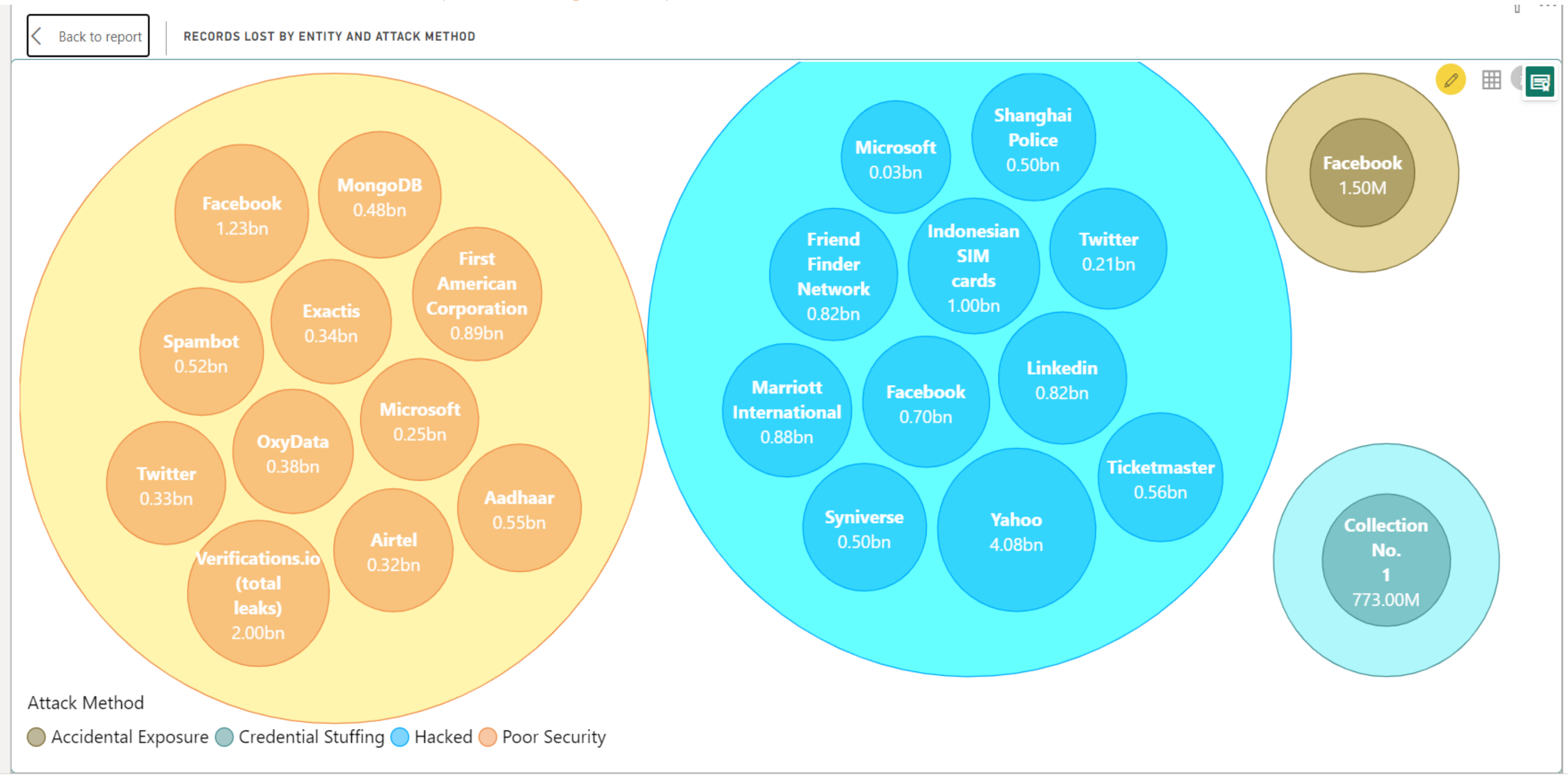

*Figure 22: Records Lost by Entity and Attack Method*

### 4.1.3. Type of Data Compromised

Figure 23 shows the type of data compromised. As shown in Figure 24, threat actors primarily targeted Personally Identifiable Information (PII) which accounted for 35% of the data breaches. Other major types of compromised data were email IDs and other non-sensitive details that are easily available online. Credit card details and health and other personal records accounted for nearly 15 - 16% of the data breaches. Data breaches compromising full details of employees/clients and other company confidential information were relatively low.

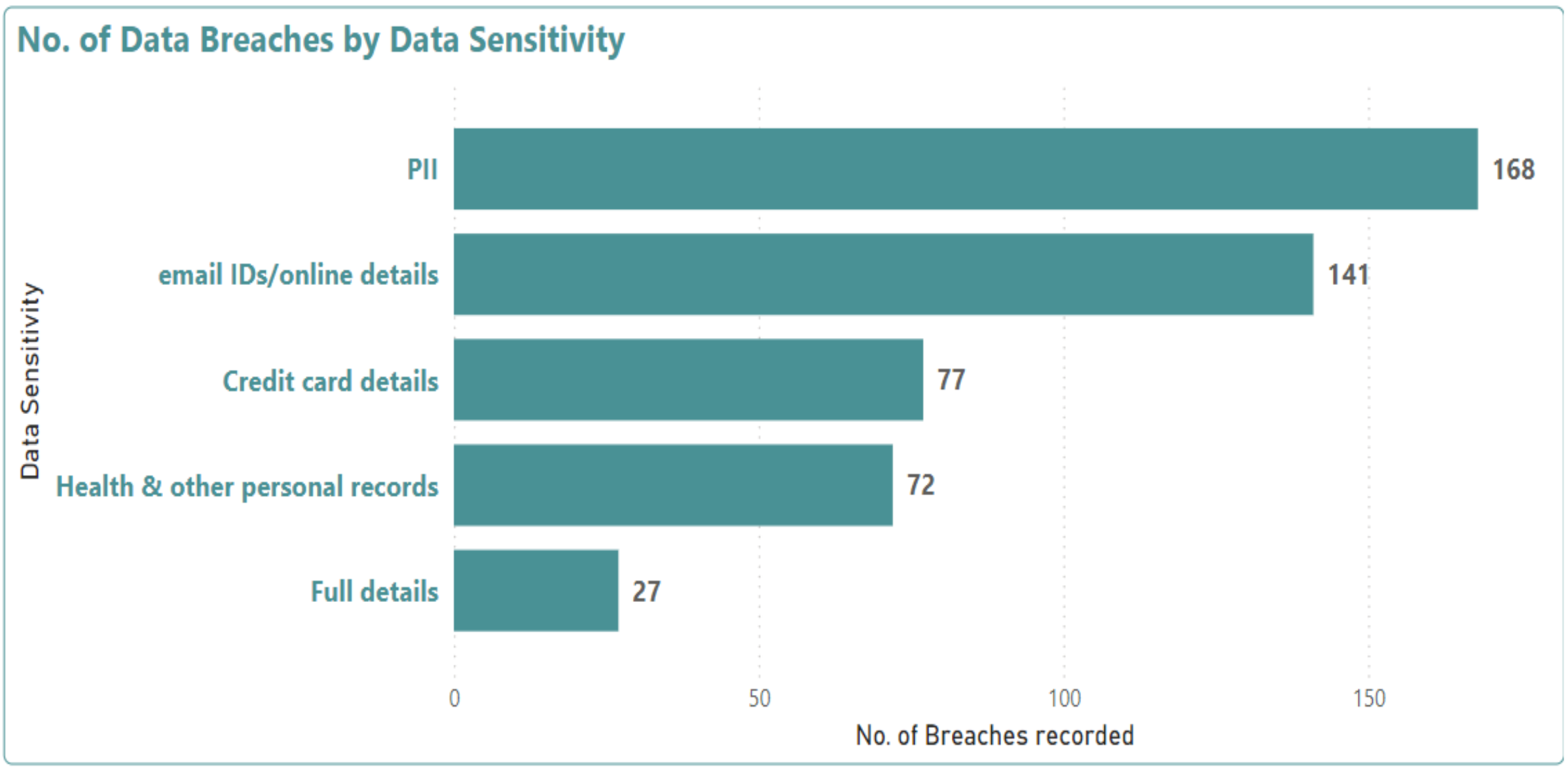

*Figure 23: No. of Data Breaches by Data Sensitivity*

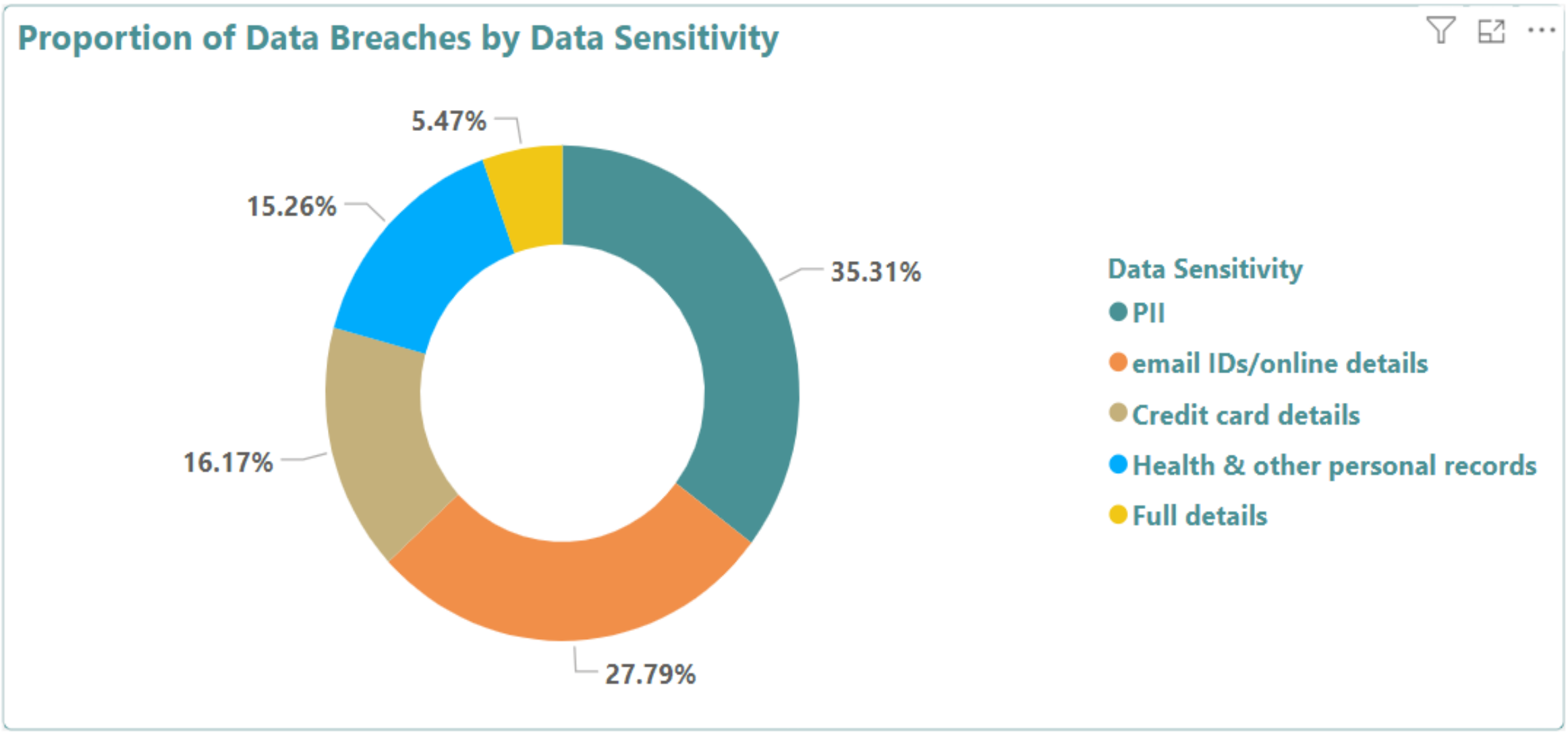

*Figure 24: Proportion of Data Breaches by Data Sensitivity*

The top 3 categories for the most records lost by data sensitivity included email IDs/online details, PII and credit card details followed by health and other personal details. It's worth noting that though data breaches compromising full details were infrequent, the records lost in such cases were significant as indicated in Figure 25.

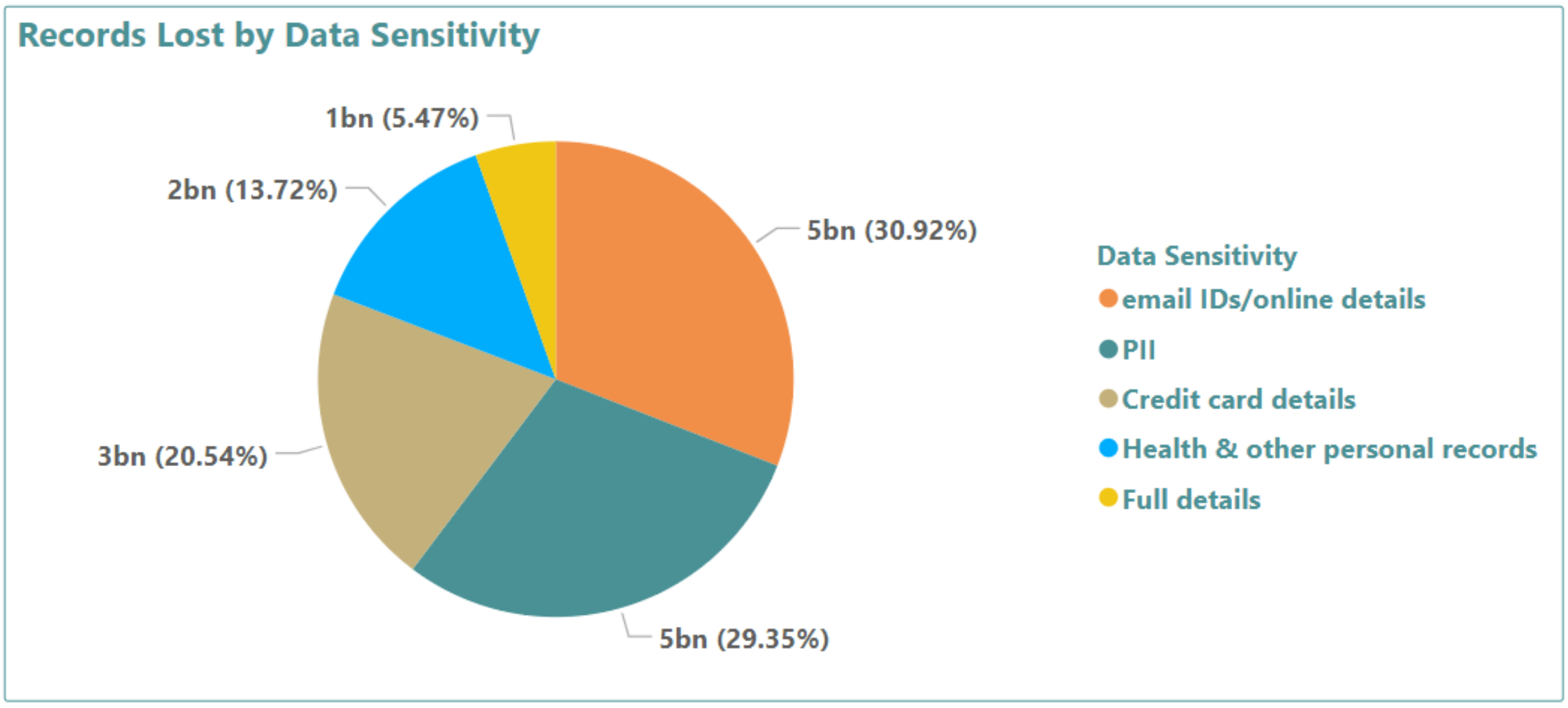

*Figure 25: Record Lost by Data Sensitivity*

Major data breaches have targeted entities processing, storing, or transmitting PII data as shown in Figure 26. Social media platforms such as Facebook, LinkedIn and Twitter topped the list for compromise of email IDs and other online details. The Shanghai Police breach was particularly severe, exposing full details of Chinese residents including their names, addresses, contact numbers, resident cards and information on cases registered against them.

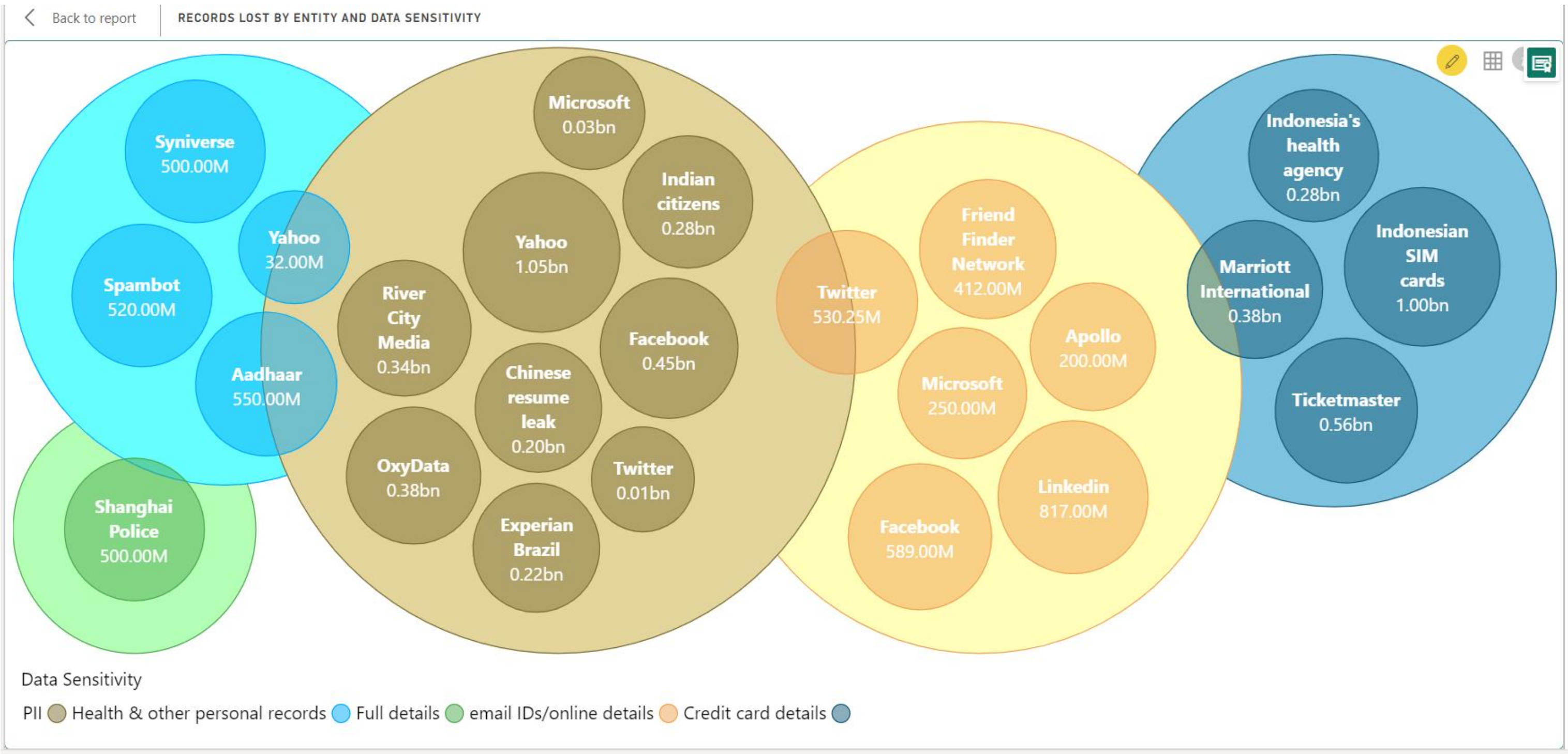

*Figure 26: Records Lost by Entity and Data Sensitivity*

### 4.1.4. Regional Biases in Cyberattacks

The findings are based on the analysis of data breaches in government and public entities. Company breaches are not included in the regional analysis due to a lack of information on data breach location for companies.

Figure 27 shows that North America, South America, Europe, Africa, Asia, and Australia were the six continents most affected by data breaches over the years, with the United States having the highest number of breaches followed by the United Kingdom.

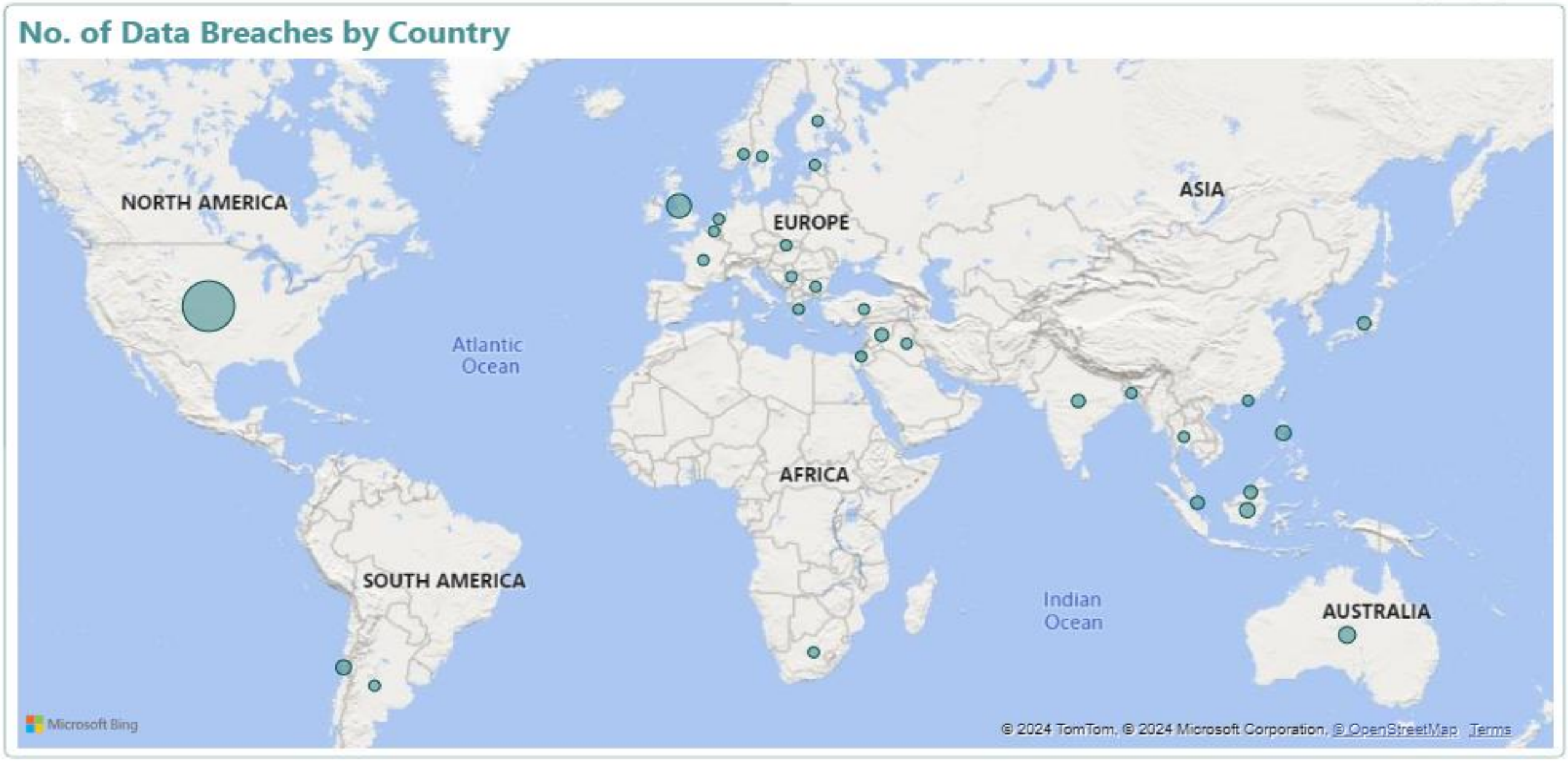

*Figure 27: No. of Data Breaches by Country*

Figure 28 shows that North America earned the top spot as the most impacted region. North America represented 55% of the breaches, while Europe saw 19%, Southeast Asia 9% and Asia 7% of the incidents.

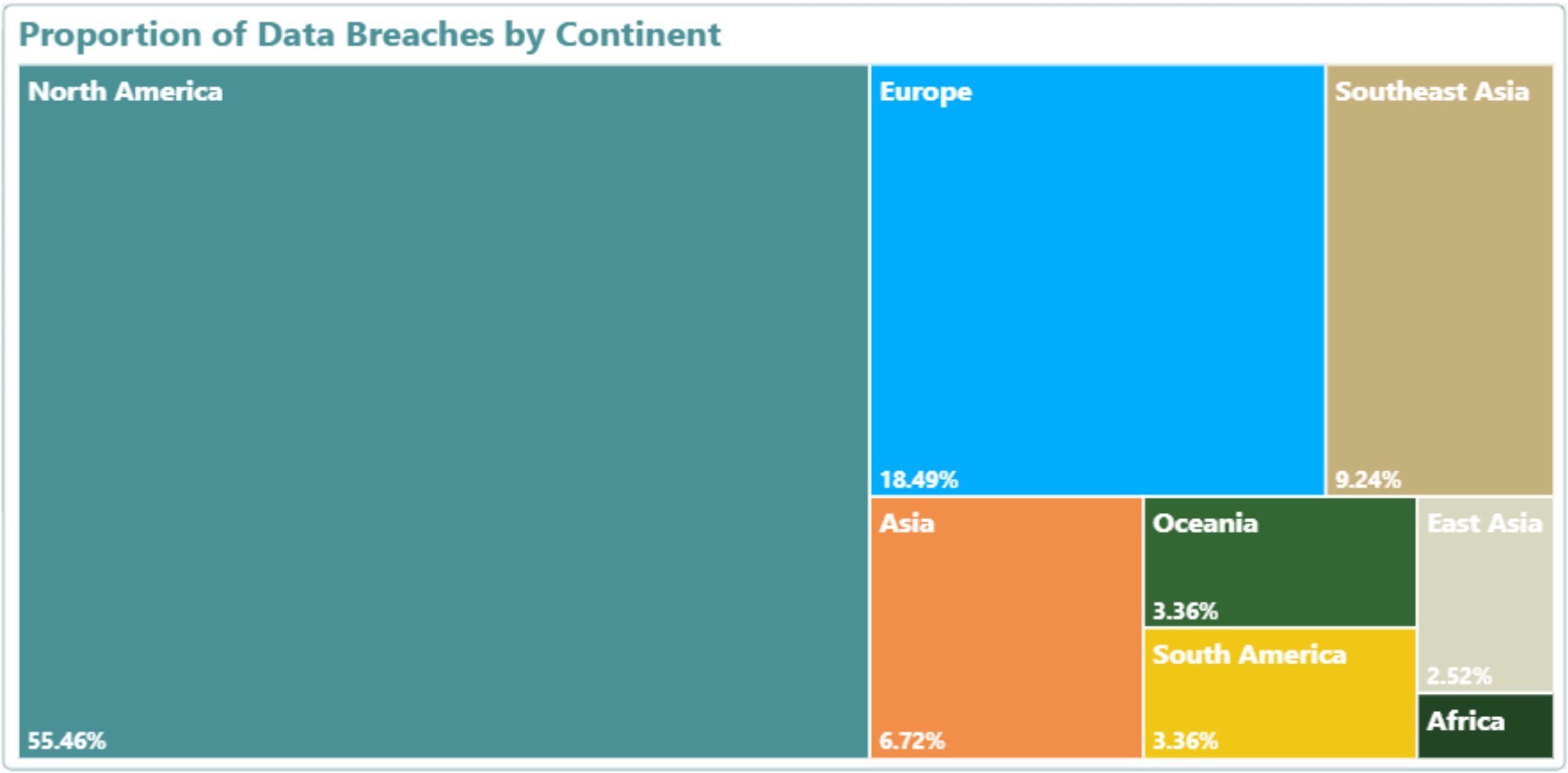

*Figure 28: Proportion of Data Breaches by Continent*

Most breaches in North America were mainly due to incidents in the United States and in South America, Chile was the most impacted country. The United Kingdom contributed to the largest share of breaches in Europe. Indonesia and the Philippines dominated Southeast Asia, while India and Syria had the highest number of breaches in Asia (refer to Figure 29).

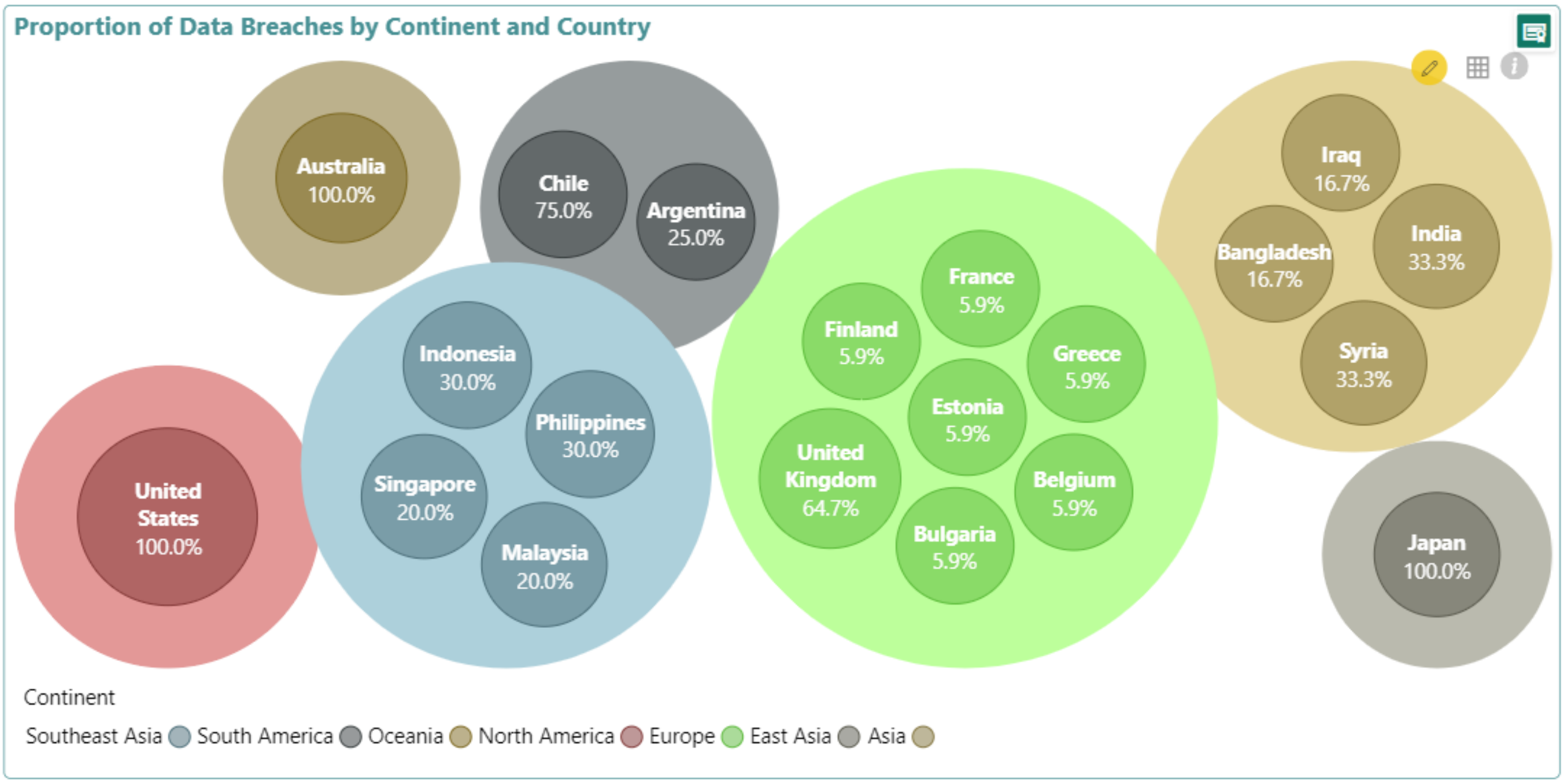

*Figure 29: Proportion of Data Breaches by Continent and Country*

North America had the most breaches between 2004 and 2020. However, a recent shift shows that Europe has become the most impacted region. In Europe, the United Kingdom is the most affected country. The most common attack method in Europe has been hacking. In 2023, North America was the second most impacted region globally (refer to Figure 30).

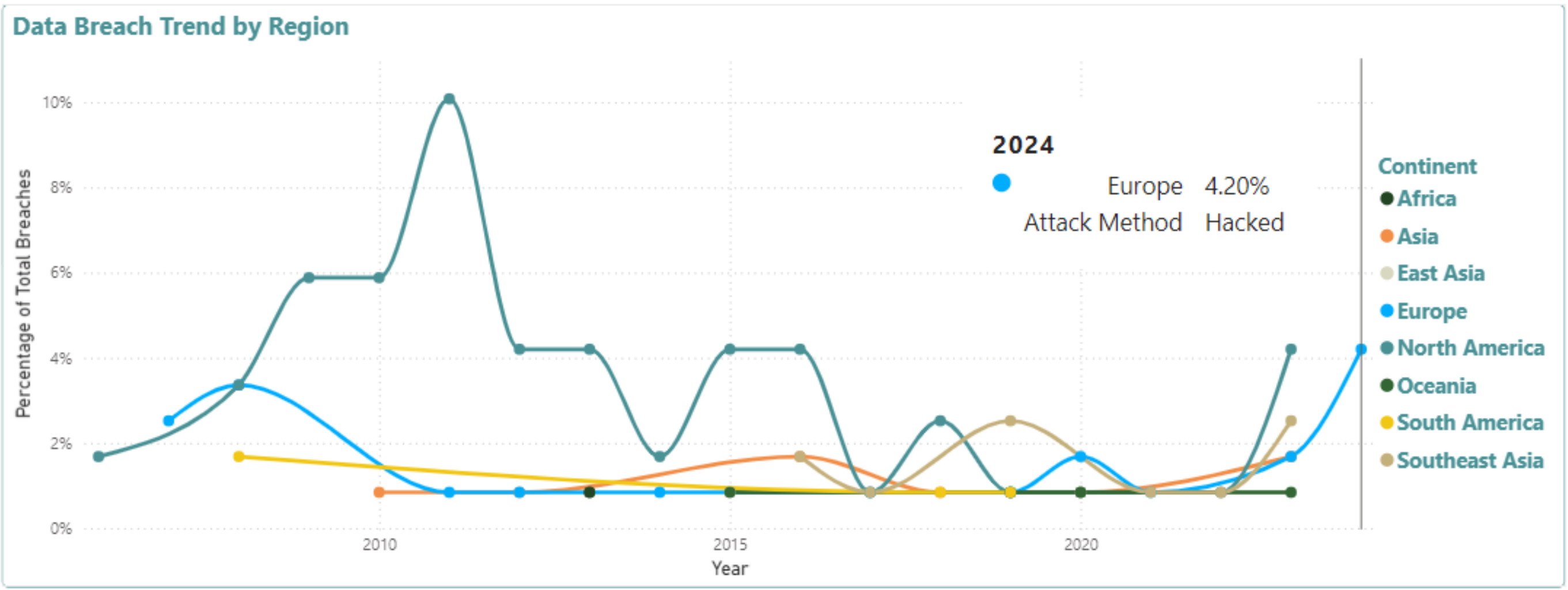

*Figure 30: Data Breach Trend by Region from 2004 to 2024*

Government sectors are more vulnerable in countries such as India and Indonesia. The United States is increasingly targeted for academic, military, and financial data breaches. Most healthcare breaches are seen in the United States and the United Kingdom. Lastly, the United Kingdom topped the list for breaches in the media/entertainment sector, including incidents involving the BBC and Ofcom (refer to Figure 31).

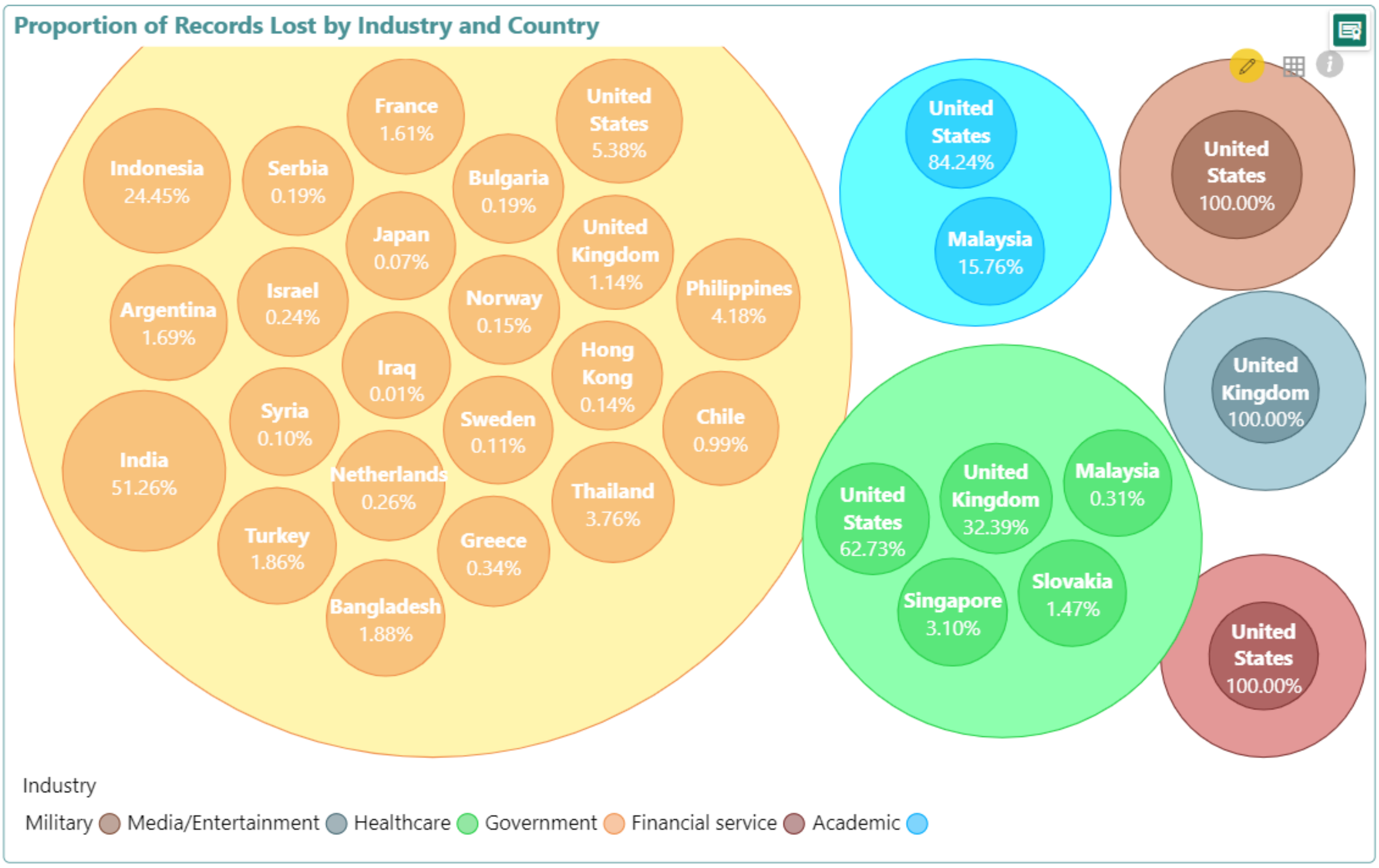

*Figure 31: Proportion of Records Lost by Industry and Country*

As shown in Figure 32, accidental exposures were more common in Indonesia and Norway. Hacking breaches affected countries like India, Indonesia, the Philippines and the United States. Lost or stolen devices were major causes of healthcare breaches mainly affecting the US and UK. Poor security measures were affected mainly India, Thailand, Bangladesh and the United States. Misconfiguration breaches were severe in Chile and insider threats were more severe in the US.

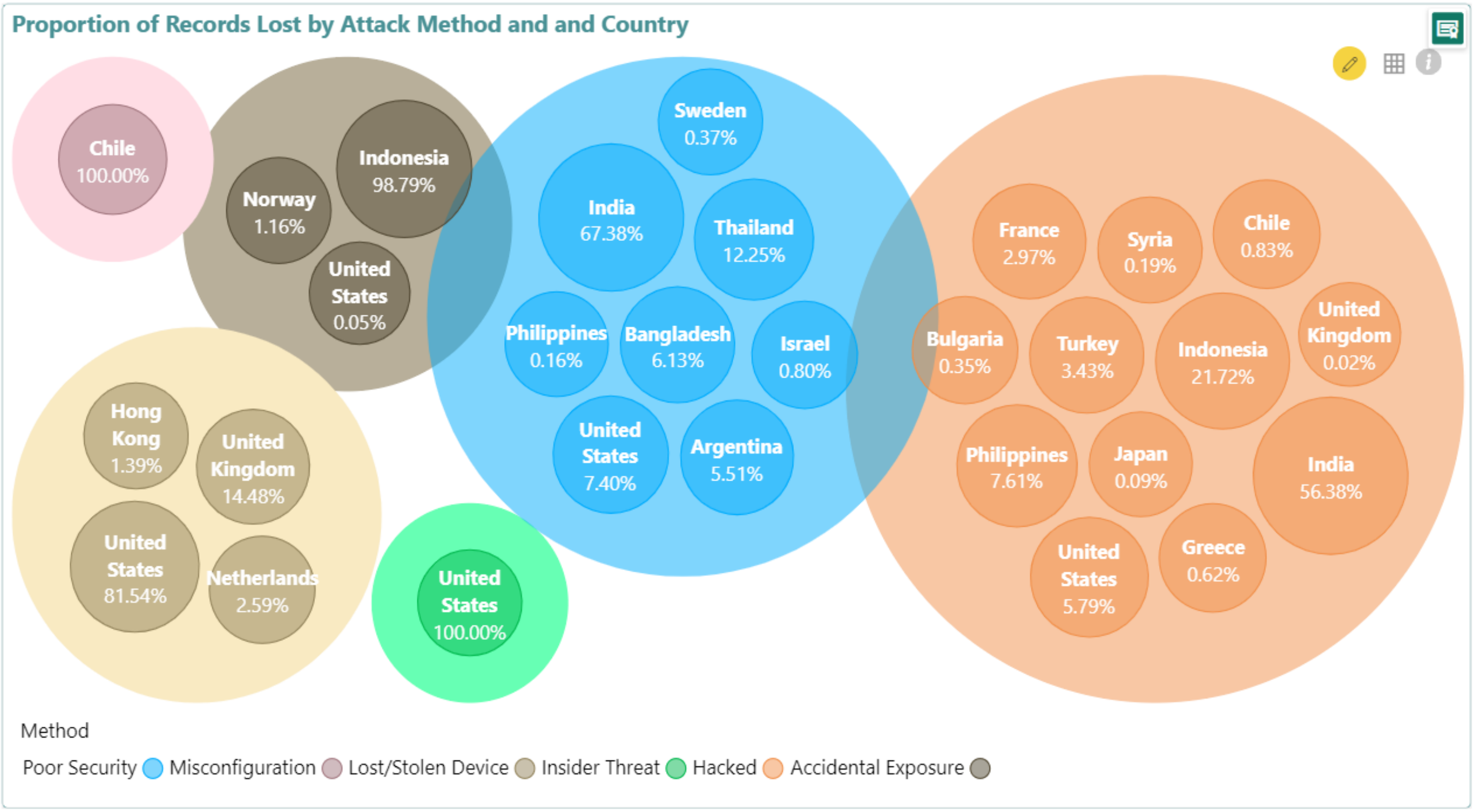

*Figure 32: Proportion of Records Lost by Attack Method and Country*

As shown in Figure 33, PII records were compromised on a large scale in countries like the United States, Thailand, Turkey and France. India tops the list for the largest number of health and personal records lost, attributed to the Aadhaar breach [80]. This is followed by the United States and the United Kingdom. Credit card detail compromises have been more severe in countries such as Japan, Indonesia, and the United States. For email ID and online details, the United Kingdom tops the list, accounting for 72% of data breaches. For full-detail exposures, the Philippines faced a more devastating impact, accounting for 66% of data breaches.

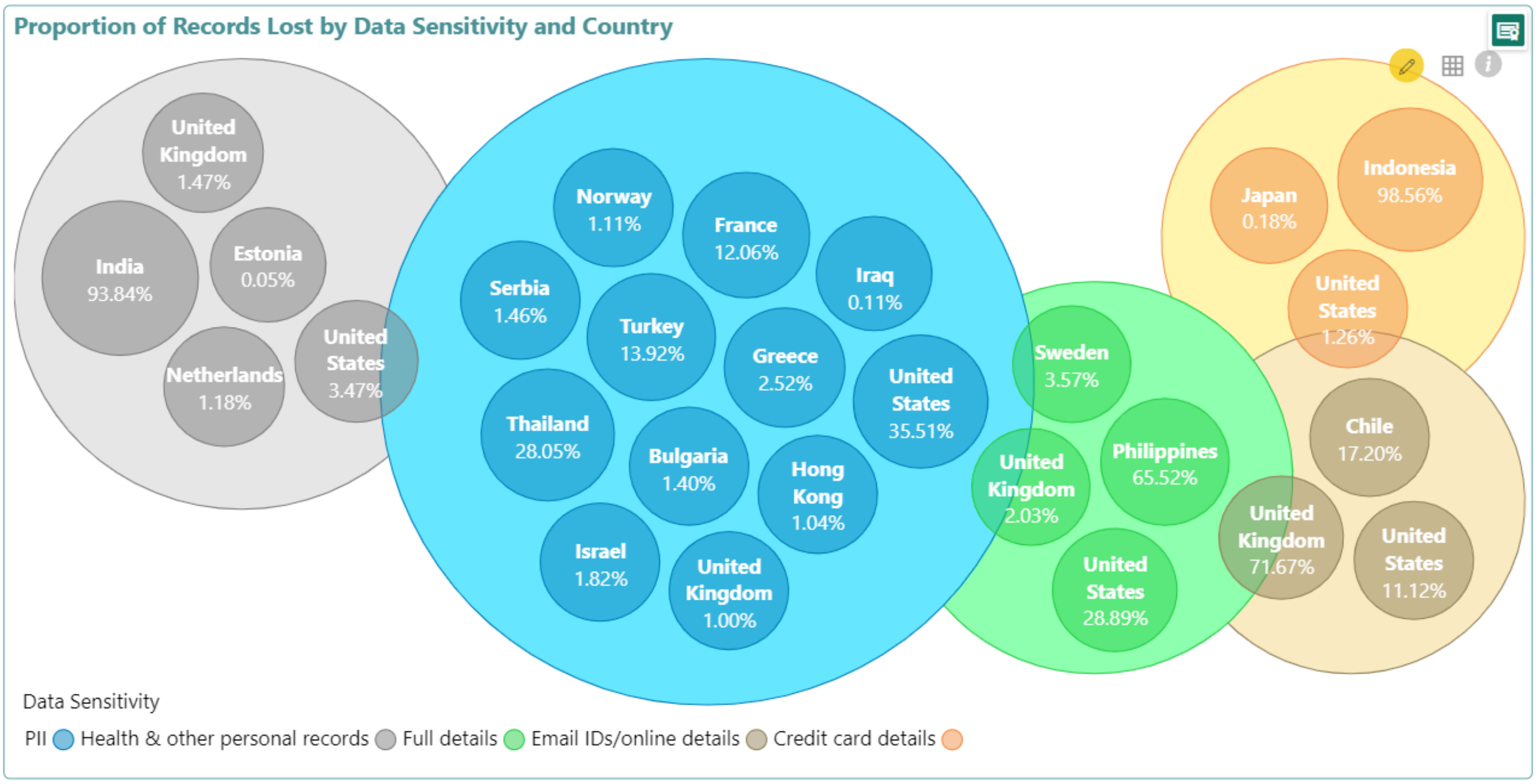

*Figure 33: Proportion of Records Lost by Data Sensitivity and Country*

## 4.2. Objective 2 – Time Series Analysis of Data Breaches

This section presents the findings related to the second objective of this study: to examine data breach categories over time to identify dominant, emerging and declining trends.

Figure 34 shows the year-wise breakdown of data breaches by industry. The data breaches involving financial services, government and healthcare sectors showed an upward trend between 2004 and 2013. Web services became dominant over the years reaching their peak in 2016. After 2016, web services, healthcare and financial services become dominant. However, retail breaches remained consistent throughout with a significant rise in 2020 and healthcare breaches peaked in 2011.

Overall, web services experienced the highest percentage of data breaches accounting for 23% of the total breaches. Technology and telecommunication breaches have become more emergent in recent years, while data breaches affecting the gaming and academic sectors remained less frequent.

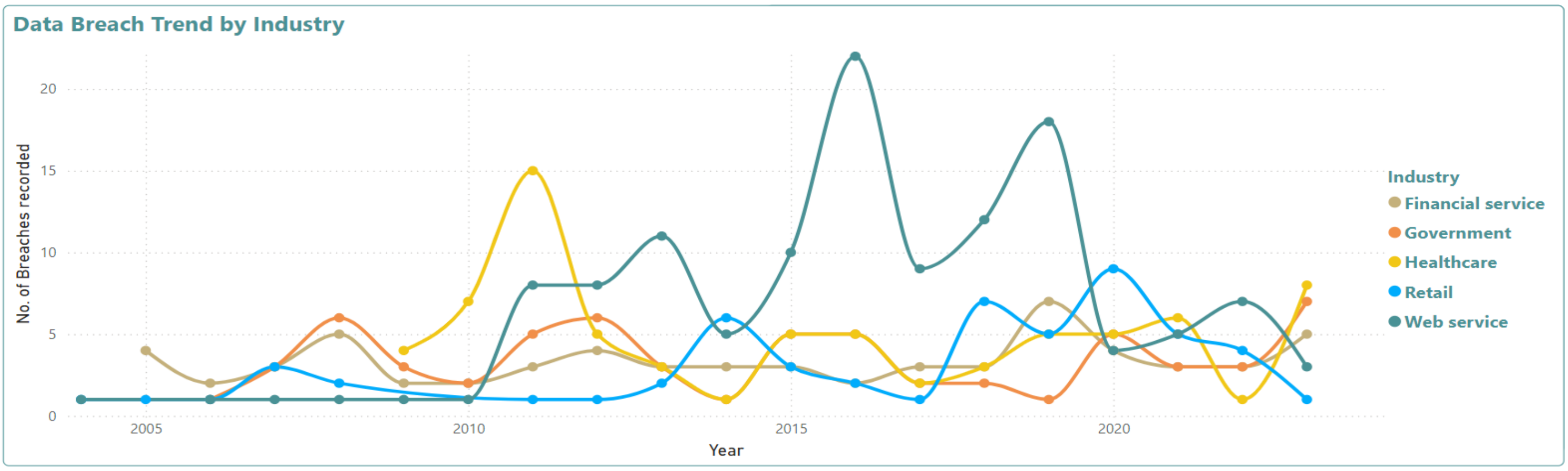

Back to report YEAR WISE BREAKDOWN OF DATA BREACHES BY INDUSTRY

| Industry | 2004 | 2005 | 2006 | 2007 | 2008 | 2009 | 2010 | 2011 | 2012 | 2013 | 2014 | 2015 | 2016 | 2017 | 2018 | 2019 | 2020 | 2021 | 2022 | 2023 | 2024 | Total |
|---|---|---|---|---|---|---|---|---|---|---|---|---|---|---|---|---|---|---|---|---|---|---|
| Academic | | | | | 0.54% | 0.18% | 0.18% | 0.36% | | 0.36% | 0.54% | | 0.36% | | | 0.36% | | | | 0.54% | | **3.45%** |
| Financial service | | 0.73% | 0.36% | 0.54% | 0.91% | 0.36% | 0.36% | 0.54% | 0.73% | 0.54% | 0.54% | 0.54% | 0.36% | 0.54% | 0.54% | 1.27% | 0.73% | 0.54% | 0.54% | 0.91% | 0.18% | **11.80%** |
| Gaming | | | | | | 0.18% | | 0.73% | 0.18% | 0.36% | | | 0.18% | | 0.18% | 0.91% | 0.91% | 0.18% | | 0.36% | | **4.17%** |
| Government | | | 0.18% | 0.54% | 1.09% | 0.54% | 0.36% | 0.91% | 1.09% | 0.54% | 0.18% | 0.91% | 0.91% | 0.36% | 0.36% | 0.18% | 0.91% | 0.54% | 0.54% | 1.27% | 0.73% | **12.16%** |
| Healthcare | | | | | | 0.73% | 1.27% | 2.72% | 0.91% | 0.54% | 0.18% | 0.91% | 0.91% | 0.36% | 0.54% | 0.91% | 0.91% | 1.09% | 0.18% | 1.45% | 0.36% | **13.97%** |
| Retail | 0.18% | 0.18% | 0.18% | 0.54% | 0.36% | | | 0.18% | 0.18% | 0.36% | 1.09% | 0.54% | 0.36% | 0.18% | 1.27% | 0.91% | 1.63% | 0.91% | 0.73% | 0.18% | 0.18% | **10.16%** |
| Technology | | | | | | | | | 0.18% | 1.09% | 0.18% | 0.36% | 0.18% | 1.09% | 1.09% | 1.45% | 0.54% | 1.27% | 0.36% | 1.27% | 0.91% | **9.98%** |
| Telecommunication | | | 0.18% | | 0.36% | | 0.18% | | 0.18% | 0.18% | | 0.54% | 0.36% | 0.36% | 0.54% | 0.36% | 0.54% | 0.91% | 0.36% | 1.09% | 0.36% | **6.53%** |
| Transportation | | | | | | | | | | | 0.36% | 0.18% | | | 1.27% | 0.36% | 0.54% | 0.54% | 0.54% | 0.54% | | **4.36%** |
| Web service | 0.18% | | 0.18% | 0.18% | 0.18% | 0.18% | 0.18% | 1.45% | 1.45% | 2.00% | 0.91% | 1.81% | 3.99% | 1.63% | 2.18% | 3.27% | 0.73% | 0.91% | 1.27% | 0.54% | 0.18% | **23.41%** |
| Total | **0.36%** | **0.91%** | **1.09%** | **1.81%** | **3.45%** | **2.18%** | **2.54%** | **6.90%** | **4.90%** | **5.99%** | **3.99%** | **5.81%** | **7.62%** | **4.54%** | **7.99%** | **9.98%** | **7.44%** | **6.90%** | **4.54%** | **8.17%** | **2.90%** | **100.00%** |

*Figure 34: Year-wise breakdown of the Proportion of Data Breaches recorded by Industry*

Figure 35 shows the year-wise breakdown of data breaches by attack method. The data breaches involving lost/stolen data declined after 2009 and hacking breaches became more prevalent. However, there is a rise in breaches due to misconfiguration and poor security.

Overall, the most dominant attack method over the years is hacking accounting for 65% of all breaches. Lost/stolen device breaches have decreased significantly over time. Misconfiguration and credential stuffing appear to be an emerging threat, while data breaches due to insider threats remain stable. Data breaches due to other attack methods like accidental exposure and web scrapping remain consistently low in occurrence.

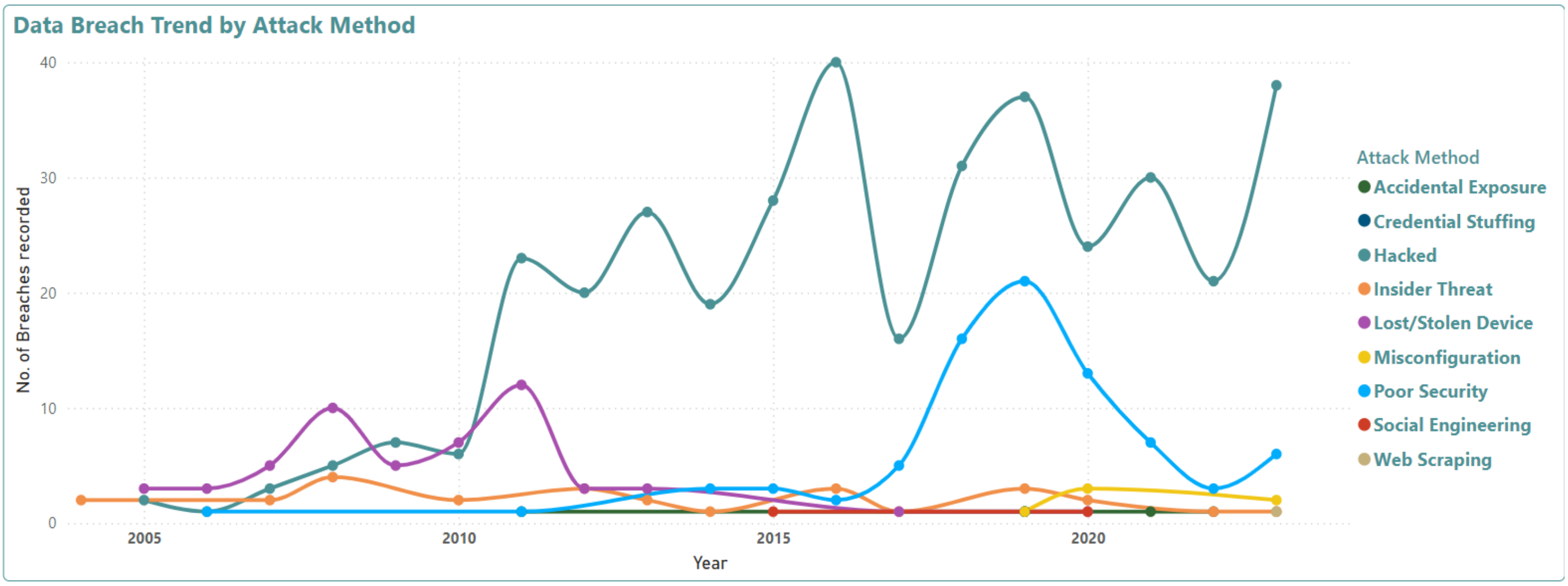

Back to report | YEAR WISE BREAKDOWN OF DATA BREACHES BY ATTACK METHOD

| Attack Method | 2004 | 2005 | 2006 | 2007 | 2008 | 2009 | 2010 | 2011 | 2012 | 2013 | 2014 | 2015 | 2016 | 2017 | 2018 | 2019 | 2020 | 2021 | 2022 | 2023 | 2024 | Total |
|---|---|---|---|---|---|---|---|---|---|---|---|---|---|---|---|---|---|---|---|---|---|---|
| Accidental Exposure | | | | | | | | 0.17% | | | | | | | | 0.17% | | 0.17% | 0.17% | | | **0.67%** |
| Credential Stuffing | | | | | | | | | | | | | | | | 0.17% | | | | | 0.17% | **0.33%** |
| Hacked | | 0.33% | 0.17% | 0.50% | 0.83% | 1.17% | 1.00% | 3.83% | 3.33% | 4.50% | 3.17% | 4.67% | 6.67% | 2.67% | 5.17% | 6.17% | 4.00% | 5.00% | 3.50% | 6.33% | 2.33% | **65.33%** |
| Insider Threat | 0.33% | | | 0.33% | 0.67% | | 0.33% | | 0.50% | 0.33% | 0.17% | | 0.50% | 0.17% | | 0.50% | 0.33% | | 0.17% | 0.17% | | **4.50%** |
| Lost/Stolen Device | | 0.50% | 0.50% | 0.83% | 1.67% | 0.83% | 1.17% | 2.00% | 0.50% | 0.50% | | | | 0.17% | | | 0.17% | | | | | **8.83%** |
| Misconfiguration | | | | | | | | | | | | | | | | 0.17% | 0.50% | | | 0.33% | 0.33% | **1.33%** |
| Poor Security | | | 0.17% | | | | | 0.17% | | | 0.50% | 0.50% | 0.33% | 0.83% | 2.67% | 3.50% | 2.17% | 1.17% | 0.50% | 1.00% | | **13.50%** |
| Social Engineering | | | | | | | | | | | | 0.17% | | | | | 0.17% | | | | | **0.33%** |
| Unknown | | | 0.17% | | 0.17% | | | 0.33% | 0.33% | 0.50% | 0.17% | 0.17% | 0.17% | 0.33% | 0.33% | 0.33% | 0.67% | 0.17% | 0.33% | 0.33% | 0.50% | **5.00%** |
| Web Scraping | | | | | | | | | | | | | | | | | | | | 0.17% | | **0.17%** |
| Total | **0.33%** | **0.83%** | **1.00%** | **1.67%** | **3.33%** | **2.00%** | **2.50%** | **6.50%** | **4.67%** | **5.83%** | **4.00%** | **5.50%** | **7.67%** | **4.17%** | **8.17%** | **11.00%** | **8.00%** | **6.50%** | **4.67%** | **8.33%** | **3.33%** | **100.00%** |

*Figure 35: Year-wise breakdown of the Proportion of Data Breaches recorded by Attack Method*

Figure 36 illustrates the breakdown of data breaches by type of compromised data over the years. The data breaches involving credit card details, email IDs and PII data showed an upward trend between 2004 and 2024 with a significant rise in PII breaches.

Also, the percentage of breaches associated with credit card details was substantial yet lesser compared to PII and email IDs. One of the emerging trends seen is frequent attacks aimed at obtaining healthcare records and other personal data. Overall, the most frequently compromised data types in breaches are PII (35%) and email IDs (29%). Credit card details (16%) and health records (15%) also saw considerable breaches.

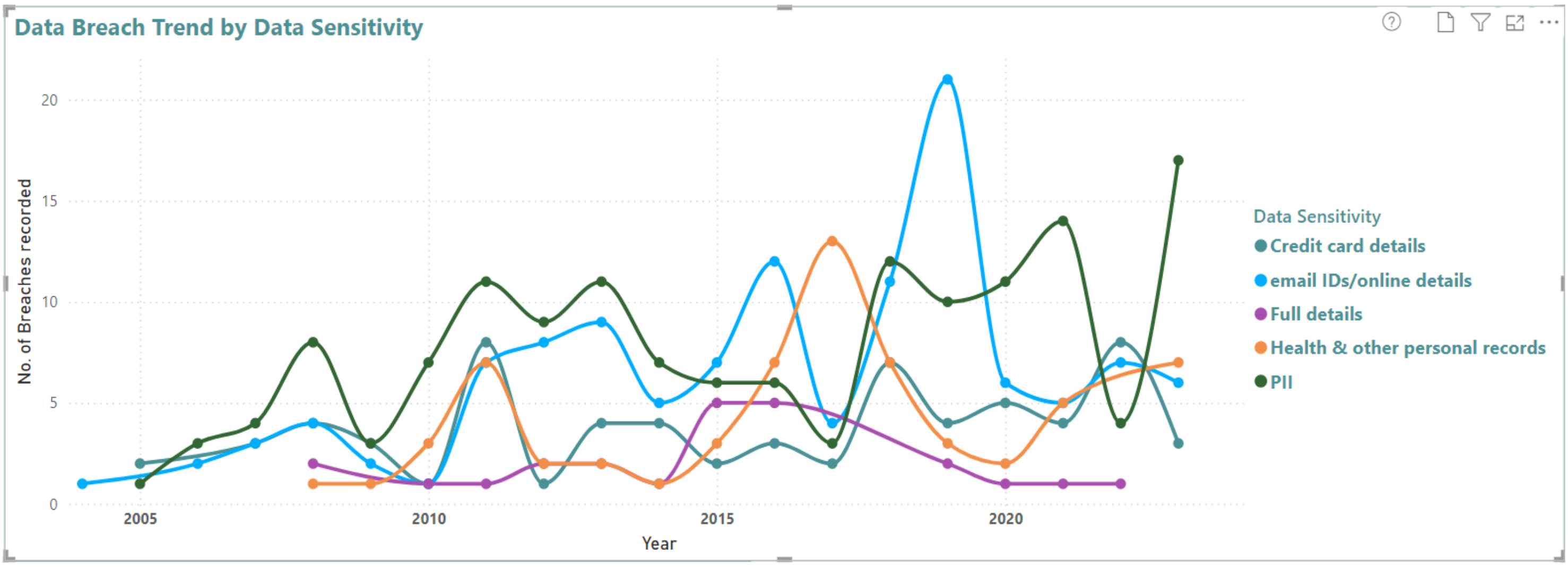

Back to report | YEAR WISE BREAKDOWN OF DATA BREACHES BY DATA SENSITIVITY

| Data Sensitivity | 2004 | 2005 | 2006 | 2007 | 2008 | 2009 | 2010 | 2011 | 2012 | 2013 | 2014 | 2015 | 2016 | 2017 | 2018 | 2019 | 2020 | 2021 | 2022 | 2023 | 2024 | Total |
|---|---|---|---|---|---|---|---|---|---|---|---|---|---|---|---|---|---|---|---|---|---|---|
| Credit card details | | 0.41% | | 0.62% | 0.82% | 0.62% | 0.21% | 1.86% | 0.21% | 1.24% | 0.82% | 0.41% | 0.62% | 0.41% | 1.65% | 0.82% | 1.03% | 0.82% | 1.65% | 0.82% | 0.82% | **15.88%** |
| email IDs/online details | 0.21% | | 0.41% | 0.62% | 0.82% | 0.41% | 0.21% | 1.44% | 1.86% | 2.27% | 1.24% | 1.86% | 2.47% | 0.82% | 2.47% | 5.15% | 1.86% | 1.44% | 1.44% | 1.86% | 0.21% | **29.07%** |
| Full details | | | | | 0.41% | | 0.21% | 0.41% | 0.41% | 0.41% | 0.21% | 1.03% | 1.03% | 0.21% | 0.21% | 0.41% | 0.21% | 0.21% | 0.21% | | | **5.57%** |
| Health & other personal records | | | | | 0.21% | 0.21% | 0.62% | 1.44% | 0.41% | 0.41% | 0.21% | 0.62% | 1.44% | 3.09% | 1.44% | 0.62% | 0.62% | 1.03% | 0.21% | 1.65% | 0.62% | **14.85%** |
| PII | | 0.21% | 0.62% | 0.82% | 1.65% | 0.62% | 1.44% | 2.47% | 1.86% | 2.89% | 1.44% | 1.24% | 1.24% | 0.62% | 2.68% | 2.27% | 2.27% | 3.30% | 1.03% | 3.92% | 2.06% | **34.64%** |
| Total | **0.21%** | **0.62%** | **1.03%** | **2.06%** | **3.92%** | **1.86%** | **2.68%** | **7.63%** | **4.74%** | **7.22%** | **3.92%** | **5.15%** | **6.80%** | **5.15%** | **8.45%** | **9.28%** | **5.98%** | **6.80%** | **4.54%** | **8.25%** | **3.71%** | **100.00%** |

*Figure 36: Year-wise breakdown of the Proportion of Data Breaches recorded by Data Sensitivity*

Figure 37 shows the data breach trend by year and future forecast. The period from 2004 to 2023, shows an increase in the number of data breaches with a significant peak in 2019. The black dashed line represents the trend line over the recorded data up to 2023, whereas the solid black line represents the forecast line. Besides that, the shaded grey area around the forecast line displays the range of uncertainty around the projection.

In general, the projection shows that concern about data breaches is fast rising from 8.62% in 2023 to 9.04 % in 2024. Within the first 6 months of 2024, 20 data breaches were recorded accounting for 3.33% of the projected breaches. This is expected to continue increasing towards the end of the year. Overall, as the cases of breaches are expected to soar in future, there would be a requisite need for improved cybersecurity.

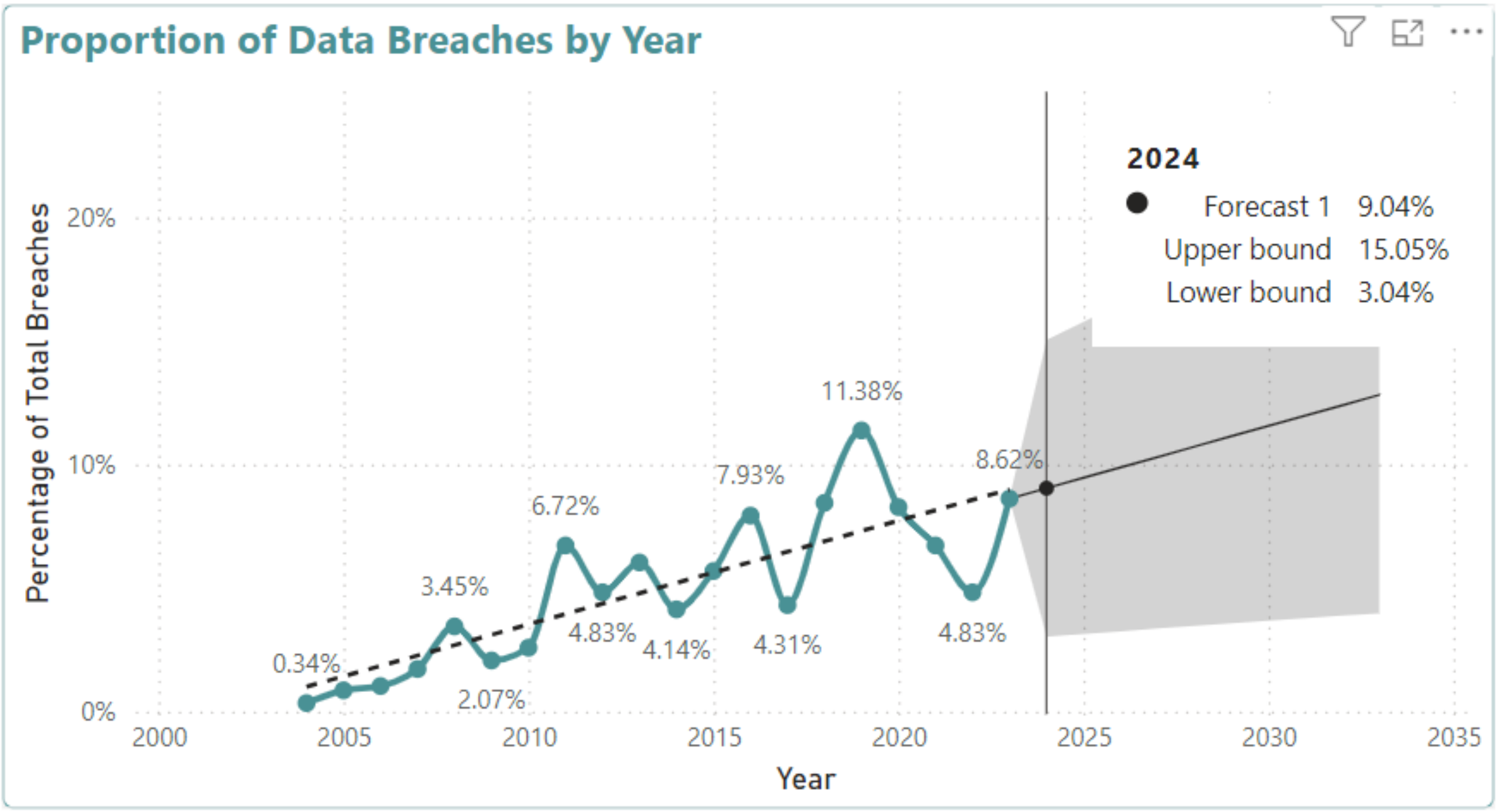

*Figure 37: Data breach trend by year and future forecast*

### 4.3. Objective 3 – Cross-Tab Analysis of Data Breaches

This section presents the findings related to the second objective of this study: to identify any relation between targeted industries, attack methods and data compromised to understand the prominent risk factors in various categories.

Figure 38 shows the proportion of data breaches by attack method and industries. The method of data breach through lost/stolen devices was commonly seen in government, healthcare and financial sectors. Web, retail, financial services, healthcare and technology are the top sectors prone to hacking breaches. Accidental exposure methods of data leakage have affected mainly government and retail industries. Financial services, government and healthcare are the top three sectors prone to data leakage through inside jobs. Lastly, poor security practices are seen widely in the web services and technology sector.

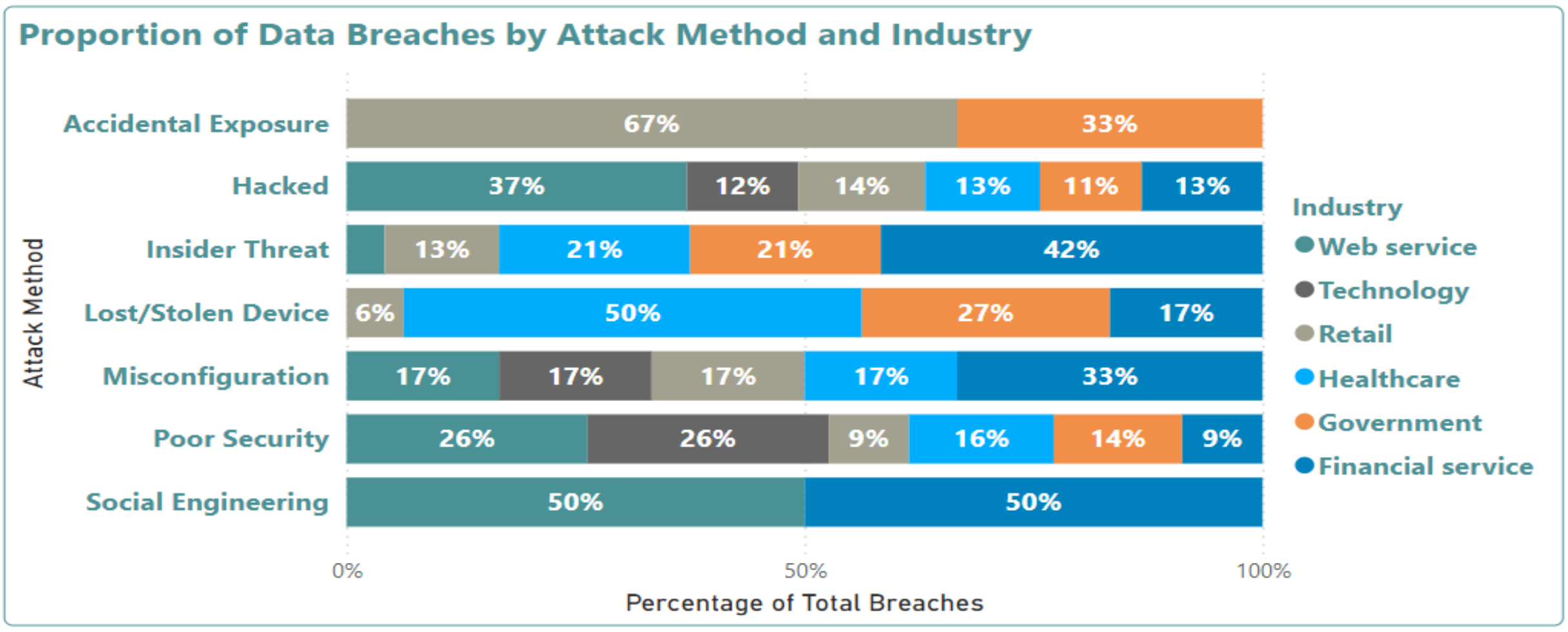

*Figure 38: Proportion of Data Breaches by Attack Method and Industry*

Figure 39 shows that web services, government and financial services are the top types of organisations that are increasingly targeted for PII. Retail and Financial services are targeted for collecting credit card details. Financial services, government and healthcare sectors need to up their security as they have more breaches with full disclosure. Healthcare sectors are targeted for health and personal records, while email IDs/ online details are compromised more in the web service sector.

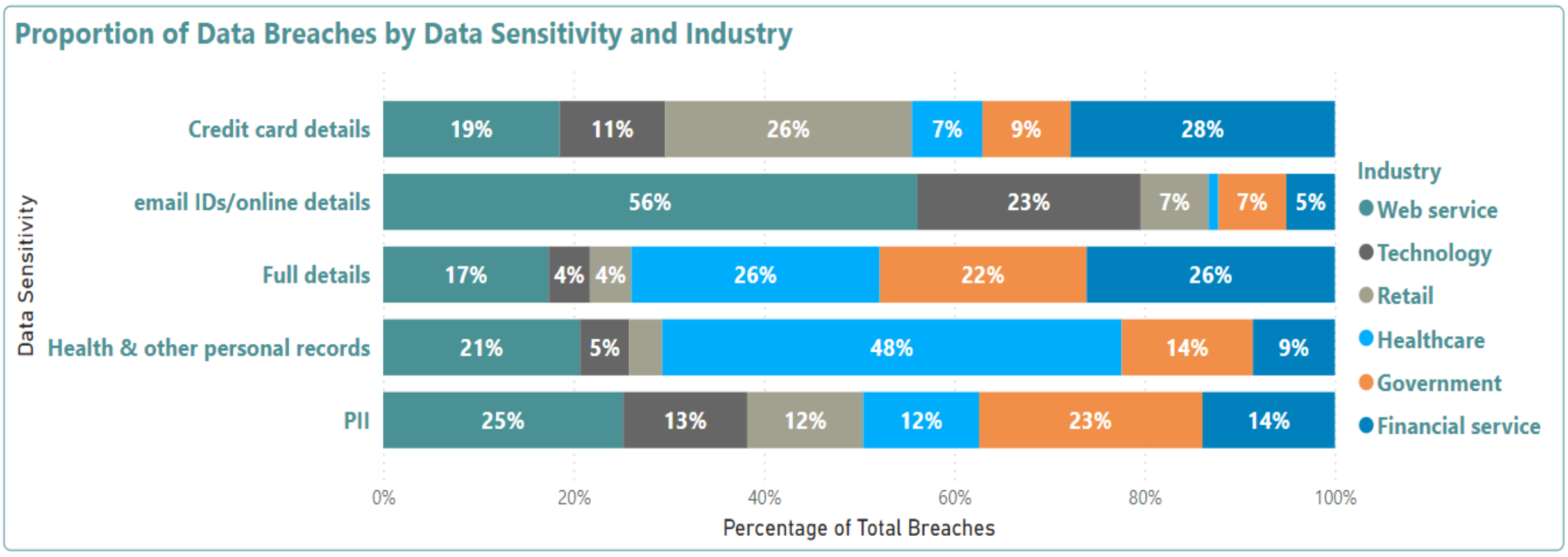

*Figure 39: Proportion of Data Breaches by Data Sensitivity and Industry*

Figure 40 shows the proportion of data breaches by attack method and data sensitivity. One of the prominent trends seen is whenever full details are compromised, it has happened because of insider threats. As insiders can easily escalate their privileges or misuse their access rights within the organisation it is comparatively easy for them to exfiltrate large amounts of data than external threats. Besides that, the hacking method is employed to obtain various types of sensitive data such as PII, email IDs, credit card details and health records. The attack method of lost/stolen devices was seen more in the healthcare or government sector putting healthcare records and PII data at the risk of compromise. Also, poor security practices have led to large-scale compromise of PII and email IDs.

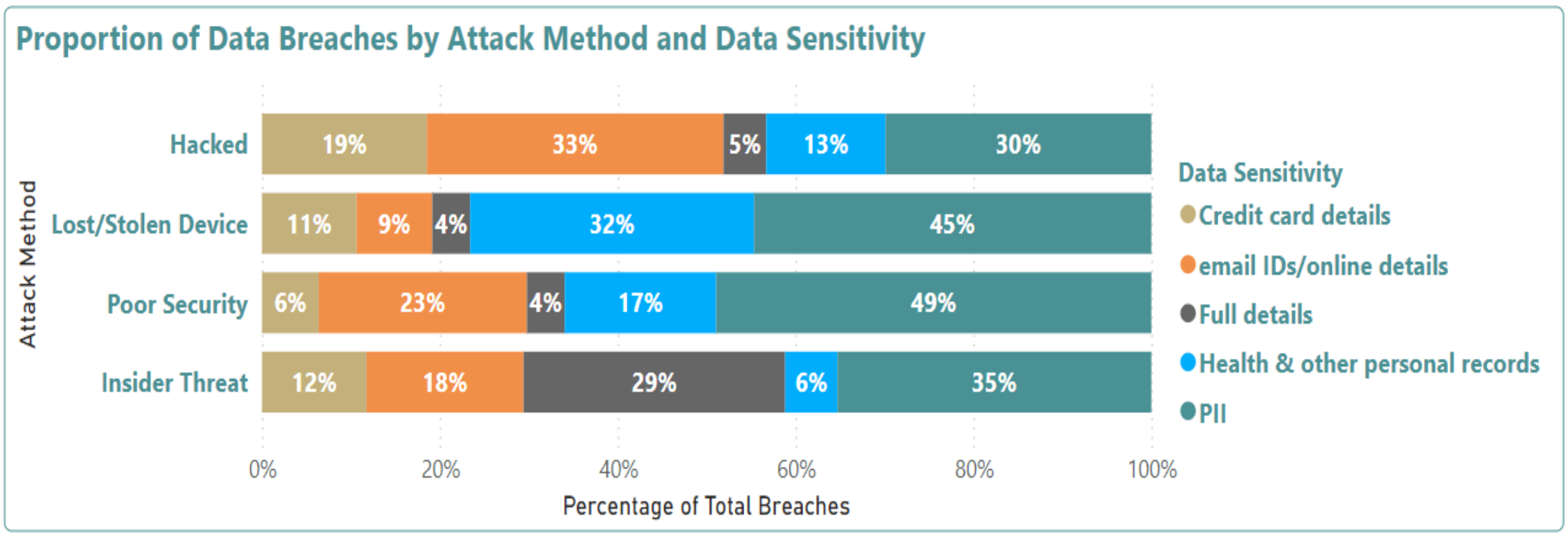

*Figure 40: Proportion of Data Breaches by Attack Method and Data Sensitivity*

## 4.4. Conclusion:

This chapter discussed the dashboard findings in detail, highlighting the prominent and emerging trends. The next chapter will compare these findings with industry reports and will find the reasons for the observed trends to develop comprehensive understanding of the data breach landscape.

# 5. Discussion and Implication

## 5.1. Interpretation of Findings:

### 5.1.1. RQ 1: Which industries or sectors are targeted the most?

Webservices, healthcare, government and financial services are the industries most affected by data breaches. Most attacks target web services because they often handle a vast amount of customer data such as email IDs, phone numbers, addresses, emails, passwords and names that could be used by cybercriminals for identity theft [78] and resale on the black market. This category also includes large multinational corporations such as Yahoo, Microsoft and Facebook whose stolen data is used for large-scale phishing attacks, spamming and advertising campaigns [78]. Web services have a complex architecture that relies on APIs, microservices, cloud services, databases, application servers and third-party integration. If not secured, these become a potential entry point for attackers.

Healthcare organisations store sensitive patient details such as medical records, national IDs, and social security numbers which are valuable in the black market. These organisations are integrated into multiple third-party systems for managing day-to-day appointments, casualties, blood tests, surgeries and emergency services. This integration spans a huge supply chain of entities including pharmacists, retailers, suppliers and distributors. Therefore, a single entry point can disrupt the entire supply chain and most importantly put people's lives at risk, creating significant pressure on the healthcare sector to meet ransom needs [101]. For example, the ransomware attack on the British NHS by Qilin was due to a vulnerability in their digital provider Synnovis which impacted London hospitals causing cancellation of appointments, delays in surgery and system and administrative failure [102].

Government sectors manage confidential data such as citizen's data, classified information and intelligence secrets making them a prime target for cyber espionage and state-sponsored attacks. Citizen's data such as national IDs, tax records, passports and biometric verification can be used for identity theft and forged documents. State-sponsored attackers target sensitive databases to steal secrets and also maintain access for strategic purposes. For instance, Chinese hackers infiltrated the critical infrastructure of the United States over five years, not to steal any information but to gain strategic control for future operations [103]. They targeted not just the military but also other key infrastructures such as aviation, water supply, power grid, transport systems and energy sectors. According to NBC News, China could exploit the access gained in critical US infrastructure to prevent intervention of US in future conflicts with Taiwan [103].

Finally, financial services are prime targets for cybercriminals who commit fraud and steal money. These institutions store sensitive financial information such as credit card details, PII data, account numbers and other banking details which are valuable for attackers. A successful breach would give direct access to funds and can severely damage customer trust and reputation. Moreover, outsourced financial services are at risk of supply chain attacks and data leaks. For example, Santander's data breach was due to unauthorised access to their database managed by a third-party vendor [104].

### 5.1.2. RQ 2: Which attack methods are more prevalent?

Hacking and poor security were the more prevalent types of attack methods compared to others (refer to Figure 20). As hacking accounts for 69% of data breaches, it is worth further exploring hacking breaches in depth, especially zero-day exploits. A zero-day bug [105] is one in which the company does not even know it exists i.e., the company had known about the vulnerability for zero days. Such a secret vulnerability is extremely difficult to find and once discovered it can be used to breach the security of a device or even an entire ecosystem [105].

While the dark web and other market forums help cybercriminals purchase stolen credentials and other ransomware and DDoS services, the author draws attention to the zero-day market, which is much darker and deeper. In this market, governments, large corporations and criminals compete over information that can have a significant impact, driven by motivations such as life-changing money, geopolitical positioning or hacktivism. [105].

Most transactions in this market happen in cryptocurrency and using escrow services; and trust is crucial which has given rise to middlemen and shady companies who work like the brokers or the matchmakers for the buyers and sellers. Their role is to hold funds in escrow and then confirm the vulnerability effectiveness i.e. vouching for the services and taking some percentage of the deal as their income [105].

Due to the usefulness of zero-day vulnerabilities, prices are generally high. Figure 41 [106] shows a list of zero-day vulnerabilities for mobile devices provided by Zerodium, a broker company. Access to Phone PIN or passcode is up to $100,000 while zero-day vulnerabilities that allow to access more than one application such as chat, email or browser cost up to $500,000. Often hackers try to bypass multiple vulnerabilities and take complete control over the device. However, a zero-day exploit allowing a full chain attack with persistence, without any interaction from the hacker costs around 2 to 2.5 million dollars [106].

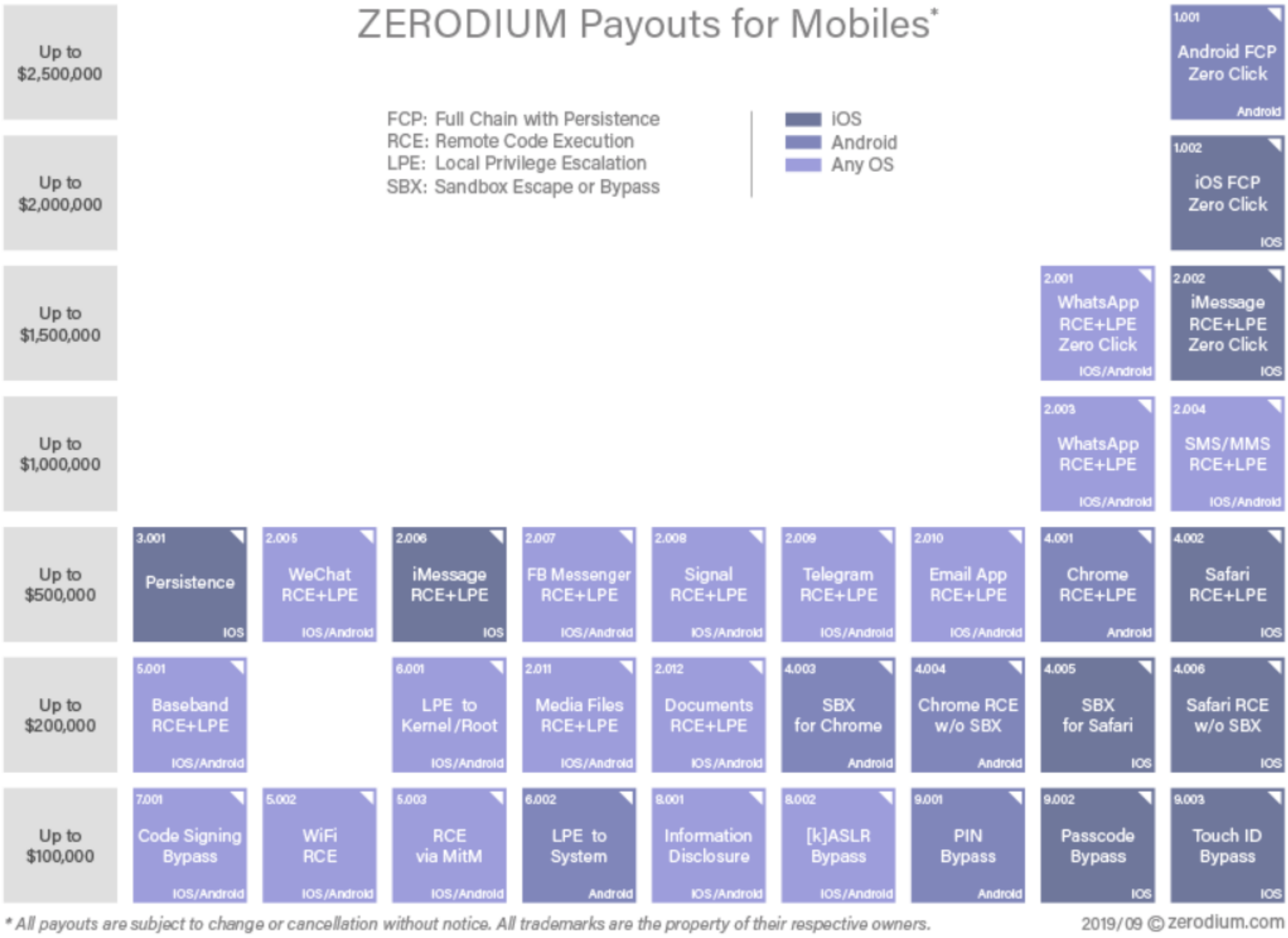

*Figure 41: Zerodium Payouts for Mobiles* [106]

A zero-day exploit can cause huge damage by infiltrating government secrets and other security services leading to cyber warfare. However, all this continue to happen whilst law enforcement and regulators are not able to do much to stop it.

From the policy perspective, the author argues that regulating zero-day market may not be quite feasible as the zero-day market is quite complex and most importantly very anonymous and secret where no one can be held accountable. It is a three-tier ecosystem consisting of white market, grey market and black market. White market could still be regulated to some extent as hackers here obtains money legally, often referred to as the bug bounty program where the companies pay them good sums for finding zero-day exploits in their system and responsibly disclosing them for patching. The grey market is in middle, not completely legal or illegal. This is where government hires hackers often with security clearance to find zero-days exploits mainly for spying or cyber warfare. As even the government is involved as the buyers or users of the market, it is difficult to impose strict regulations as this may hinder their own security intelligence process. Well, the black part of the market is known for data theft and extortion [105], [107]. An example of this is the extortion of nation state actors by the CLOP ransomware group using a zero-day vulnerability in MOVEit software, peaking hacking crimes in 2023 (refer to section 2.4.5).

Moreover, regulating attempts to force disclosure may drive away researchers from findings zero days in the first place due to incentive issues. Mandatory disclosures can jeopardise security and also put public safety at risk, if vulnerabilities are exploited before a patch is available [105].

Other reasons include geographical implications making it difficult for a single country to impose regulations., different countries might have their own priorities based on legality and morality. This often leads to the argument that zero-day market is morally dubious with different entities crossing lines i.e. government can reach out to black market of another country for geopolitical advancement and everything is so anonymous and invisible that it is hard to prosecute anyone or regulate the market [107].

As an alternative, the author calls for an international co-operation from different government security services and intelligence agencies to make the global infrastructure safe for everyone. A unified consortium of law enforcement security specialists who can collectively work together to find zero-day exploits and use it for the betterment of everyone. For instance, in 2023, a vulnerability was found in PHP known as CVE-2023-3824. PHP is the backbone of internet, and this vulnerability could have been used to gain unauthorised access to any server in the world " [108]. Instead, it was used to exploit the website of LockBit and take down their several severs and accounts causing their operations to collapse completely. This was done by a joint taskforce of law enforcement from 10 countries, referred to as "Operation Cronos" [108].

### 5.1.3. RQ 3: What type of data or data sensitivity is at risk of compromise?

PII data including email IDs, credit card details and health records are the data increasingly targeted by various threat actors. Threat actors exfiltrate such data and sell it on the dark web for profit.  The frequent breaches of personal data strongly suggest that organisations do not have full control over their digital assets [50, p. 6]. There is a lack of visibility into the entire data lifecycle, which may be attributed to the fact that data is not just confined within the organisation but is regularly transferred between different "systems, people and third-party suppliers" for business needs [50, p. 6]. Despite the existence of data privacy regulations with heavy fines for neglecting security measures, organisations lack basic data security controls such as encryption, network segmentation, MFAs and role-based access controls as seen in most data breaches. Organisations first need to gain a complete understanding of their digital estate including visibility into hosting services, third-party entities, customers and the data retention period. This will help organisations to discover and classify data based on their sensitivity and put security controls in place to protect it [50, p. 6].

Previous studies do not explore the reasons behind the large-scale exploitation of PII data in depth. As the world is getting more digitalised, data becomes invaluable. Table 6 summarises the cost of sensitive data from a study conducted by Privacy Affairs on the Dark Web Price Index in 2023 [109].

| Data Category | Cost in USD |
|---|---|
| Online banking login information | $10 - $2400 |
| Full credit card details | $10 - $110 |
| Email database dumps | $100 - £120 |
| Full package (documents + accounts information) | $1000 |
| Hacked Services such as Netflix, Bet365, Kaspersky, Canva, Uber account | $4 -$40 |
| Driver's license | $150 -$1500 |
| Compromised social media accounts such as Facebook, Instagram, Twitter and Gmail | $25 -$65 |
| Forged identification card | $150 - $500 |
| Forged passports | $3000 -$4000 |

*Table 6: Cost of PII data in USD* [109]

It is quite shocking to see that credit card details, compromised accounts or services and some online banking information are available for less than $50. It only takes $1000 to purchase the full package including personal details and necessary documents, attributing to a notable increase in identity theft crimes in recent years [109]. The price range differs due to various reasons such as the attack sophistication, brand value, seller reputation and freshness of the data. For instance, if a bank suffered recent large-scale breach, the market would be temporarily flooded with stolen logins driving down the price, however reputed sellers might remain unaffected as trust is crucial in this anonymous marketplace. The success rate for fresh PII is higher, fresh logins are more likely to be valid and allow unauthorised access to the account. Especially for time sensitive attacks like account take over, it is crucial to weaponise the data before the victim notices or reacts. Other factors affecting PII price is the attack sophistication. Larger global banks might have more sophisticated security measures making their logins valuable to cybercriminals. On the other hand, smaller institutions might be an easier target affecting price. Lastly, volume and demand also affect the price; logins for bank with a larger customer base might be higher, rising its price [105].

#### 5.1.4. RQ 4: Are there any regional biases in data breaches especially for government and public entities?

All continents except Antarctica were affected. Antarctica mainly carries out scientific research on environment monitoring and climate change. These operations do not involve any sensitive data that could become a potential target of cybercriminals.

North America recorded the highest number of breaches historically attributed to the fact that North America has a large digital infrastructure with numerous multinational corporations, financial institutions and tech companies holding sensitive data that are high-value targets for threat actors. North America additionally faces threats from organised crime groups [110] targeting credit card details, Social Security Numbers (SSN) and Personal health information (PHI) to monetise in the illegal market. Furthermore, strict regulations especially in the United States such as the California Consumer Privacy Act (CCPA) and Health Insurance Portability and Accountability Act (HIPAA) mandate data breach disclosures leading to higher reported incidents [94, p. 176].

In 2023, Europe overtook North America in the number of breaches reported, with the United Kingdom experiencing a surge in ransomware attacks [46]. The high internet penetration in Europe has expanded the digital landscape making it more vulnerable to cyber-attacks [111]. Some of the critics have attributed the increase in reported breaches to the 72-hour mandatory disclosure period under GDPR [80, p. 7] [112], but the author suggests that the changing threat landscape and the rapid evolution of sophisticated attacks are driving factors for it. Further, the GDPR requirement for EU citizen's data to be kept in Europe is forcing foreign companies to deploy their cloud servers in the EU itself. This has made cloud services massively adopted, leading to an increase in incidents through misconfigurations and inadequate security practices.

India suffered the highest personal data loss due to the Aadhar breach (refer to Figure 16 and Figure 33). This could be attributed to the lack of data protection law, transfer of personal data without user consent and poor security practices. India's Digital and Personal Data Protection Act (DPDPA) came into effect only in 2023 [113]. The author expects more breaches to be reported in coming years with the Data Protection Act in place.

Africa experienced fewer data breaches compared to North America and Europe (see Figure 28). African countries heavily rely on paper-based systems and have not fully digitalised minimising the attack surface for cybercriminals. Poor economic conditions and smaller digital infrastructures make them the least favourable target for extortion activities [114]. Also, language barriers further make it difficult to execute widespread phishing and social engineering attacks. However, the author predicts that with the increasing use of weaponised AI, such attacks will become easy to craft allowing cybercriminals to target various linguistic groups across Africa.

Finally, attention must be drawn to countries known for being on the attacking side causing data breaches and extortions through state-sponsored APTs. A few such examples include the Chinese APT group Stone Panda (APT 10) [115] linked to various attacks in healthcare, government and technology sectors; Russian APT group Fancy Bear (APT 28) [116] and Cozy Bear (APT 29) [43] known for targeting political entities, critical infrastructure and various industries ; and North Korean Lazarus group famous for Wanna Cry attack in 2017 and Bangladesh Bank heist in 2016 resulting in billions of loss to business and organisations worldwide [117].

The United Nations reveals that North Korea's cyber-attacks on crypto companies from 2017- 2023 brought in a revenue of $3 billion which was used to support weapons of mass destruction (WMD)

[118], [119]. As a response to violating sanctions, the UN imposed a ban on the trade of luxury goods and restrictions on North Korean workers from being employed in other countries to stop the inflow of foreign currency used to support the North Korean government's illicit programs [120]. However, North Korea continues to fund their nuclear programme through cryptocurrency and bitcoins extorted through cyber-attacks [117].

The author feels that the irony of war and peace points to a mutual deterrence, cyber warfare is just like nuclear weapons deterring enemies from attacking [121]. Nations use cyber weapons not just to steal data or extort money but also to gain strategic access for future operations. This includes targeting critical infrastructure, healthcare, power grids and banking systems to undermine the enemy's defences and ability to retaliate during a cyber war [103].

While the United Nations propose solutions to neutralise cyber threats and maintain global peace through open talks and collaboration it is hard to do so because countries do not trust each other. Consequently, the deeply rooted rivalries and conflicting interests between nations hinder effective cooperation on cybersecurity issues making it difficult to implement unified strategies for addressing these global threats.

### 5.1.5. RQ 5: How have data breaches evolved and are there any specific trends in various industries, methods used and types of data compromised?

As seen in Figure 34, the healthcare sector showed a significant peak in 2011 due to breaches in Tricare management, NHS and Healthnet compromising millions of records. All three high-profile breaches were due to lost devices (see Figure 42 below). However, Figure 35 showed a decline in breaches caused by lost devices after 2011. This could be attributed to the adoption of data encryption and Mobile Device Management (MDM) solutions by many organisations.

Webservices peaked in 2016 (refer to Figure 34) attributing to digitalisation and insecure IT infrastructure. Moreover, the rise in ransomware as a service model enabled threat actors without much technical skills to infiltrate IT systems. Retail breaches peaked in 2020 (refer to Figure 34) due to breaches in Marriot hotels, Rakuten and MGM hotels (see Figure 43 below). Retail industries have always had a consistent number of breaches as they were targeted by Magecart, a group of cybercriminals [34, p. 72]. Their attack tactics involve inserting malicious JavaScript code into the retailer's website to extract customer information or payment details [42]. According to Verizon, the sudden peak in 2020 [122], could be attributed to more online presence after COVID–19, especially in retail sectors relying on e-commerce and cloud-based workflows [122]. The least affected industry is NGOs may be because they are low-value targets compared to larger organisations and most of the data are publicly available and are non-sensitive in nature [78].

**Drill down view for factors contributing the 2011 peak in healthcare breaches**

| Entity | Industry | Attack Method | Year | Sum of Records |
|---|---|---|---|---|
| Health Net - IBM | Healthcare | Lost/Stolen Device | 2011 | 1900000 |
| Nemours Foundation | Healthcare | Lost/Stolen Device | 2011 | 2655489 |
| NHS | Healthcare | Lost/Stolen Device | 2011 | 17230000 |
| Sutter Medical Foundation | Healthcare | Lost/Stolen Device | 2011 | 4243434 |
| Tricare | Healthcare | Lost/Stolen Device | 2011 | 4901432 |
| **Total** | | | | **30930355** |

*Figure 42: Drill down view for factors contributing to the 2011 peak in healthcare breaches (Top 5)*

**Drill down view for factors contributing to the 2020 peak in retail breaches**

| Entity | Industry | Attack Method | Year | Sum of Records |
|---|---|---|---|---|
| Marriott Hotels | Retail | Insider Threat | 2020 | 5200000 |
| MGM Hotels | Retail | Hacked | 2020 | 10600000 |
| Rakuten | Retail | Misconfiguration | 2020 | 1381735 |
| Tokopedia | Retail | Hacked | 2020 | 91000000 |
| Wawa (company) | Retail | Hacked | 2020 | 30000000 |
| **Total** | | | | **138181735** |

*Figure 43: Drill down view for factors contributing to the 2020 peak in retail breaches (Top 5)*

Referring to Figure 35, Hacking breaches are growing attributed to the rise in ransomware attacks and zero-day exploits. Misconfiguration and credential stuffing appear to be an emerging threat. This could be attributed to increasing data migration from on-premises to the cloud for enhancing operational efficiency [69]. Mergers, acquisitions and data migrations introduce new complexities related to architectural design, risk assessments and oversight of sub-processors. Gaining complete visibility across diverse environments including private cloud, public cloud and hybrid models is challenging and contributes to delayed detection and response to unauthorised access and slow and low-intensity attacks [69]. The shortage of skilled cloud security personnel further makes it difficult for organisations to safeguard their data [123, pp. 6, 8].

Figure 36 shows a surge in PII breaches. The author argues that there are two major reasons for PII breaches, firstly cybercriminals compile large volumes of PII data from the dark web for identity theft and scam people for money laundering and fraud activities such as Facebook account takeovers, impersonation and blackmailing to pay the ransom if the data is of embarrassing nature [50, p. 6]. Secondly, PII data such as name, date of birth, address, phone no., email IDs, national IDs, passport no. etc when stolen can be used for years as these details do not change frequently. For the same reason, it is seen that the percentage of credit card data breaches is relatively lower compared to PII because as soon as the breach is reported, the credit card numbers can be blocked or accounts can be deactivated and even passwords can be changed swiftly, affecting the longevity of the data before hackers weaponise it [50, p. 6].

The significant peak in 2019 (refer to Figure 37) could be attributed to the disruptions caused by the COVID-19 pandemic. Verizon's 2020 data breach report reveals [122] that out of 474 data breaches, 36 were directly linked to COVID 19 with human error being the major cause. Companies were presented with the security challenge of keeping their corporate network safe with an increase in remote working. More attacks were targeted at stealing login credentials using phishing and social engineering tactics. It was found that phishing emails related to COVID-19 experienced a higher click rate than unrelated ones. A phishing simulation run on 16,000 people during COVID-19 revealed that people not only clicked the phishing email but also gave out credentials by signing into the fake websites, this number was three times more than the simulation phishing run in the previous year before COVID [122].

### 5.1.6. RQ 6: Is there any relation between industries, attack methods and type of data compromised?

Comparing Figure 38 and Figure 39 it is evident that hacking breaches are predominant in the web services sector due to the large volume of PII data involved. Lost devices were seen mostly in healthcare sectors attributed to poor security practices. According to a recent report published by HIPAA, the main targeted locations for healthcare breaches were network servers and email [124]. CIS states that cybercriminals target PHI because they are of more value than PII data, an average cost for PII is $158 whereas for PHI it is $335 [124]. PHI information can be used to claim medical settlements, file health insurance or even use or resale of prescriptions [124]. Another attack method targeted at the healthcare sector is ransomware. As disruption in operations can put people's lives at risk, healthcare sectors are more likely to pay ransom compared to other sectors. Moreover, many of the hospitals are dependent on legacy equipment for patient care forcing them to stick to older versions of the operating systems due to interoperability issues [101]. This opens room for unpatched vulnerabilities which are exploited by ransomware groups peaking healthcare breaches in 2023, according to an FBI report [18]. Zero-day exploits also peaked in 2023. These attacks are often targeted at critical infrastructure, cryptocurrency systems and wallets where the cybercriminals can get huge amounts of money. For example, the NotPetya cyberattack, the worst in history utilised a single zero-day vulnerability - EternalBlue along with Mimikatz to collapse operations in Ukraine causing damages in billions [125]. Financial sectors reported the highest number of full disclosures caused by insider threats. This could be attributed to the fact that employees working in banks can easily find security loopholes to gain access to account details and customer information and use it to create fraudulent accounts or reset bank cards for monetary gain. Wells Fargo's [126] cross-selling account scam shows the extent to which insider threats can gain control over customer information and misuse it for personal gain. Finally, misconfigurations were mainly seen in financial services owing to factors such as shadow IT, lack of visibility in complex environments, human error and lack of expertise/training [123, pp. 6, 8]. The most frequently seen types were unsecured AWS S3 bucket and improper server-side configuration leading to unauthorised access to the database compromising millions of records.

## 5.2. Conclusion

This chapter explored the arguments and counterarguments for the trends identified in Chapter 4. The chapter highlighted the implications of various data breach risks enabling organisations to identify their vulnerabilities better and manage their security spending effectively. The findings from the research can also be fed into their threat intelligence process to identify internal and external threats and improve their security awareness training to reflect a more current threat landscape. The next chapter restates the research objectives and critically reviews them to see to what extent the research objectives are met. The chapter further provides recommendations for prevention and mitigation strategies and concludes by analysing the limitations and future directions.

# 6. Conclusion

## 6.1. Summary and Self Reflection

This section provides a summary of the work done so far and reflects on the research objectives to see whether they have been met or not and to what extent. Limitations are identified in section 6.3.

**Research Topic and its Significance:** The research aimed at providing key insights into data breach trends across different industries, regions, attack methods and data sensitivity to help improve cyber resilience which is the ability of the organisation to prepare, respond and recover from cyber-attacks and security breaches.

**Research motivation:** The research theme was inspired by the ICO's Data Security Incident Dashboard developed in Power BI.

**Literature review and gap analysis:** The research explored industry-wide data breach reports from sources such as IBM, Verizon, Identity Theft Resource Center, GOV.UK, DTEX, FBI, Ponemon Insitute, SecurityScorecard, Black Kite, Imperva and many others to understand the current threat landscape and emerging threats. Also, academic papers were reviewed to identify gaps and extend previous work. Many of the previous studies analysed data breaches using the Privacy Rights Clearing House database which only contains breaches in the United States. This research combined two data sources to conduct a global analysis of data breaches from 2004 to 2024.

**Methodology:** The research provided a descriptive analysis of data breach trends using visual analytics techniques in Power BI. The data sources identified were Information is Beautiful: World's Biggest Data Breaches & Hacks - Source A and Wikipedia: List of Data Breaches - Source B. Of this, Source B had two data tables one for companies and the other for government and public entities. All three data tables were combined to produce a complete data set. Source A was filtered out for government entities and combined with the government entities of Source B for regional analysis. Classification tables were created for attack methods, industries and regions to remove redundancy and map values as best fit.

**Novel aspects:** The data sources are connected in real-time implying that whenever data updates on the designated websites, it will reflect in the Power BI dashboard upon refresh.

**Challenges:** Integration challenges for three data tables due to inconsistencies in data structures and data types, missing fields, typos and errors were addressed in the data cleaning and data modelling stages within Microsoft Power BI.

**Ethical issues:** The dataset used is available publicly and does not contain any personal data. Power BI loads data in a read-only mode, any manipulations performed are not reflected in the source.

**Research objectives and research questions:** This section restates the research objectives and research question highlighting the key findings and arguments made.

**OBJ1: to provide an overview of affected industries, attack methods, data sensitivity and regions to understand the overall landscape of data breaches.**

**RQ 1: Which industries or sectors are targeted the most?**

Web services, healthcare, government and financial services are the top four industries with the greatest number of incidents recorded, of which the Web services sector is the most impacted accounting for over 23% of all breaches.

Complex IT environments, insecure APIs, third-party integrations and third-party vendors combined with poor security practices were identified as the major contributors to data breaches. In the government sector, it was observed that state-sponsored attackers not only aimed to steal data but also maintain access for future strategic operations.

**RQ 2: Which attack methods are more prevalent?**

Hacking accounts for most data breaches at 69% followed by poor security at 14%. Although lost or stolen devices were historically the third most common cause of data breaches, it is now observed that credential stuffing and insider threats compromise more records compared to lost or stolen devices.

Further exploration of hacking breaches caused by zero-day exploits highlighted the operational aspects of the zero-day market and regulatory challenges. The challenges identified were the morally dubious and anonymous nature of the zero-day market, geographical implications and incentive issues. The author highlighted the need for international cooperation in making global IT infrastructure safe for everyone.

**RQ 3: What type of data or data sensitivity is at risk of compromise?**

PII accounted for 35% of data breaches, whereas credit card details and PHI accounted for 15-16% of data breaches.

The factors contributing to PII breaches were identified as lack of visibility into the digital estate, lack of basic data security controls such as encryption, network segmentation, MFAs and role-based access controls and failure to identify, classify and protect data based on sensitivity. A further exploration of the cost of PII in the dark web revealed that some of the compromised accounts of online banking services were available for less than $50. The price range differed due to various reasons such as the attack's sophistication, brand value, seller reputation, volume and demand and freshness of the data.

**RQ 4: Are there any regional biases in data breaches, especially for government and public entities?**

Over the years, the United States had the greatest number of data breaches. However, Europe overtook the United States as the most affected region in 2023. One of the contributing factor identified was large-scale breaches in United Kingdom due to hacking.

North America faced the highest number of breaches in the United States. The factors identified were large digital expansion, high-value targets in MNCs and financial services and organised crimes targeted at stealing credit card details. For Europe, cloud migrations and ransomware attacks resulted in a high number of incidents. Africa experienced fewer breaches due to smaller infrastructure and language barriers. Countries on the attacking side such as North Korea used the money extorted from cyber-attacks to fund their nuclear programs. United Nations attempts to neutralise cyber threats suffers due to conflicting interests between nations and ongoing rivalries. The author highlighted that the paradox of war and peace points to a mutual deterrence, cyber warfare is just like nuclear weapons deterring enemies from attacking

**OBJ 2: to examine data breach categories over time to identify dominant, emerging and declining trends.**

**RQ 5: How have data breaches evolved and are there any specific trends in various industries, methods used and types of data compromised?**

*Industries:* Healthcare breaches peaked over the years due to increase in ransomware attacks and poor security practices, retail experienced attacks from organised crime groups aimed at stealing credit card details during the checkout process and cloud misconfiguration breaches were identified to be more common in web, technology and financial services. The sudden peak in 2019 was identified to be related to COVID-19 with people giving out credentials signing into fake websites.

Attack method: Hacking was the dominant attack method observed due to the rise in ransomware attacks, phishing, zero-day exploits, application vulnerabilities and third-party exploits. Misconfiguration and credential stuffing appeared to be an emerging threat owing to factors such as cloud migration, mergers and acquisitions, lack of skilled security experts and lack of oversight of subprocesses causing delayed detection and response. Lost devices showed a declining trend due to the adoption of data encryption and other endpoint controls by organisations.

Type of data compromised: The percentage of breaches associated with credit card details was substantial yet lesser compared to PII due to longevity issues. The reasons for PII breaches were identified as identity theft, money laundering and fraud.

The projection of data breaches showed an increasing trend underscoring the need for improved cybersecurity.

Emerging Threats: The emerging threats highlighted were as follows:

- Ransomware attacks have moved away from just locking away data to double extortion where data is exfiltrated and threatened to release if the ransom is not paid, and triple extortion where the victim's customer or partners are attacked to build pressure into paying ransom. The RaaS model has enabled less sophisticated attackers to launch successful attacks.

- Cloud misconfigurations are on the rise. A shift to secure-by-design models is a necessity with an increase in cloud migrations.
- AI is being used to supercharge cyber-attacks. Identified use cases include phishing, deep fake videos, automated vulnerability detection and malicious LLM models like fraud GPT and Worm GPT.
- Supply chain risks have become easy entry points for attackers, especially for ransomware groups to infiltrate data and steal credentials without being noticed until a ransom is demanded.
- Quantum computing is a ticking time bomb that can decrypt the current data extorted by ransomware groups.

**OBJ 3: to identify any relation between targeted industries, attack methods and data compromised to understand the prominent risk factors in various categories.**

**RQ 6: Is there any relation between industries, attack methods and type of data compromised?**

Hacking was the dominant method used for web service breaches. Lost devices and ransomware affected the healthcare sectors severely compared to other attack methods. Zero-day exploits were mainly targeted at critical infrastructure and cryptocurrency systems. Financial sectors had the highest number of full disclosure breaches caused by insider threats. The common types of misconfiguration breaches seen were unsecured AWS S3 buckets, improper handling of session tokens and server-side configurations.

Overall, the author evaluated multiple industry reports and academic reports and visually analysed data sets in Power BI to address the research objectives in depth. This project helped the author gain a good understanding of the threat landscape and its wider implications on organisations and policymakers. The research objectives were mainly addressed in Chapter 4 and Chapter 5. Refer to Table 1 for a comprehensive mapping of chapters with respect to research objectives.

## 6.2. Recommendation:

Despite numerous high-profile cases in the past, many organisations are still unprepared to manage data breaches. Table 4 identified the hacking types in the section 2.4.2. This section provides recommendation strategies to improve cyber resilience as seen in Table 7. References are drawn from ISACA Journal, Volume 2, 2024 –The (r)evolution of data ecosystem [127], and previous experience.

| Incident Type | Detect | Prevent | Respond | Recover |
|---|---|---|---|---|
| Application Vulnerability | Vulnerability Scanning, Penetration testing, Code Review | Secure Coding, Regular patching | Fix Vulnerability, Disable or isolate affected components. | Restore application from backup, Update incident register. |
| Third-Party (Supply chain) | Monitor Vendor activities, Vendor audits | Vendor Due Diligence, DPA (Data Processing Agreement), MSA (Master Service Agreement) | Terminate compromised third-party connection,<br>Review and revoke affected credentials or permissions | Restore systems from backups,<br>Replace Vendor if necessary |
| Malicious Insider | SIEM (Security Information and event management) to monitor user activity | Background checks, DLP (data Loss prevention), employee training, RBAC (Role based access control) | Restrict access,<br>Follow breach notification policy if necessary | Access audits,<br>Disciplinary action |
| Phishing | Email and Web Filtering, incident reporting tickets | Security awareness training, phishing simulations | Password reset, quarantine emails, takedown phishing website | Update email security filters, conduct follow-up training and simulations. |
| Network Sniffing | Firewall, IDS/IPs (Intrusion detection and prevention) | Encrypt data at rest and transit, use security protocols- HTTPs, SSH, VPNs, network segmentation, disable unnecessary services and ports | Block malicious IP, Identify and isolate malicious device | Review Firewall rules |
| Leaked/Exposed Credentials | Unusual login activities and risky sign-ins,<br>Dark web monitoring | Password Manager, MFA, Strong password policy | Reset exposed credential | Access review, notify and advise user on password hygiene |

| Incident Type | Detect | Prevent | Respond | Recover |
|---|---|---|---|---|
| Brute Force | IDS/IPS alerts | Strong password policy, MFA, CAPTCHA, implement rate limiting | Block IP addresses, Lock account | Account recovery, reset password. |
| Malware, Ransomware | Antivirus software, EDR (Endpoint Detection and Response) | Regular backups, Regular patching and software update, use antivirus and anti-malware software, security awareness training | Block erroneous connection, Isolate infected system | Rebuild affected system to remove malware<br><br>Restore from clean backups |
| SQL Injection [128] | WAF (Web Application Firewall), input validation, DAST (Dynamic Application Security Testing) scans, monitor database errors | Use parameterised queries, input sanitisation | Reset password, fix the underlying SQL injection flaw, Database Audit to check for backdoors | Restore from clean backups, conduct regular security audit. |
| Cloud Misconfiguration | CSPM (Cloud Security Posture Management) | Identity and Access Management,<br><br>cloud audit trail, cloud config management,<br><br>IaC (Infrastructure as a code), establish security baseline and use predefined templates | Correct misconfiguration,<br><br>Update security policies | Autoscaling and System restoration<br><br>Configuration hardening, |

*Table 7: Data breach prevention and remediation strategies*

## 6.3. Limitations:

The dataset starts at 2004, data breaches before this period is not included. All data breaches that occurred over time might not be captured especially the ones that were never reported or discovered. The first dataset [21] only includes data breaches involving personal or sensitive data. It does not cover breaches involving other types of data. The form of attacks captured in the dataset is limited to poor policies, hacked, lost devices and inside jobs. For the second dataset [22] no data sensitivity field is captured, but the attack method captures additional forms of attack such as misconfiguration, social engineering, and accidentally leaked or published. To overcome these limitations, both datasets are combined to gain better insights into data breaches over time.

The varying levels of detail about each data breach make it difficult to compare and analyse the breaches. Some entries include only basic information about the breach, while others include detailed information making it difficult to draw accurate conclusions.

The dataset may be skewed due to corporate under reporting. Comparatively, more data breaches are reported by regulated industries such as finance and hospitals than by unregulated industries such as non-profit organisations and less regulated institutions such as education and retail [13]. The data may be more biased towards bigger organisations as small to medium-sized businesses are reluctant to report breaches [11].

While regional analysis of data breaches in government and public entities offers insights into general trends, it should not be considered comprehensive as some countries are missing such as Canada and Germany. Thus, caution is advised when interpreting results due to unavailability of regional data for broader analysis.

Although efforts have been made to identify and correct discrepancies some inaccuracies may remain such as incorrect dates or record compromised due to human error or misreporting by the organisations involved.

Furthermore, these datasets are compiled from multiple sources such as the New York Times, Forbes, Guardian, Tech Radar, BBC, PC Mag, Tech Crunch, IdTheftCentre, DataBreaches.net and other news reports [21] . Hence the accuracy and reliability of the data may vary. Although it is possible to verify the source of the data, the datasets may not be up-to-date and accurate.

## 6.4. Future direction for research

It is found that hacking breaches are growing mainly due to ransomware attacks and zero-day exploits [34]. Though this research investigated the dark secrets and motivations behind zero-day exploits, it did not deep dive into the evolution of ransomware attacks due to time constraints. Figure 44 shows ransomware attacks by sector. The "Ransomware Attacks" data [129] from the website – "Information is beautiful" can be combined with the datasets used in this research to provide a broader analysis of data breach trends and insights. The dataset can provide insights on:

1. industries targeted by ransomware
2. entities affected
3. time series analysis of ransomware
4. prevalent ransomware groups
5. cost of ransomware breach
6. locations targeted

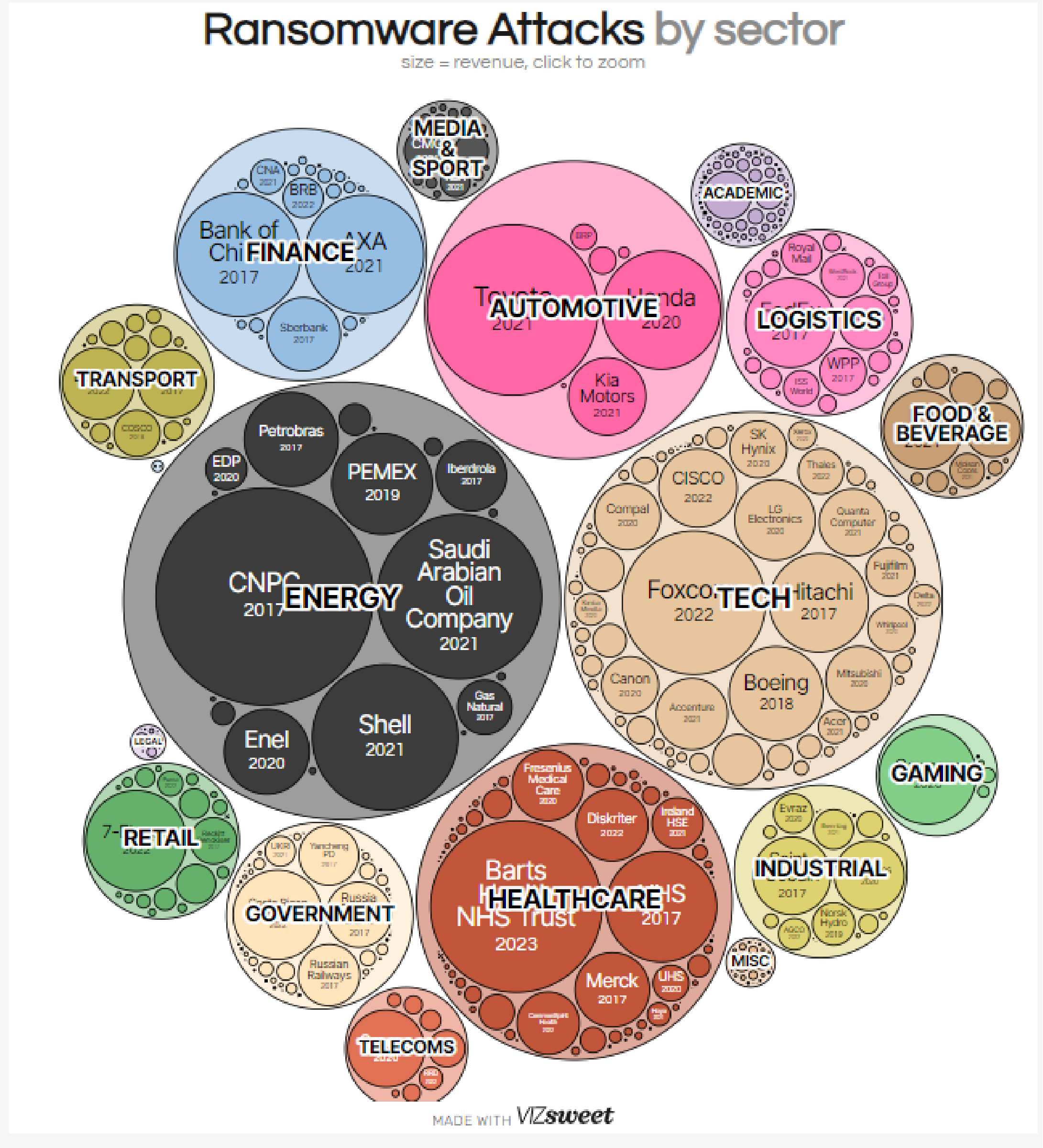

*Figure 44: Ransomware attacks by Sector* [129]

Another approach to studying data breaches is incorporating real-life case studies and drawing lessons to improve prevention and response strategies. Looking at failures might be rather unusual as many researchers focus on how security should be implemented correctly. However, examining such failures can be more informative and can help avoid the errors that led to the failure in the first place.

Bitdefender Cyber Security assessment report [123] suggests that attacks powered by AI will increase in future. Though there are legislations to regulate the use of AI, the author feels that these regulations may not be able to keep up with the changes in AI technology. Future research can investigate the regulatory and compliance risks for data breaches caused by AI systems.

An IBM report states that 82% of data breaches are due to data stored in the cloud with 39% of breaches in multi-cloud environments [1, p. 6]. The main causes of cloud breaches are misconfiguration, credential stuffing via insecure API, third-party integration and weak authentication controls [46]. An organised study focusing on the security posture of cloud-migrated applications will help minimise the cost of cloud breaches.

Finally, future studies can focus on the risks associated with AI technology and how it can be integrated with the incident response process to detect, contain, eradicate and recover from breaches while ensuring its fair, responsible and transparent use.

## 6.5. Concluding thoughts

The threat landscape is ever evolving and there is no certainty on how future threats would look like or from where they will come from. Attackers won't stop innovating and will continue to use new technologies such as AI to supercharge their attacks. Traditional reliance on rules, signatures and playbooks to combat security breaches may not work in future as new and novel attacks will easily bypass these rules and allow attackers to infiltrate the system. What is important here is to keep up to date with emerging risks and adopt security practices to protect digital assets and third-party ecosystems. Also, though most data breaches are motivated by financial gain, the attacks are targeted not only at large corporations and well-known organisations but also at those who are not prepared [78]. Attackers do not see any difference between the development or production environment. Relating this to the castle analogy, a small crack in the wall can be exploited to enter the castle and cause huge devastation. Hence it is crucial to have complete visibility and context of the digital estate rather than looking at it as part of siloed solutions.

The author compares data breach attacks to medical viruses that can affect any organism irrespective of their status. The human body is immune to viruses or diseases that have been researched and for which vaccines have been developed. However, as witnessed during COVID-19, when a new form of virus enters the human body, the body immediately detects it, and then antigens fight back to restore immunity. Similarly, future security defences should replicate the human immune system that can detect new threats and respond to them quickly rather than relying on rule-based solutions. This would not be possible without integrating new technologies such as AI to have complete visibility over the entire digital estate including emails, network, endpoint and cloud. But above all, people are at the heart of security and the front-line defenders [127]. Management must promote a security culture to have a vigilant staff who are trained and aware of the security risks. Cyber resilience can only be achieved through joint efforts of management, the security team and every employee in the organisation. A proactive approach to data security is a must. Organisations should consider that the threat is already in and mitigate it while addressing security, privacy and operational risks [127].

# Appendix A – Dashboard Screenshot

As mentioned in section 3.5, Appendix A provides screenshots of the Power BI dashboard components.

HOME PAGE

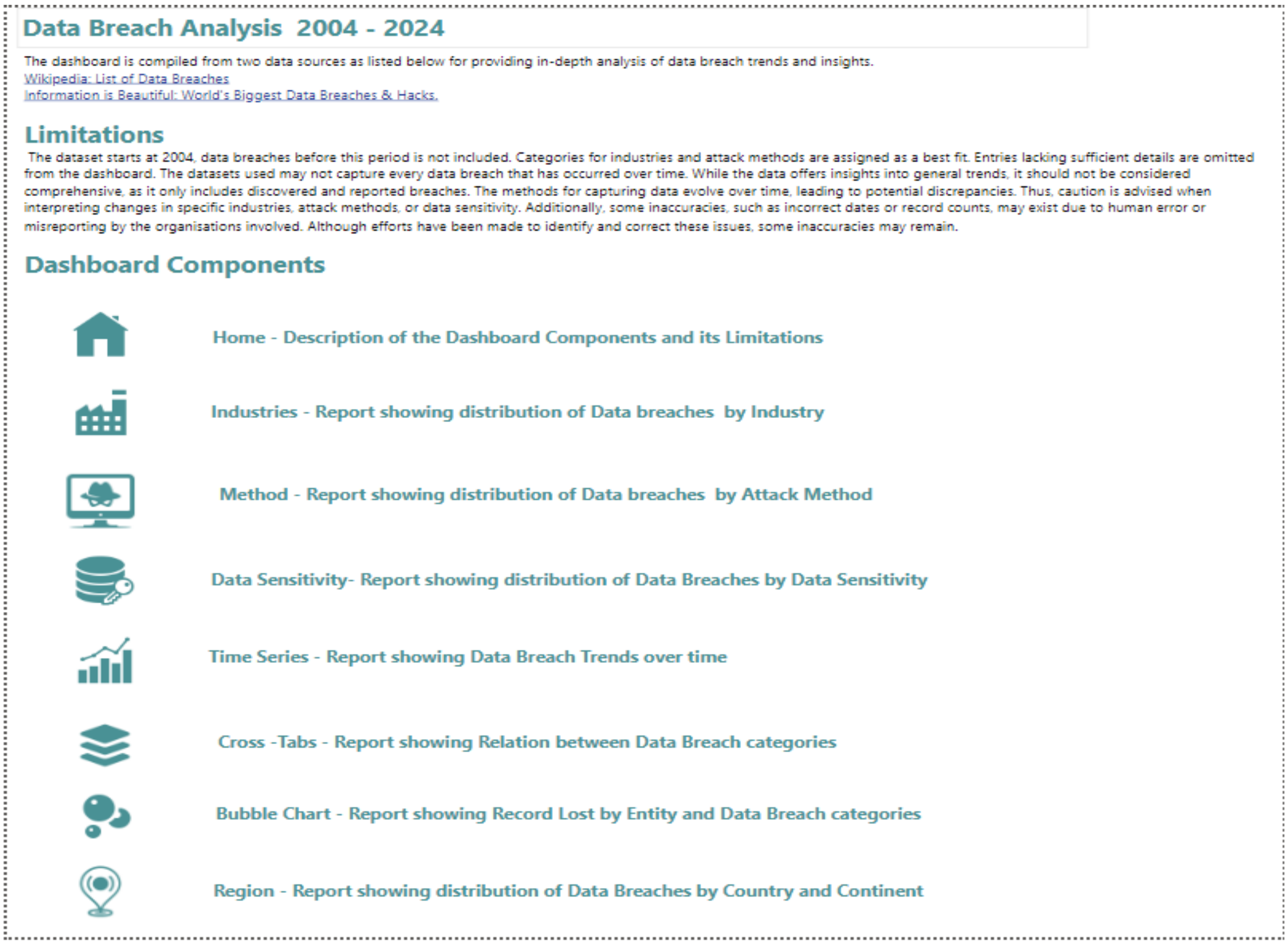

INDUSTRIES

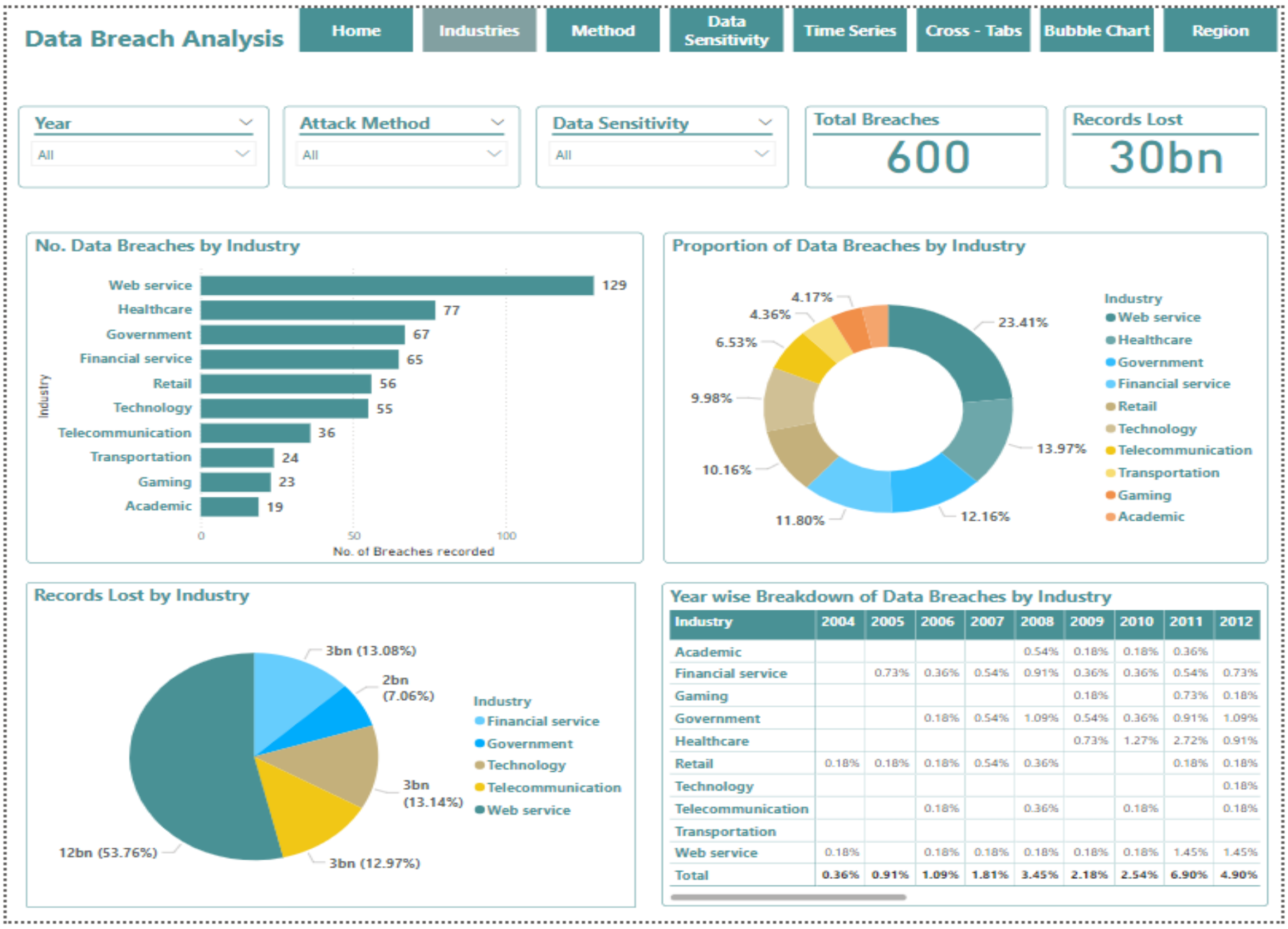

ATTACK METHOD

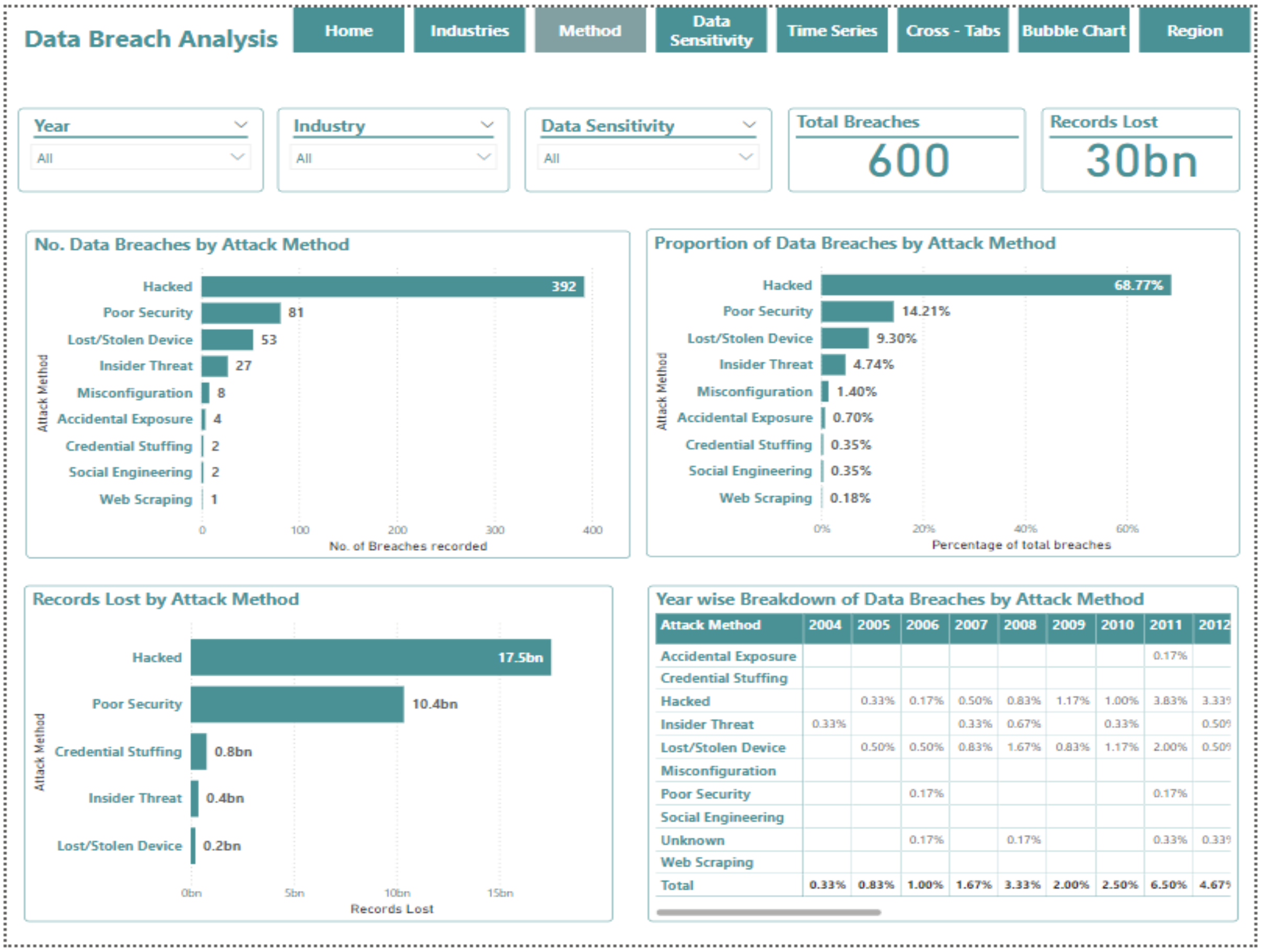

DATA SENSITIVITY

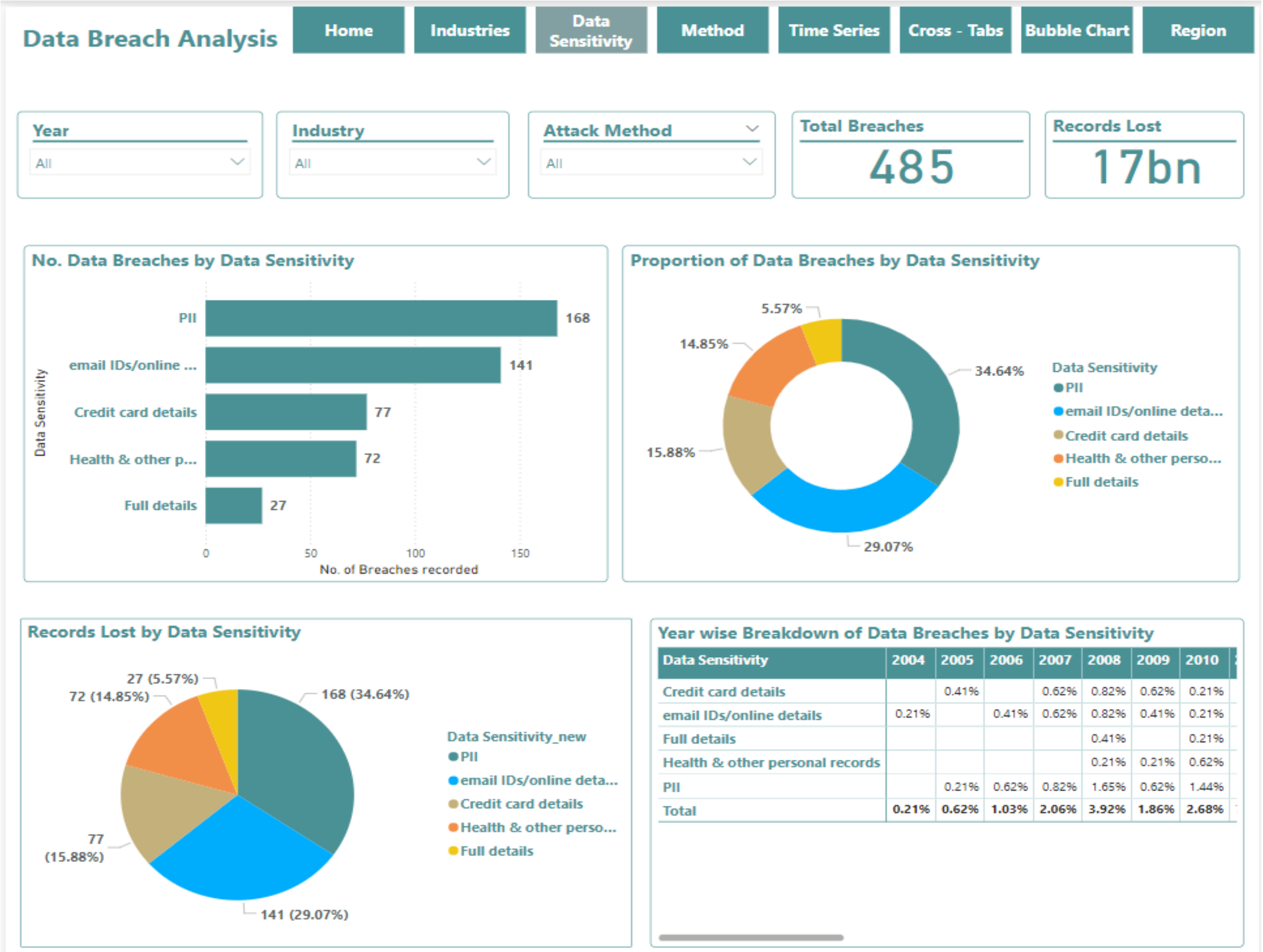

TIME SERIES

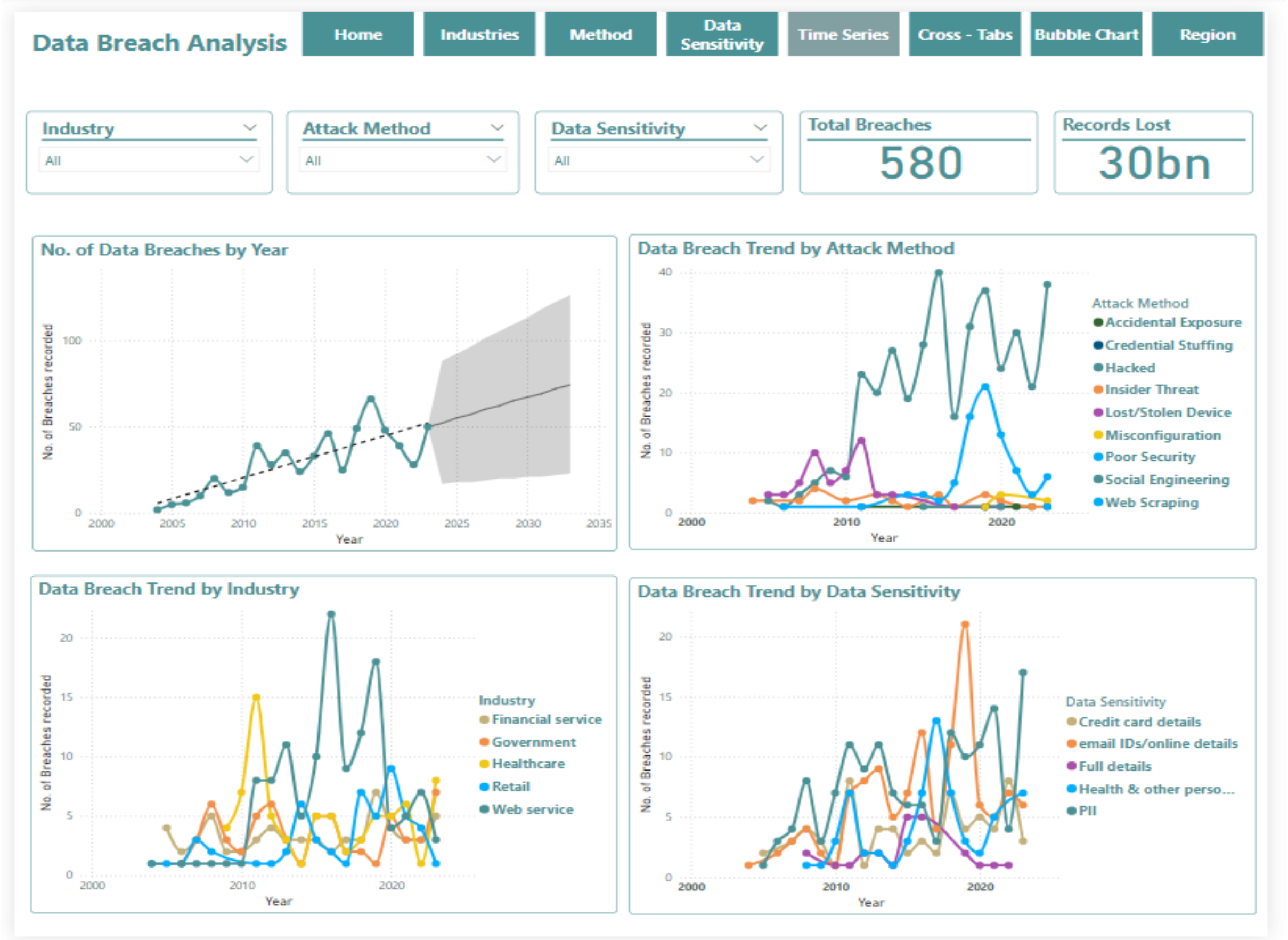

CROSS TABS

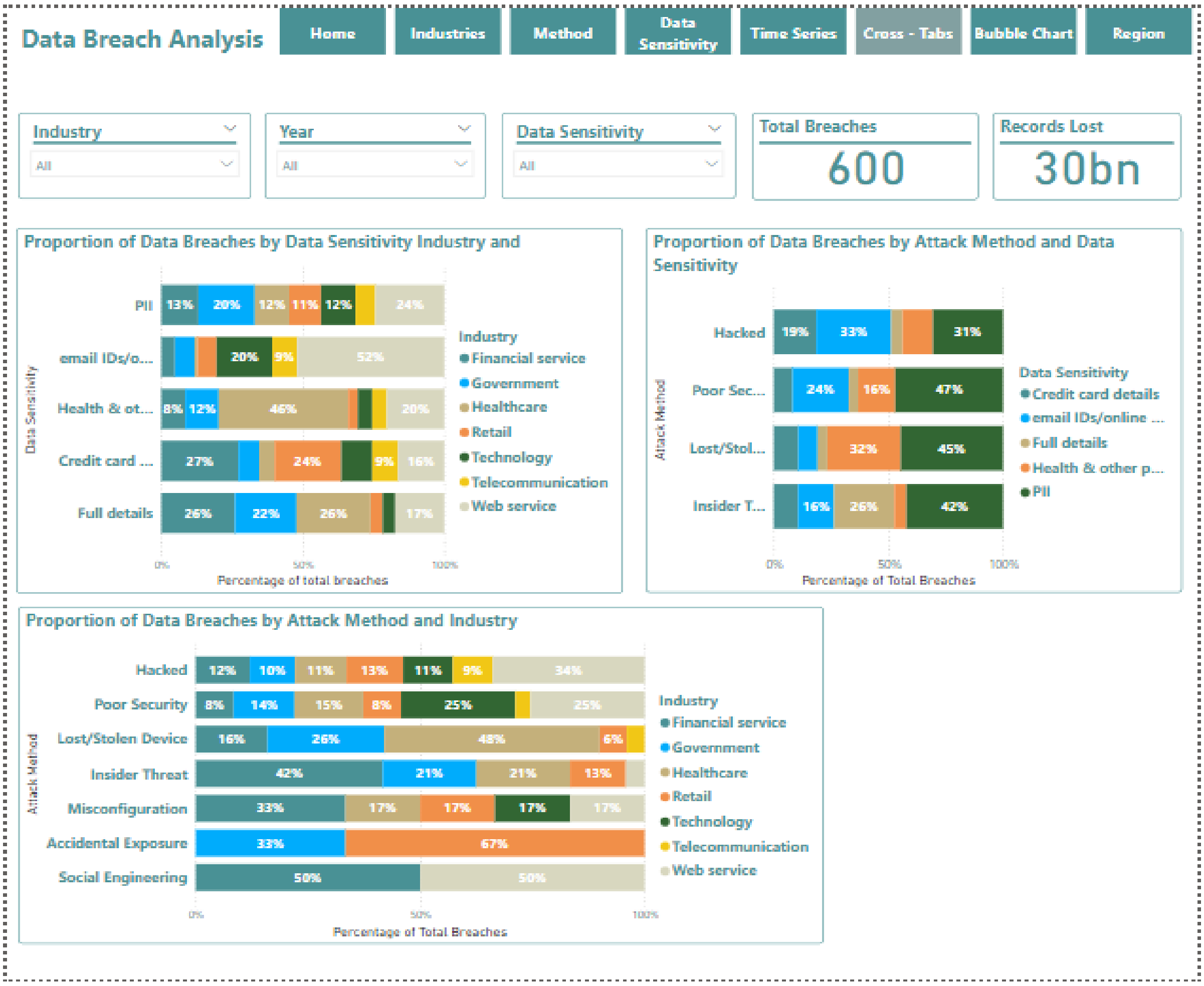

BUBBLE CHART

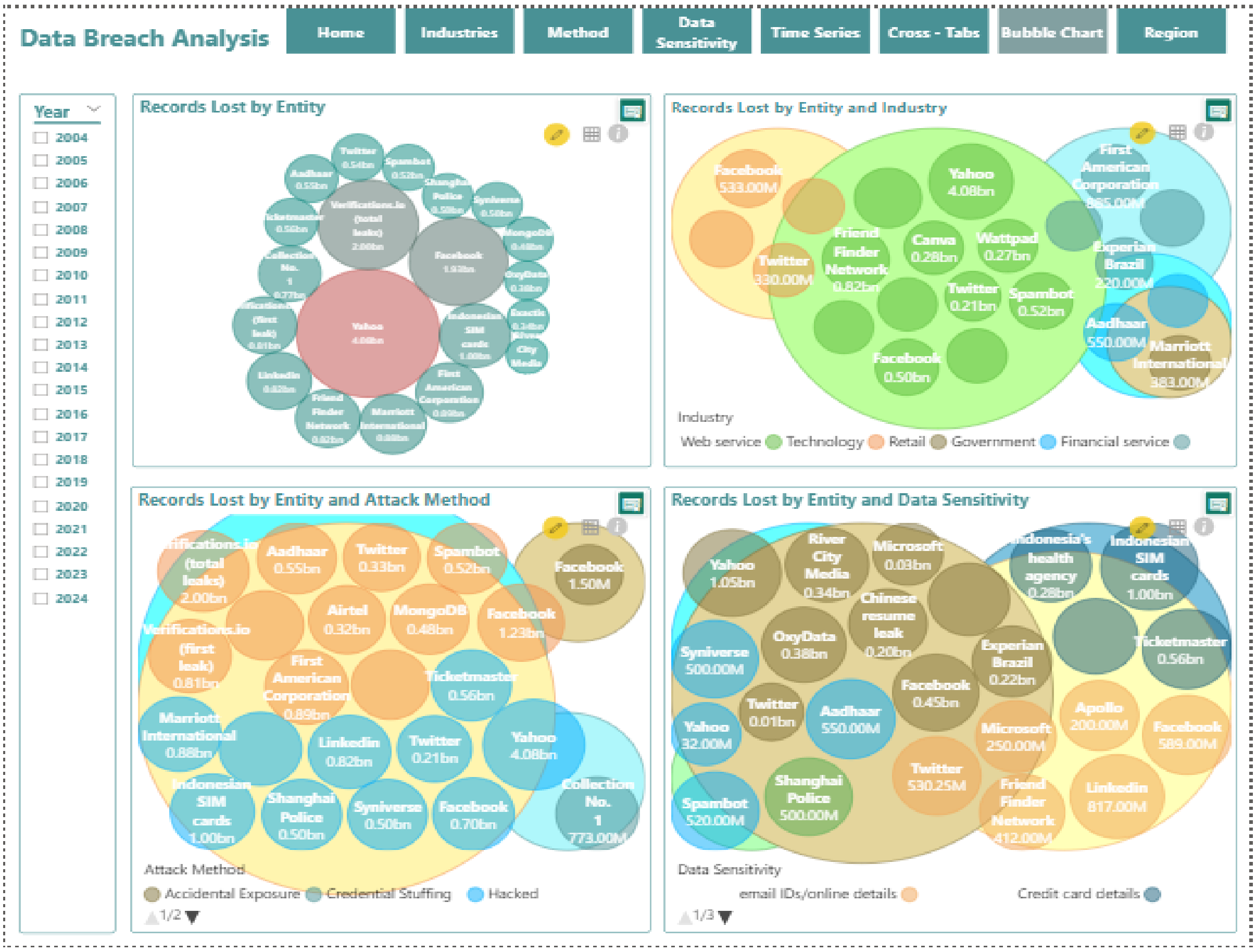

REGION

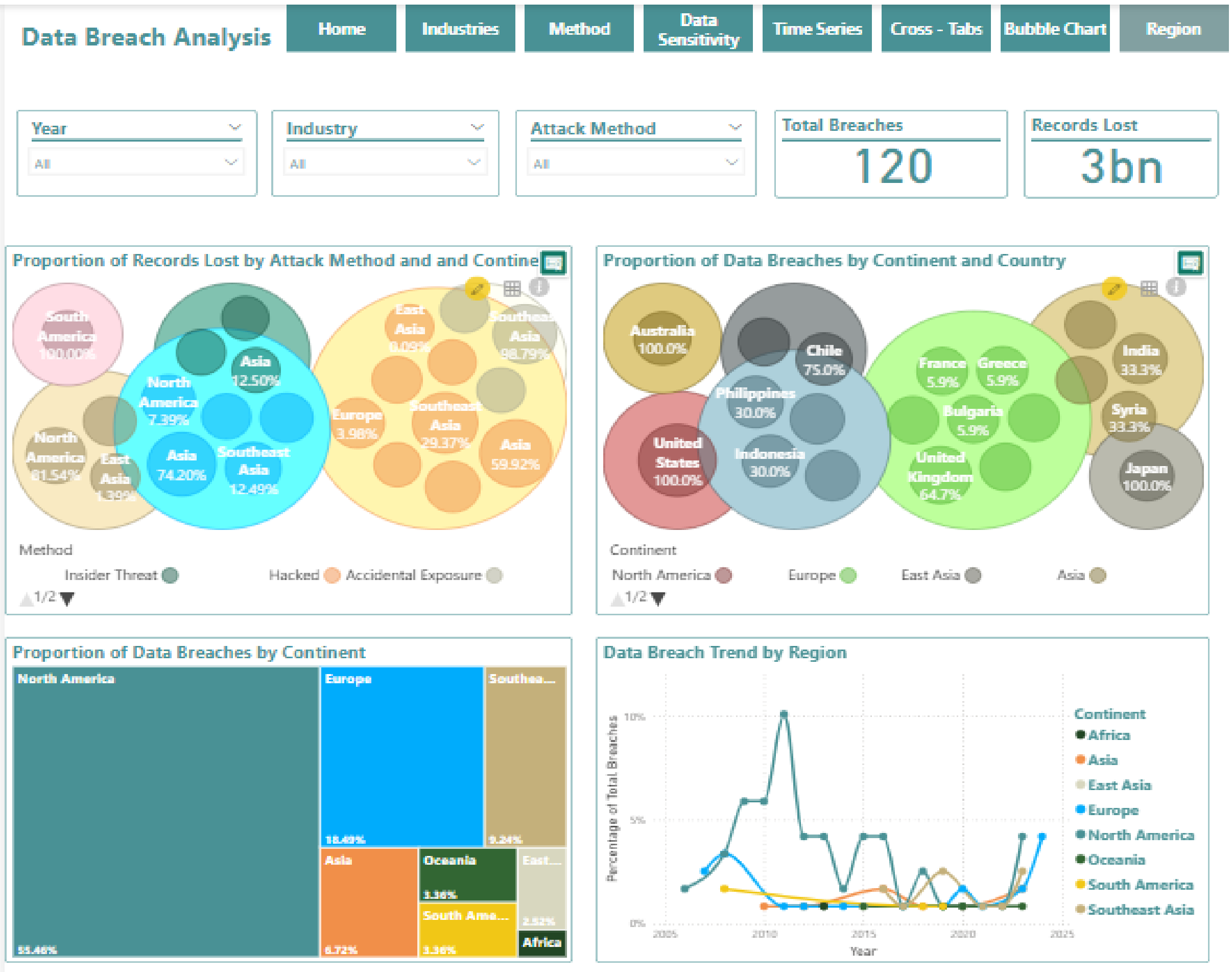

# Appendix B – Dataset and Preprocessing steps

As mentioned in section 3.5, appendix B provides snapshots of data preprocessing steps and dataset used.

## Preprocessing Steps

1. Information is Beautiful -Worlds's Biggest Breaches and Hacks

Advanced Editor

informationisbeautiful

Display Options

```
let
    Source = GoogleSheets.Contents("https://docs.google.com/spreadsheets/d/1i0oIJJMRG-7t1GT-mr4smaTTU7988yXVz8nPlwaJ8Xk/edit?gid=2"),
    breaches_Table = Source{[name="breaches",ItemKind="Table"]}[Data],
    #"Promoted Headers" = Table.PromoteHeaders(breaches_Table, [PromoteAllScalars=true]),
    #"Changed Type" = Table.TransformColumnTypes(#"Promoted Headers",{{"organisation", type text}, {"alternative name", type text},
        {"records lost", type any}, {"year   ", type any}, {"date", type any}, {"story", type text}, {"sector", type text}, {"method",
        type text}, {"interesting story", type text}, {"data sensitivity", type any}, {"displayed records", type text}, {"", type any},
        {"source name", type text}, {"1st source link", type text}, {"2nd source link", type text}, {"ID", Int64.Type}}),
    #"Removed Top Rows" = Table.Skip(#"Changed Type",1),
    #"Reordered Columns" = Table.ReorderColumns(#"Removed Top Rows",{"organisation", "year   ", "records lost", "date", "sector",
        "method", "data sensitivity"}),
    #"Renamed Columns" = Table.RenameColumns(#"Reordered Columns",{{"organisation", "Entity"}, {"year   ", "Year"}, {"records lost",
        "Records"}, {"sector", "Organisation Type"}, {"method", "Method"}}),
    #"Renamed Columns1" = Table.RenameColumns(#"Renamed Columns",{{"data sensitivity", "Data sensitivity"}}),
    #"Changed Type1" = Table.TransformColumnTypes(#"Renamed Columns1",{{"date", type date}}),
    #"Removed Errors" = Table.RemoveRowsWithErrors(#"Changed Type1", {"date"}),
    #"Removed Duplicates" = Table.Distinct(#"Removed Errors", {"Entity", "Year", "Records"}),
    #"Added Custom" = Table.AddColumn(#"Removed Duplicates", "Source", each "information is beautiful"),
    #"Renamed Columns2" = Table.RenameColumns(#"Added Custom",{{"Organisation Type", "Sector"}}),
    #"Filtered Rows" = Table.SelectRows(#"Renamed Columns2", each ([Entity] <> "") and ([Data sensitivity] <> "")),
    #"Removed Columns" = Table.RemoveColumns(#"Filtered Rows",{"alternative name", "story", "interesting story", "displayed records", "",
        "source name", "1st source link", "2nd source link", "ID"})
in
    #"Removed Columns"
```

✓ No syntax errors have been detected.

Done Cancel

2. Wikipedia – List of data breaches involving company

Advanced Editor

wikipedia - Companies

Display Options

```
let
   Source = Web.BrowserContents("https://en.wikipedia.org/wiki/List_of_data_breaches"),
    #"Extracted Table From Html" = Html.Table(Source, {{"Column1", "TABLE.wikitable.sortable:nth-child(14) > * > TR > :nth-child(1)"},
        {"Column2", "TABLE.wikitable.sortable:nth-child(14) > * > TR > :nth-child(2)"}, {"Column3", "TABLE.wikitable.sortable:nth-child
        (14) > * > TR > :nth-child(3)"}, {"Column4", "TABLE.wikitable.sortable:nth-child(14) > * > TR > :nth-child(4)"}, {"Column5",
        "TABLE.wikitable.sortable:nth-child(14) > * > TR > :nth-child(5)"}, {"Column6", "TABLE.wikitable.sortable:nth-child(14) > * > TR
        > :nth-child(6)"}}, [RowSelector="TABLE.wikitable.sortable:nth-child(14) > * > TR"]),
    #"Changed Type" = Table.TransformColumnTypes(#"Extracted Table From Html",{{"Column1", type text}, {"Column2", type text}, {"Column3",
         type text}, {"Column4", type text}, {"Column5", type text}, {"Column6", type text}}),
    #"Promoted Headers" = Table.PromoteHeaders(#"Changed Type", [PromoteAllScalars=true]),
    #"Changed Type1" = Table.TransformColumnTypes(#"Promoted Headers",{{"Entity", type text}, {"Year", type text}, {"Records", type text},
         {"Organization type", type text}, {"Method", type text}, {"Sources", type text}}),
    #"Added Conditional Column" = Table.AddColumn(#"Changed Type1", "Data Sensitivity", each if [Year] = "hkklo" then "" else null),
    #"Removed Duplicates" = Table.Distinct(#"Added Conditional Column", {"Entity", "Year", "Records"}),
    #"Removed Columns" = Table.RemoveColumns(#"Removed Duplicates",{"Sources"}),
    #"Added Custom" = Table.AddColumn(#"Removed Columns", "Data Sensitivity.1", each ""),
    #"Removed Columns1" = Table.RemoveColumns(#"Added Custom",{"Data Sensitivity"}),
    #"Renamed Columns" = Table.RenameColumns(#"Removed Columns1",{{"Data Sensitivity.1", "Data Sensitivity"}}),
    #"Added Custom1" = Table.AddColumn(#"Renamed Columns", "Source", each "list of databreaaches-wikipedia"),
    #"Renamed Columns1" = Table.RenameColumns(#"Added Custom1",{{"Organization type", "Sector"}}),
    #"Removed Columns2" = Table.RemoveColumns(#"Renamed Columns1",{"Data Sensitivity"})
in
    #"Removed Columns2"
```

✓ No syntax errors have been detected.

Done Cancel

3. Wikipedia- List of data breaches involving government and public entity

Advanced Editor

## wikipedia - Government/Public Entity

Display Options

```
let
    Source = Web.BrowserContents("https://en.wikipedia.org/wiki/List_of_data_breaches"),
    #"Extracted Table From Html" = Html.Table(Source, {{"Column1", "TABLE.wikitable.sortable:nth-child(12) > * > TR > :nth-child(1)"},
        {"Column2", "TABLE.wikitable.sortable:nth-child(12) > * > TR > :nth-child(2)"}, {"Column3", "TABLE.wikitable.sortable:nth-child
        (12) > * > TR > :nth-child(3)"}, {"Column4", "TABLE.wikitable.sortable:nth-child(12) > * > TR > :nth-child(4)"}, {"Column5",
        "TABLE.wikitable.sortable:nth-child(12) > * > TR > :nth-child(5)"}, {"Column6", "TABLE.wikitable.sortable:nth-child(12) > * > TR
        > :nth-child(6)"}, {"Column7", "TABLE.wikitable.sortable:nth-child(12) > * > TR > :nth-child(7)"}},
        [RowSelector="TABLE.wikitable.sortable:nth-child(12) > * > TR"]),
    #"Promoted Headers" = Table.PromoteHeaders(#"Extracted Table From Html", [PromoteAllScalars=true]),
    #"Changed Type" = Table.TransformColumnTypes(#"Promoted Headers",{{"Government", type text}, {"Agency", type text}, {"Year",
        Int64.Type}, {"Records", type text}, {"Organization type", type text}, {"Method", type text}, {"Sources", type text}}),
    #"Renamed Columns" = Table.RenameColumns(#"Changed Type",{{"Government", "Region"}, {"Agency", "Entity"}, {"Organization type",
        "Sector"}}),
    #"Removed Columns" = Table.RemoveColumns(#"Renamed Columns",{"Sources"}),
    #"Added Custom" = Table.AddColumn(#"Removed Columns", "Source", each "Wikipedia Region"),
    #"Removed Duplicates" = Table.Distinct(#"Added Custom", {"Region", "Entity", "Year"}),
    #"Filtered Rows" = Table.SelectRows(#"Removed Duplicates", each ([Entity] <> "?") and ([Method] <> "") and ([Sector] <> "")),
    #"Merged Queries" = Table.NestedJoin(#"Filtered Rows", {"Region"}, #"region classification", {"Region"}, "region classification",
        JoinKind.LeftOuter),
    #"Expanded region classification" = Table.ExpandTableColumn(#"Merged Queries", "region classification", {"Region", "Country",
        "Continent"}, {"region classification.Region", "region classification.Country", "region classification.Continent"}),
    #"Filtered Rows1" = Table.SelectRows(#"Expanded region classification", each ([region classification.Region] <> null))

in
    #"Added Conditional Column1"
```

✓ No syntax errors have been detected.

Done Cancel

4. Complete dataset (Result dataset after preprocessing steps)

Advanced Editor

## complete dataset

Display Options

```
let
    Source = Table.Combine({informationisbeautiful, #"wikipedia - Companies", #"wikipedia - Government/Public Entity"}),
    #"Changed Type" = Table.TransformColumnTypes(Source,{{"Year", Int64.Type}}),
    #"Filtered Rows" = Table.SelectRows(#"Changed Type", each true),
    #"Changed Type1" = Table.TransformColumnTypes(#"Filtered Rows",{{"Records", Int64.Type}}),
    #"Removed Errors" = Table.RemoveRowsWithErrors(#"Changed Type1", {"Records"}),
    #"Removed Errors1" = Table.RemoveRowsWithErrors(#"Removed Errors", {"Year"}),
    #"Added Conditional Column2" = Table.AddColumn(#"Removed Errors1", "Custom", each if [Data sensitivity] = 1 then "email IDs/online
        details" else if [Data sensitivity] = 2 then "PII" else if [Data sensitivity] = "2" then "PII" else if [Data sensitivity] = 3
        then "Credit card details" else if [Data sensitivity] = "3" then "Credit card details" else if [Data sensitivity] = 4 then
        "Health & other personal records" else if [Data sensitivity] = 5 then "Full details" else if [Data sensitivity] = "5" then "Full
        details" else "unknown"),
    #"Renamed Columns2" = Table.RenameColumns(#"Added Conditional Column2",{{"Data sensitivity", "Data sensitivity_old"}, {"Custom",
        "Data Sensitivity"}}),
    #"Merged Queries" = Table.NestedJoin(#"Renamed Columns2", {"Sector"}, #"sector classification", {"Sector_old"}, "sector
        classification", JoinKind.LeftOuter),
    #"Expanded sector classification" = Table.ExpandTableColumn(#"Merged Queries", "sector classification", {"Industry"}, {"Industry"}),
    #"Merged Queries1" = Table.NestedJoin(#"Expanded sector classification", {"Method"}, #"method classification", {"method_old"},
        "method classification", JoinKind.LeftOuter),
    #"Expanded method classification" = Table.ExpandTableColumn(#"Merged Queries1", "method classification", {"Attack Method"}, {"Attack
        Method"}),
    #"Filtered Rows1" = Table.SelectRows(#"Expanded method classification", each ([Industry] <> "")),
    #"Removed Duplicates" = Table.Distinct(#"Filtered Rows1", {"Entity", "Year", "Records"}),
    #"Removed Duplicates1" = Table.Distinct(#"Removed Duplicates", {"Entity", "Records"}),
    #"Removed Duplicates2" = Table.Distinct(#"Removed Duplicates1", {"Year", "Records"}),
    #"Filtered Rows2" = Table.SelectRows(#"Removed Duplicates2", each true),
    #"Removed Columns" = Table.RemoveColumns(#"Filtered Rows2",{"date", "Sector", "Method", "Data sensitivity_old", "Region"})
in
    #"Removed Columns"
```

✓ No syntax errors have been detected.

Done Cancel

Snapshots of the classification table are provided below after removing redundancy and assigning value as best fit.

1. Classification of Attack Method

| 1 | method_old | Attack Method |
|---|---|---|
| 2 | poor security | Poor Security |
| 3 | hacked and published | Hacked |
| 4 | internal job | Insider Threat |
| 5 | data leak due to security vulnerabilities | Poor Security |
| 6 | social engineering | Social Engineering |
| 7 | publicly accessible Amazon Web Services (AWS) server | Misconfiguration |
| 8 | misconfigured web server | Misconfiguration |
| 9 | compromised credentials | Credential Stuffing |
| 10 | misconfigured API | Misconfiguration |
| 11 | improper setting, hacked | Misconfiguration |
| 12 | rogue contractor | Insider Threat |
| 13 | security | Misconfiguration |
| 14 | intentionally lost | Lost/Stolen Device |
| 15 | data exposed by misconfiguration | Misconfiguration |
| 16 | poor security/inside job | Insider Threat |
| 17 | unprotected api | Misconfiguration |
| 18 | poor security / hacked | Hacked |
| 19 | hacked by ALPHV | Hacked |
| 20 | accidentally uploaded | Accidental Exposure |
| 21 | data misuse | Hacked |
| 22 | ransomware hacked | Hacked |
| 23 | misconfigured website | Misconfiguration |
| 24 | compilation of multiple data breaches | Credential Stuffing |
| 25 | web scraping | Web Scraping |
| 26 | unsecured S3 bucket | Misconfiguration |
| 27 | lost / stolen computer | Lost/Stolen Device |
| 28 | unknown | Unknown |
| 29 | accidentally published | Accidental Exposure |
| 30 | lost / stolen media | Lost/Stolen Device |
| 31 | lost device | Lost/Stolen Device |
| 32 | lost device | Lost/Stolen Device |
| 33 | inside job | Insider Threat |
| 34 | poor security | Poor Security |
| 35 | hacked, poor security | Hacked |
| 36 | poor security | Poor Security |
| 37 | hacked | Hacked |
| 38 | hacked | Hacked |
| 39 | oops! | Unknown |

## 2. Classification of Sector

| 1 | Sector_old | Industry |
|---|---|---|
| 2 | health | Healthcare |
| 3 | media | Media/Entertainment |
| 4 | academic | Academic |
| 5 | tech | Technology |
| 6 | financial | Financial service |
| 7 | finance | Financial service |
| 8 | retail | Retail |
| 9 | government | Government |
| 10 | gaming | Gaming |
| 11 | telecoms | Telecommunication |
| 12 | transport | Transportation |
| 13 | health | Healthcare |
| 14 | web | Web service |
| 15 | web | Web service |
| 16 | NGO | Non-Profit Organisation |
| 17 | misc | misc |
| 18 | legal | legal |
| 19 | app | Technology |
| 20 | misc, health | Healthcare |
| 21 | tech, health | Healthcare |
| 22 | tech, app | Technology |
| 23 | web, tech | Technology |
| 24 | tech, web | Technology |
| 25 | government, health | Healthcare |
| 26 | web, military | Government |
| 27 | tech, retail | Retail |
| 28 | military | Government |
| 29 | military, health | Healthcare |
| 30 | web, gaming | Gaming |
| 31 | government, military | Government |
| 32 | various | various |
| 33 | social network | Social Media |
| 34 | healthcare | Healthcare |
| 35 | advertising | Advertising |
| 36 | telecommunications | Telecommunication |
| 37 | genealogy | Healthcare |
| 38 | dating | dating |
| 39 | educational services | Academic |
| 40 | gambling | gambling |
| 41 | background check | Government |
| 42 | hosting provider | Web service |
| 43 | restaurant | Hospitality |
| 44 | charity | Non-Profit Organisation |
| 45 | e-commerce | Retail |
| 46 | nonprofit, financial | Financial service |
| 47 | healthcare, retail | Healthcare |
| 48 | computer services for charities | Non-Profit Organisation |
| 49 | data broker | data broker |
| 50 | hotel/casino | Hospitality |
| 51 | energy | energy |
| 52 | humanitarian | Non-Profit Organisation |
| 53 | shopping | Retail |
| 54 | local search | local search |
| 55 | publisher (magazine) | Media/Entertainment |
| 56 | food | Retail |
| 57 | QR code payment | Financial service |
| 58 | clinical laboratory | clinical laboratory |
| 59 | phone accessories | Retail |
| 60 | telecom | Telecommunication |
| 61 | market analysis | market analysis |
| 62 | ticket distribution | Media/Entertainment |
| 63 | telephone directory | Telecommunication |
| 64 | consumer goods | Retail |
| 65 | online marketing | online marketing |
| 66 | malware tools | Government |
| 67 | bureau | Government |
| 68 | political | political |
| 69 | government | Government |
| 70 | personal and demographic data about residents and | Government |
| 71 | corporation | Government |

## 3. Classification of Region

| 1 | Region | Country | Continent |
|---|---|---|---|
| 2 | India | India | Asia |
| 3 | Australia | Australia | Oceania |
| 4 | Sydney, Australia | Australia | Oceania |
| 5 | Bangladesh | Bangladesh | Asia |
| 6 | United Kingdom | United Kingdom | Europe |
| 7 | Bulgaria | Bulgaria | Europe |
| 8 | California | United States | North America |
| 9 | United States | United States | North America |
| 10 | Colorado, US | United States | North America |
| 11 | Philippines | Philippines | Southeast Asia |
| 12 | South Korea | South Korea | East Asia |
| 13 | Indonesia | Indonesia | Southeast Asia |
| 14 | England/Wales | United Kingdom | Europe |
| 15 | European Union | European Union | Europe |
| 16 | Florida | United States | North America |
| 17 | Singapore | Singapore | Southeast Asia |
| 18 | Ireland | Ireland | Europe |
| 19 | London, UK | United Kingdom | Europe |
| 20 | Japan | Japan | East Asia |
| 21 | Cedar Rapids, Iowa | United States | North America |
| 22 | Chile | Chile | South America |
| 23 | Slovakia | Slovakia | Europe |
| 24 | Norway | Norway | Europe |
| 25 | Texas, US | United States | North America |
| 26 | Columbus, Ohio | United States | North America |
| 27 | Oregon | United States | North America |
| 28 | Puerto Rico | United States | North America |
| 29 | Argentina | Argentina | South America |
| 30 | Russia | Russia | Europe and Asia |
| 31 | San Francisco, California | United States | North America |
| 32 | New South Wales, AU | Australia | Oceania |
| 33 | South Africa | Africa | Africa |
| 34 | South Carolina, US | United States | North America |
| 35 | Stanford, California | United States | North America |
| 36 | Syrian government (Syria Files) | Syria | Asia |
| 37 | Texas | United States | North America |
| 38 | Berkeley, California | United States | North America |
| 39 | College Park, Maryland | United States | North America |
| 40 | Orange County, Florida | United States | North America |
| 41 | Miami, Florida | United States | North America |
| 42 | Salt Lake City, Utah | United States | North America |
| 43 | Milwaukee, Wisconsin | United States | North America |
| 44 | Shah Alam, Malaysia | Malaysia | Southeast Asia |
| 45 | Virginia, US | United States | North America |
| 46 | Washington, US | United States | North America |
| 47 | New Haven, Connecticut | United States | North America |

**Dataset**

Below screenshots are the data set obtained after the preprocessing steps.

| | Entity | Year | Records | Source | Data Sensitivity | Industry | Attack Method |
|---|---|---|---|---|---|---|---|
| 2 | Go Daddy | 2022 | 1228000 | information is beautiful | Credit card details | Web service | Hacked |
| 3 | Microsoft | 2023 | 30000000 | information is beautiful | PII | Web service | Hacked |
| 4 | LastPass | 2022 | 33000000 | information is beautiful | PII | Web service | Hacked |
| 5 | Twitter | 2021 | 5400000 | information is beautiful | PII | Web service | Hacked |
| 6 | Twitter | 2022 | 200000000 | information is beautiful | email IDs/online details | Web service | Hacked |
| 7 | Plex | 2022 | 15000000 | information is beautiful | email IDs/online details | Web service | Hacked |
| 8 | Open Subtitles | 2022 | 100000 | information is beautiful | email IDs/online details | Web service | Hacked |
| 9 | FlexBooker | 2022 | 3700000 | information is beautiful | Credit card details | Web service | Hacked |
| 10 | GoDaddy | 2021 | 1200000 | information is beautiful | email IDs/online details | Web service | Hacked |
| 11 | Linkedin | 2021 | 700000000 | information is beautiful | email IDs/online details | Web service | Hacked |
| 12 | db8151dd | 2020 | 22000000 | information is beautiful | PII | Web service | Hacked |
| 13 | Canva | 2019 | 139000000 | information is beautiful | PII | Web service | Hacked |
| 14 | Dubsmash | 2019 | 162000000 | information is beautiful | email IDs/online details | Web service | Hacked |
| 15 | ShareThis | 2019 | 41000000 | information is beautiful | email IDs/online details | Web service | Hacked |
| 16 | Animoto | 2019 | 25000000 | information is beautiful | email IDs/online details | Web service | Hacked |
| 17 | EyeEm | 2019 | 22000000 | information is beautiful | email IDs/online details | Web service | Hacked |
| 18 | Whitepages | 2019 | 18000000 | information is beautiful | email IDs/online details | Web service | Hacked |
| 19 | Fotolog | 2019 | 16000000 | information is beautiful | email IDs/online details | Web service | Hacked |
| 20 | BookMate | 2019 | 8000000 | information is beautiful | email IDs/online details | Web service | Hacked |
| 21 | CoffeeMeetsBagel | 2019 | 6000000 | information is beautiful | email IDs/online details | Web service | Hacked |
| 22 | Artsy | 2019 | 1000000 | information is beautiful | email IDs/online details | Web service | Hacked |
| 23 | DataCamp | 2019 | 700000 | information is beautiful | email IDs/online details | Web service | Hacked |
| 24 | YouNow | 2019 | 40000000 | information is beautiful | email IDs/online details | Web service | Hacked |
| 25 | 500px | 2019 | 14800000 | information is beautiful | PII | Web service | Hacked |
| 26 | BriansClub | 2019 | 26000000 | information is beautiful | Credit card details | Web service | Hacked |
| 27 | Quora | 2018 | 100000000 | information is beautiful | email IDs/online details | Web service | Hacked |
| 28 | Facebook | 2018 | 50000000 | information is beautiful | email IDs/online details | Web service | Hacked |
| 29 | MyHeritage | 2018 | 92283889 | information is beautiful | email IDs/online details | Web service | Hacked |
| 30 | Ticketmaster | 2018 | 40000 | information is beautiful | Credit card details | Web service | Hacked |
| 31 | Orbitz | 2018 | 880000 | information is beautiful | Credit card details | Web service | Hacked |
| 32 | TicketFly | 2018 | 27000000 | information is beautiful | PII | Web service | Hacked |
| 33 | High Tail Hall | 2018 | 411000 | information is beautiful | PII | Web service | Hacked |
| 34 | Facebook | 2018 | 29000000 | information is beautiful | PII | Web service | Hacked |
| 35 | Disqus | 2017 | 17500000 | information is beautiful | Health & other personal records | Web service | Hacked |
| 36 | Yahoo | 2017 | 32000000 | information is beautiful | Health & other personal records | Web service | Hacked |
| 37 | DaFont | 2017 | 700000 | information is beautiful | Health & other personal records | Web service | Hacked |
| 38 | Zomato | 2017 | 17000000 | information is beautiful | Health & other personal records | Web service | Hacked |
| 39 | Instagram | 2017 | 6000000 | information is beautiful | email IDs/online details | Web service | Hacked |
| 40 | Linkedin | 2016 | 117000000 | information is beautiful | email IDs/online details | Web service | Hacked |
| 41 | Tumblr | 2016 | 65000000 | information is beautiful | email IDs/online details | Web service | Hacked |
| 42 | Yahoo | 2016 | 500000000 | information is beautiful | PII | Web service | Hacked |
| 43 | Mail. ru | 2016 | 25000000 | information is beautiful | PII | Web service | Hacked |
| 44 | PayAsUGym | 2016 | 300000 | information is beautiful | email IDs/online details | Web service | Hacked |
| 45 | Lynda.com | 2016 | 9500000 | information is beautiful | email IDs/online details | Web service | Hacked |
| 46 | Linux Ubuntu forums | 2016 | 2000000 | information is beautiful | email IDs/online details | Web service | Hacked |
| 47 | uTorrent | 2016 | 35000 | information is beautiful | email IDs/online details | Web service | Hacked |
| 48 | VK | 2016 | 171000000 | information is beautiful | Health & other personal records | Web service | Hacked |
| 49 | KM.ru & Nival | 2016 | 1500000 | information is beautiful | Health & other personal records | Web service | Hacked |
| 50 | Fling | 2016 | 40000000 | information is beautiful | Health & other personal records | Web service | Hacked |
| 51 | MySpace | 2016 | 164000000 | information is beautiful | email IDs/online details | Web service | Hacked |
| 52 | Dailymotion | 2016 | 85200000 | information is beautiful | email IDs/online details | Web service | Hacked |
| 53 | Weebly | 2016 | 43000000 | information is beautiful | Health & other personal records | Web service | Hacked |
| 54 | Interpark | 2016 | 10000000 | information is beautiful | PII | Web service | Hacked |
| 55 | Friend Finder Network | 2016 | 412000000 | information is beautiful | email IDs/online details | Web service | Hacked |
| 56 | Brazzers | 2016 | 790724 | information is beautiful | Health & other personal records | Web service | Hacked |
| 57 | ClixSense | 2016 | 6600000 | information is beautiful | Full details | Web service | Hacked |
| 58 | Kromtech | 2015 | 13000000 | information is beautiful | email IDs/online details | Web service | Hacked |
| 59 | VTech | 2015 | 6400000 | information is beautiful | Full details | Web service | Hacked |
| 60 | Hacking Team | 2015 | 500000 | information is beautiful | Full details | Web service | Hacked |
| 61 | AshleyMadison.com | 2015 | 37000000 | information is beautiful | email IDs/online details | Web service | Hacked |
| 62 | Adult Friend Finder | 2015 | 3900000 | information is beautiful | email IDs/online details | Web service | Hacked |
| 63 | Securus Technologies | 2015 | 70000000 | information is beautiful | Full details | Web service | Hacked |
| 64 | AOL | 2014 | 2400000 | information is beautiful | email IDs/online details | Web service | Hacked |
| 65 | Ebay | 2014 | 145000000 | information is beautiful | email IDs/online details | Web service | Hacked |
| 66 | GMail | 2014 | 5000000 | information is beautiful | email IDs/online details | Web service | Hacked |
| 67 | Tianya | 2013 | 40000000 | information is beautiful | email IDs/online details | Web service | Hacked |
| 68 | Scribd | 2013 | 500000 | information is beautiful | email IDs/online details | Web service | Hacked |
| 69 | Living Social | 2013 | 50000000 | information is beautiful | email IDs/online details | Web service | Hacked |
| 70 | Yahoo | 2013 | 550000000 | information is beautiful | PII | Web service | Hacked |
| 71 | Twitter | 2013 | 250000 | information is beautiful | email IDs/online details | Web service | Hacked |
| 72 | OVH | 2013 | 200000 | information is beautiful | PII | Web service | Hacked |
| 73 | Drupal | 2013 | 1000000 | information is beautiful | email IDs/online details | Web service | Hacked |
| 74 | MacRumours.com | 2013 | 860000 | information is beautiful | email IDs/online details | Web service | Hacked |
| 75 | China Software Developer Network | 2012 | 6000000 | information is beautiful | email IDs/online details | Web service | Hacked |
| 76 | Zappos | 2012 | 24000000 | information is beautiful | PII | Web service | Hacked |
| 77 | Formspring | 2012 | 420000 | information is beautiful | email IDs/online details | Web service | Hacked |
| 78 | Last.fm | 2012 | 43500000 | information is beautiful | email IDs/online details | Web service | Hacked |
| 79 | LinkedIn, eHarmony, Last.fm | 2012 | 8000000 | information is beautiful | email IDs/online details | Web service | Hacked |
| 80 | Dropbox | 2012 | 68700000 | information is beautiful | email IDs/online details | Web service | Hacked |
| 81 | Betfair | 2011 | 2300000 | information is beautiful | Credit card details | Web service | Hacked |
| 82 | Epsilon | 2011 | 3000000 | information is beautiful | email IDs/online details | Web service | Hacked |
| 83 | Chinese gaming sites | 2011 | 10000000 | information is beautiful | email IDs/online details | Web service | Hacked |
| 84 | Writerspace.com | 2011 | 62000 | information is beautiful | email IDs/online details | Web service | Hacked |
| 85 | Sony Pictures | 2011 | 1000000 | information is beautiful | email IDs/online details | Web service | Hacked |
| 86 | Steam | 2011 | 35000000 | information is beautiful | Credit card details | Web service | Hacked |
| 87 | Nexon Korea Corp | 2011 | 13200000 | information is beautiful | PII | Web service | Hacked |
| 88 | Gawker.com | 2010 | 1500000 | information is beautiful | PII | Web service | Hacked |
| 89 | Network Solutions | 2009 | 573000 | information is beautiful | Credit card details | Web service | Hacked |
| 90 | Auction.co.kr | 2008 | 18000000 | information is beautiful | Credit card details | Web service | Hacked |
| 91 | Monster.com | 2007 | 1600000 | information is beautiful | PII | Web service | Hacked |
| 92 | Cencora | 2024 | 100000 | information is beautiful | Health & other personal records | Healthcare | Hacked |
| 93 | Singing River | 2023 | 895000 | information is beautiful | Health & other personal records | Healthcare | Hacked |
| 94 | Ticketmaster | 2024 | 560000000 | information is beautiful | Credit card details | Media/Entertainment | Hacked |
| 95 | Philadelphia Inquirer | 2023 | 25000 | information is beautiful | Health & other personal records | Media/Entertainment | Hacked |
| 96 | Western Sydney University | 2023 | 7500 | information is beautiful | email IDs/online details | Academic | Hacked |
| 97 | Everbridge | 2024 | 5600000 | information is beautiful | email IDs/online details | Technology | Hacked |
| 98 | OmniVision | 2023 | 100000 | information is beautiful | Credit card details | Technology | Hacked |
| 99 | Okta | 2023 | 134 | information is beautiful | email IDs/online details | Technology | Hacked |
| 100 | Acer | 2023 | 10000000 | information is beautiful | email IDs/online details | Technology | Hacked |
| 101 | CoinSquare | 2022 | 50000 | information is beautiful | email IDs/online details | Technology | Hacked |
| 102 | Santander | 2024 | 30000000 | information is beautiful | Credit card details | Financial service | Hacked |
| 103 | Shanghai Police | 2022 | 500000000 | information is beautiful | Full details | Financial service | Hacked |
| 104 | City of Helsinki | 2024 | 80000 | information is beautiful | Health & other personal records | Government | Hacked |
| 105 | Maine Government | 2023 | 1300000 | information is beautiful | Health & other personal records | Government | Hacked |
| 106 | NATO | 2023 | 8000 | information is beautiful | Health & other personal records | Government | Hacked |
| 107 | agency | 2022 | 279000000 | information is beautiful | Credit card details | Government | Hacked |
| 108 | Discord.io | 2023 | 760000 | information is beautiful | email IDs/online details | Gaming | Hacked |
| 109 | AT&T | 2024 | 73000000 | information is beautiful | Health & other personal records | Telecommunication | Hacked |
| 110 | Indonesian SIM cards | 2022 | 1000000000 | information is beautiful | Credit card details | Telecommunication | Hacked |
| 111 | Uber | 2022 | 20000000 | information is beautiful | email IDs/online details | Transportation | Hacked |
| 112 | Welltok | 2023 | 8500000 | information is beautiful | Health & other personal records | Healthcare | Hacked |
| 113 | Delta Dental | 2023 | 7000000 | information is beautiful | Credit card details | Healthcare | Hacked |
| 114 | PharMerica | 2023 | 5800000 | information is beautiful | Health & other personal records | Healthcare | Hacked |
| 115 | 23andMe | 2023 | 6900000 | information is beautiful | Health & other personal records | Healthcare | Hacked |
| 116 | Mailchimp | 2022 | 106586 | information is beautiful | email IDs/online details | Technology | Hacked |
| 117 | PayHere | 2022 | 1580249 | information is beautiful | Credit card details | Financial service | Hacked |
| 118 | CDEK | 2022 | 18218203 | information is beautiful | Credit card details | Retail | Hacked |
| 119 | Washington State Dpt of Licensing | 2022 | 257000 | information is beautiful | Credit card details | Government | Hacked |
| 120 | Acer | 2021 | 3000000 | information is beautiful | email IDs/online details | Technology | Hacked |
| 121 | Avvo | 2019 | 4101101 | information is beautiful | email IDs/online details | legal | Hacked |
| 122 | Aimware | 2019 | 305470 | information is beautiful | Credit card details | Gaming | Hacked |
| 123 | Twitch | 2021 | 10000000 | information is beautiful | Health & other personal records | Gaming | Hacked |
| 124 | Syniverse | 2021 | 500000000 | information is beautiful | Health & other personal records | Telecommunication | Hacked |
| 125 | Pandora Papers | 2021 | 11900000 | information is beautiful | Health & other personal records | Government | Hacked |
| 126 | Neiman Marcus | 2021 | 4600000 | information is beautiful | Credit card details | Retail | Hacked |
| 127 | Epik | 2021 | 15000000 | information is beautiful | Full details | Retail | Hacked |
| 128 | T-Mobile | 2021 | 76000000 | information is beautiful | Credit card details | Telecommunication | Hacked |
| 129 | Contact tracing data | 2021 | 38000000 | information is beautiful | Credit card details | Telecommunication | Hacked |
| 130 | Estonian gov | 2021 | 280000 | information is beautiful | Health & other personal records | Government | Hacked |
| 131 | Meet Mindful | 2021 | 2240000 | information is beautiful | Health & other personal records | Technology | Hacked |
| 132 | Gab | 2021 | 4000000 | information is beautiful | Credit card details | Technology | Hacked |
| 133 | Facebook | 2021 | 533000000 | information is beautiful | email IDs/online details | Technology | Hacked |
| 134 | T-Mobile | 2020 | 200000 | information is beautiful | email IDs/online details | Telecommunication | Hacked |
| 135 | The Hospital Group | 2020 | 1000000 | information is beautiful | Health & other personal records | Healthcare | Hacked |
| 136 | SolarWinds | 2020 | 50000000 | information is beautiful | Credit card details | Technology | Hacked |
| 137 | GEDmatch | 2020 | 1400000 | information is beautiful | Full details | Healthcare | Hacked |
| 138 | Nintendo | 2020 | 300000 | information is beautiful | Credit card details | Gaming | Hacked |
| 139 | EasyJet | 2020 | 9000000 | information is beautiful | Credit card details | Transportation | Hacked |
| 140 | Boots Advantage Card | 2020 | 150000 | information is beautiful | email IDs/online details | Retail | Hacked |
| 141 | Tesco Clubcard | 2020 | 600000 | information is beautiful | email IDs/online details | Retail | Hacked |
| 142 | Wawa | 2019 | 30000000 | information is beautiful | Credit card details | Retail | Hacked |
| 143 | Australian National University | 2019 | 200000 | information is beautiful | Health & other personal records | Academic | Hacked |
| 144 | HauteLook | 2019 | 28000000 | information is beautiful | email IDs/online details | Retail | Hacked |
| 145 | Armor Games | 2019 | 11000000 | information is beautiful | email IDs/online details | Gaming | Hacked |
| 146 | Coinmama | 2019 | 450000 | information is beautiful | email IDs/online details | Financial service | Hacked |
| 147 | Roll20 | 2019 | 4000000 | information is beautiful | email IDs/online details | Gaming | Hacked |
| 148 | Stronghold Kingdoms | 2019 | 5000000 | information is beautiful | email IDs/online details | Gaming | Hacked |
| 149 | Blank Media Games | 2019 | 7600000 | information is beautiful | email IDs/online details | Gaming | Hacked |
| 150 | Capital One | 2019 | 100000000 | information is beautiful | Credit card details | Financial service | Hacked |
| 151 | Click2Gov | 2018 | 300000 | information is beautiful | Credit card details | Financial service | Hacked |
| 152 | SingHealth | 2018 | 1500000 | information is beautiful | Health & other personal records | Healthcare | Hacked |
| 153 | Cathay Pacific Airways | 2018 | 94000000 | information is beautiful | Credit card details | Transportation | Hacked |
| 154 | Marriott International | 2018 | 383000000 | information is beautiful | Credit card details | Retail | Hacked |
| 155 | Dixons Carphone | 2018 | 10000000 | information is beautiful | email IDs/online details | Telecommunication | Hacked |
| 156 | Saks and Lord & Taylor | 2018 | 5000000 | information is beautiful | Credit card details | Retail | Hacked |
| 157 | British Airways | 2018 | 380000 | information is beautiful | Health & other personal records | Transportation | Hacked |
| 158 | T-Mobile | 2018 | 2000000 | information is beautiful | email IDs/online details | Telecommunication | Hacked |
| 159 | MyFitnessPal | 2018 | 150000000 | information is beautiful | email IDs/online details | Technology | Hacked |
| 160 | Helse Sør-Øst RHF | 2018 | 3000000 | information is beautiful | Health & other personal records | Healthcare | Hacked |
| 161 | Dell | 2018 | 100000 | information is beautiful | email IDs/online details | Technology | Hacked |
| 162 | Vision Direct | 2018 | 16300 | information is beautiful | Health & other personal records | Retail | Hacked |
| 163 | Newegg | 2018 | 45000000 | information is beautiful | Credit card details | Retail | Hacked |
| 164 | Uber | 2017 | 57000000 | information is beautiful | email IDs/online details | Technology | Hacked |
| 165 | Wonga | 2017 | 270000 | information is beautiful | Health & other personal records | Financial service | Hacked |
| 166 | Snapchat | 2017 | 1700000 | information is beautiful | email IDs/online details | Technology | Hacked |
| 167 | Bell | 2017 | 1900000 | information is beautiful | email IDs/online details | Telecommunication | Hacked |
| 168 | TIO Networks | 2017 | 1600000 | information is beautiful | Health & other personal records | Financial service | Hacked |
| 169 | Malaysian telcos & MVNOs | 2017 | 46200000 | information is beautiful | Health & other personal records | Telecommunication | Hacked |
| 170 | Malaysian medical practitioners | 2017 | 81309 | information is beautiful | Health & other personal records | Healthcare | Hacked |
| 171 | Equifax | 2017 | 143000000 | information is beautiful | Health & other personal records | Financial service | Hacked |
| 172 | Mossack Fonseca | 2016 | 11500000 | information is beautiful | Full details | misc | Hacked |
| 173 | Philippines' Commission on Elections | 2016 | 55000000 | information is beautiful | Full details | Government | Hacked |
| 174 | Syrian government | 2016 | 274477 | information is beautiful | email IDs/online details | Government | Hacked |
| 175 | Minecraft | 2016 | 7000000 | information is beautiful | email IDs/online details | Gaming | Hacked |
| 176 | Banner Health | 2016 | 3700000 | information is beautiful | Credit card details | Healthcare | Hacked |
| 177 | Wendy's | 2016 | 1025 | information is beautiful | Credit card details | Retail | Hacked |
| 178 | Telegram | 2016 | 15000000 | information is beautiful | email IDs/online details | Technology | Hacked |
| 179 | Quest Diagnostics | 2016 | 34000 | information is beautiful | Health & other personal records | Healthcare | Hacked |
| 180 | Carefirst | 2015 | 1100000 | information is beautiful | email IDs/online details | Healthcare | Hacked |
| 181 | Premera | 2015 | 11000000 | information is beautiful | Full details | Healthcare | Hacked |
| 182 | Invest Bank | 2015 | 40000 | information is beautiful | Health & other personal records | Financial service | Hacked |
| 183 | Management (2nd Breach) | 2015 | 21500000 | information is beautiful | Full details | Government | Hacked |
| 184 | IRS | 2015 | 100000 | information is beautiful | email IDs/online details | Government | Hacked |
| 185 | Experian / T-mobile | 2015 | 15000000 | information is beautiful | Credit card details | Telecommunication | Hacked |
| 186 | CarPhone Warehouse | 2015 | 2400000 | information is beautiful | Credit card details | Telecommunication | Hacked |
| 187 | British Airways | 2015 | 10000 | information is beautiful | email IDs/online details | Transportation | Hacked |
| 188 | UCLA Health | 2015 | 4500000 | information is beautiful | Health & other personal records | Healthcare | Hacked |
| 189 | UPS | 2014 | 4000000 | information is beautiful | Credit card details | Retail | Hacked |
| 190 | JP Morgan Chase | 2014 | 76000000 | information is beautiful | Credit card details | Financial service | Hacked |
| 191 | HSBC Turkey | 2014 | 2700000 | information is beautiful | Health & other personal records | Financial service | Hacked |
| 192 | Staples | 2014 | 1160000 | information is beautiful | Credit card details | Retail | Hacked |
| 193 | Home Depot | 2014 | 56000000 | information is beautiful | Credit card details | Retail | Hacked |
| 194 | (France) | 2014 | 600000 | information is beautiful | email IDs/online details | Retail | Hacked |
| 195 | Massive American business hack | 2013 | 160000000 | information is beautiful | Full details | Financial service | Hacked |
| 196 | Central Hudson Gas & Electric | 2013 | 110000 | information is beautiful | Credit card details | misc | Hacked |
| 197 | Apple | 2013 | 275000 | information is beautiful | email IDs/online details | Technology | Hacked |
| 198 | Ubuntu | 2013 | 2000000 | information is beautiful | Credit card details | Technology | Hacked |
| 199 | Yahoo Japan | 2013 | 22000000 | information is beautiful | email IDs/online details | Technology | Hacked |
| 200 | Adobe | 2013 | 38000000 | information is beautiful | Credit card details | Technology | Hacked |
| 201 | Target | 2013 | 70000000 | information is beautiful | Credit card details | Retail | Hacked |
| 202 | Global Payments | 2012 | 1500000 | information is beautiful | Credit card details | Financial service | Hacked |
| 203 | Three Iranian banks | 2012 | 3000000 | information is beautiful | Full details | Financial service | Hacked |
| 204 | Medicaid | 2012 | 780000 | information is beautiful | Full details | Healthcare | Hacked |
| 205 | Yahoo Voices | 2012 | 450000 | information is beautiful | email IDs/online details | Technology | Hacked |
| 206 | Militarysingles.com | 2012 | 163792 | information is beautiful | email IDs/online details | Government | Hacked |
| 207 | South Carolina State Dept. of Revenue | 2012 | 3600000 | information is beautiful | email IDs/online details | Government | Hacked |
| 208 | Ankle & foot Center of Tampa Bay, Inc. | 2011 | 156000 | information is beautiful | Health & other personal records | Healthcare | Hacked |
| 209 | Sony PSN | 2011 | 77000000 | information is beautiful | email IDs/online details | Gaming | Hacked |
| 210 | US Law Enforcement | 2011 | 123461 | information is beautiful | Credit card details | Government | Hacked |
| 211 | Stratfor | 2011 | 935000 | information is beautiful | Credit card details | Government | Hacked |
| 212 | Bethesda Game Studios | 2011 | 200000 | information is beautiful | email IDs/online details | Gaming | Hacked |
| 213 | Citigroup | 2011 | 210000 | information is beautiful | Credit card details | Financial service | Hacked |
| 214 | San Francisco Public Utilities Commission | 2011 | 180000 | information is beautiful | email IDs/online details | Government | Hacked |
| 215 | Sony Online Entertainment | 2011 | 24600000 | information is beautiful | Credit card details | Gaming | Hacked |
| 216 | AT&T | 2010 | 114000 | information is beautiful | email IDs/online details | Telecommunication | Hacked |
| 217 | US Federal Reserve Bank of Cleveland | 2010 | 400000 | information is beautiful | Credit card details | Financial service | Hacked |
| 218 | Heartland | 2009 | 130000000 | information is beautiful | Credit card details | Financial service | Hacked |
| 219 | RockYou! | 2009 | 32000000 | information is beautiful | email IDs/online details | Gaming | Hacked |
| 220 | CheckFree Corporation | 2009 | 5000000 | information is beautiful | email IDs/online details | Financial service | Hacked |
| 221 | University of California Berkeley | 2009 | 160000 | information is beautiful | Credit card details | Academic | Hacked |
| 222 | Hannaford Brothers Supermarket Chain | 2008 | 4200000 | information is beautiful | Credit card details | Retail | Hacked |
| 223 | RBS Worldpay | 2008 | 1500000 | information is beautiful | Full details | Financial service | Hacked |
| 224 | Chile Ministry Of Education | 2008 | 6000000 | information is beautiful | email IDs/online details | Government | Hacked |
| 225 | TK / TJ Maxx | 2007 | 94000000 | information is beautiful | Credit card details | Retail | Hacked |
| 226 | TD Ameritrade | 2007 | 6300000 | information is beautiful | email IDs/online details | Financial service | Hacked |
| 227 | KDDI | 2006 | 4000000 | information is beautiful | email IDs/online details | Telecommunication | Hacked |
| 228 | Cardsystems Solutions Inc. | 2005 | 40000000 | information is beautiful | Credit card details | Financial service | Hacked |
| 229 | Sav-Rx | 2023 | 2800000 | information is beautiful | PII | Healthcare | Hacked |
| 230 | BBC | 2024 | 25000 | information is beautiful | PII | Media/Entertainment | Hacked |
| 231 | The Post Millennial | 2024 | 26000000 | information is beautiful | PII | Media/Entertainment | Hacked |
| 232 | Cooler Master | 2024 | 500000 | information is beautiful | PII | Technology | Hacked |
| 233 | Financial Business and Consumer Solutions | 2024 | 3200000 | information is beautiful | PII | Technology | Hacked |
| 234 | WebTPA | 2023 | 240000 | information is beautiful | PII | Technology | Hacked |
| 235 | Sony | 2023 | 6800 | information is beautiful | PII | Technology | Hacked |
| 236 | ChatGPT | 2023 | 101000 | information is beautiful | PII | Technology | Hacked |
| 237 | First American | 2023 | 44000 | information is beautiful | PII | Financial service | Hacked |
| 238 | PayPal | 2023 | 34942 | information is beautiful | PII | Financial service | Hacked |
| 239 | Latitude Financial | 2023 | 14000000 | information is beautiful | PII | Financial service | Hacked |
| 240 | Panda Restaurants | 2024 | 47000 | information is beautiful | PII | Retail | Hacked |
| 241 | Topgolf Callaway | 2023 | 1114954 | information is beautiful | PII | Retail | Hacked |
| 242 | Shein | 2022 | 39000000 | information is beautiful | PII | Retail | Hacked |
| 243 | UK Ministry of Defense | 2024 | 270000 | information is beautiful | PII | Government | Hacked |
| 244 | French government | 2024 | 43000000 | information is beautiful | PII | Government | Hacked |
| 245 | Stanford University | 2023 | 27000 | information is beautiful | PII | Academic | Hacked |
| 246 | TIAA | 2023 | 2300000 | information is beautiful | PII | Financial service | Hacked |
| 247 | USG | 2023 | 800000 | information is beautiful | PII | Government | Hacked |
| 248 | Ohio Lottery | 2023 | 538000 | information is beautiful | PII | Gaming | Hacked |
| 249 | Xfinity | 2023 | 36000000 | information is beautiful | PII | Telecommunication | Hacked |
| 250 | T-Mobile | 2023 | 37000000 | information is beautiful | PII | Telecommunication | Hacked |
| 251 | T-Mobile | 2023 | 836 | information is beautiful | PII | Telecommunication | Hacked |
| 252 | CommuteAir | 2023 | 1500000 | information is beautiful | PII | Transportation | Hacked |
| 253 | Indian Railways | 2022 | 30000000 | information is beautiful | PII | Transportation | Hacked |
| 254 | Robinhood | 2021 | 5000937 | information is beautiful | PII | Financial service | Hacked |
| 255 | Travelio | 2021 | 471376 | information is beautiful | PII | misc | Hacked |
| 256 | Nvidia | 2021 | 100000 | information is beautiful | PII | Technology | Hacked |
| 257 | Guntrader | 2021 | 111000 | information is beautiful | PII | Retail | Hacked |
| 258 | VW | 2021 | 3300000 | information is beautiful | PII | Transportation | Hacked |
| 259 | Air India | 2021 | 4500000 | information is beautiful | PII | Transportation | Hacked |
| 260 | Omiai dating app | 2021 | 1710000 | information is beautiful | PII | Technology | Hacked |
| 261 | Park Mobile | 2021 | 21000000 | information is beautiful | PII | Transportation | Hacked |
| 262 | Ubiquiti | 2021 | 16000000 | information is beautiful | PII | Technology | Hacked |
| 263 | Experian Brazil | 2021 | 220000000 | information is beautiful | PII | Financial service | Hacked |
| 264 | Ledger | 2020 | 270000 | information is beautiful | PII | Financial service | Hacked |
| 265 | Ho Mobile | 2020 | 2500000 | information is beautiful | PII | Telecommunication | Hacked |
| 266 | Drizly | 2020 | 2400000 | information is beautiful | PII | Technology | Hacked |
| 267 | Pakistani mobile operators | 2020 | 115000000 | information is beautiful | PII | Telecommunication | Hacked |
| 268 | US Marshals Service | 2020 | 387000 | information is beautiful | PII | Government | Hacked |
| 269 | MGM Hotels | 2020 | 10600000 | information is beautiful | PII | Retail | Hacked |
| 270 | US Customs and Border Protection | 2019 | 100000 | information is beautiful | PII | Government | Hacked |
| 271 | Toyota | 2019 | 3100000 | information is beautiful | PII | Transportation | Hacked |
| 272 | Houzz | 2019 | 57000000 | information is beautiful | PII | Retail | Hacked |
| 273 | DoorDash | 2019 | 4900000 | information is beautiful | PII | Transportation | Hacked |
| 274 | Medicare & Medicaid | 2018 | 93689 | information is beautiful | PII | Healthcare | Hacked |
| 275 | Cellebrite | 2017 | 3000000 | information is beautiful | PII | Technology | Hacked |
| 276 | Turkish citizenship database | 2016 | 49611709 | information is beautiful | PII | Government | Hacked |
| 277 | Clinton campaign | 2016 | 5000000 | information is beautiful | PII | Government | Hacked |
| 278 | Three | 2016 | 130000 | information is beautiful | PII | Telecommunication | Hacked |
| 279 | US Office of Personnel Management | 2015 | 4000000 | information is beautiful | PII | Government | Hacked |
| 280 | MSpy | 2015 | 400000 | information is beautiful | PII | Technology | Hacked |
| 281 | TalkTalk | 2015 | 157000 | information is beautiful | PII | Telecommunication | Hacked |
| 282 | Anthem | 2015 | 80000000 | information is beautiful | PII | Healthcare | Hacked |
| 283 | Neiman Marcus | 2014 | 1100000 | information is beautiful | PII | Retail | Hacked |
| 284 | Community Health Systems | 2014 | 4500000 | information is beautiful | PII | Healthcare | Hacked |
| 285 | Sony Pictures | 2014 | 10000000 | information is beautiful | PII | misc | Hacked |
| 286 | Japan Airlines | 2014 | 750000 | information is beautiful | PII | Transportation | Hacked |
| 287 | Snapchat | 2013 | 4600000 | information is beautiful | PII | Technology | Hacked |
| 288 | University of Delaware | 2013 | 74000 | information is beautiful | PII | Academic | Hacked |
| 289 | UbiSoft | 2013 | 58000000 | information is beautiful | PII | Gaming | Hacked |
| 290 | Kirkwood Community College | 2013 | 125000 | information is beautiful | PII | Academic | Hacked |
| 291 | Washington State court system | 2013 | 160000 | information is beautiful | PII | Government | Hacked |
| 292 | Blizzard | 2012 | 14000000 | information is beautiful | PII | Gaming | Hacked |
| 293 | KT Corp. | 2012 | 8700000 | information is beautiful | PII | Telecommunication | Hacked |
| 294 | Greek government | 2012 | 9000000 | information is beautiful | PII | Government | Hacked |
| 295 | Seacoast Radiology, PA | 2011 | 231400 | information is beautiful | PII | Healthcare | Hacked |
| 296 | University of Wisconsin - Milwaukee | 2011 | 73000 | information is beautiful | PII | Academic | Hacked |
| 297 | Southern California Medical-Legal Consultants | 2011 | 300000 | information is beautiful | PII | Healthcare | Hacked |
| 298 | Sega | 2011 | 1290755 | information is beautiful | PII | Gaming | Hacked |
| 299 | Washington Post | 2011 | 1270000 | information is beautiful | PII | misc | Hacked |
| 300 | Honda Canada | 2011 | 283000 | information is beautiful | PII | Retail | Hacked |
| 301 | Ohio State University | 2010 | 760000 | information is beautiful | PII | Academic | Hacked |
| 302 | Virginia Prescription Monitoring Program | 2009 | 531400 | information is beautiful | PII | Healthcare | Hacked |
| 303 | 500px | 2020 | 14870304 | list of databreaaches- wikipedia | unknown | Social Media | Hacked |
| 304 | Adobe Systems Incorporated | 2013 | 152000000 | list of databreaaches- wikipedia | unknown | Technology | Hacked |
| 305 | AerServ (subsidiary of InMobi) | 2018 | 75000 | list of databreaaches- wikipedia | unknown | Advertising | Hacked |
| 306 | Air Canada | 2018 | 20000 | list of databreaaches- wikipedia | unknown | Transportation | Hacked |
| 307 | Animal Jam | 2020 | 46000000 | list of databreaaches- wikipedia | unknown | Gaming | Hacked |
| 308 | Ankle & Foot Center of Tampa Bay, Inc. | 2021 | 156000 | list of databreaaches- wikipedia | unknown | Healthcare | Hacked |
| 309 | Ashley Madison | 2015 | 32000000 | list of databreaaches- wikipedia | unknown | dating | Hacked |
| 310 | Bailey's Inc. | 2015 | 250000 | list of databreaaches- wikipedia | unknown | Retail | Hacked |
| 311 | Benesse | 2014 | 35040000 | list of databreaaches- wikipedia | unknown | Academic | Hacked |
| 312 | Blank Media Games | 2018 | 7633234 | list of databreaaches- wikipedia | unknown | Gaming | Hacked |
| 313 | British Airways | 2018 | 500000 | list of databreaaches- wikipedia | unknown | Transportation | Hacked |
| 314 | Canva | 2019 | 140000000 | list of databreaaches- wikipedia | unknown | Web service | Hacked |
| 315 | Capcom | 2020 | 350000 | list of databreaaches- wikipedia | unknown | Gaming | Hacked |
| 316 | Cathay Pacific Airways | 2018 | 9400000 | list of databreaaches- wikipedia | unknown | Transportation | Hacked |
| 317 | Central Coast Credit Union | 2016 | 60000 | list of databreaaches- wikipedia | unknown | Financial service | Hacked |
| 318 | Citigroup | 2011 | 360083 | list of databreaaches- wikipedia | unknown | Financial service | Hacked |
| 319 | Cox Communications | 2016 | 40000 | list of databreaaches- wikipedia | unknown | Telecommunication | Hacked |
| 320 | Cutout.Pro | 2024 | 19972829 | list of databreaaches- wikipedia | unknown | Web service | Hacked |
| 321 | CyberServe | 2021 | 1107034 | list of databreaaches- wikipedia | unknown | Web service | Hacked |
| 322 | Data Processors International (MasterCard, Visa, Discover Financial Services and American Express) | 2008 | 8000000 | list of databreaaches- wikipedia | unknown | Financial service | Hacked |
| 323 | Dropbox | 2012 | 68648009 | list of databreaaches- wikipedia | unknown | Web service | Hacked |
| 324 | DSW Inc. | 2005 | 1400000 | list of databreaaches- wikipedia | unknown | Retail | Hacked |
| 325 | EssilorLuxottica | 2021 | 77093812 | list of databreaaches- wikipedia | unknown | Healthcare | Hacked |
| 326 | Evide data breach | 2023 | 1000 | list of databreaaches- wikipedia | unknown | Non-Profit Organisation | Hacked |
| 327 | Excellus BlueCross BlueShield | 2015 | 10000000 | list of databreaaches- wikipedia | unknown | Healthcare | Hacked |
| 328 | Facebook | 2010 | 87000000 | list of databreaaches- wikipedia | unknown | Social Media | Hacked |
| 329 | Fast Retailing | 2019 | 461091 | list of databreaaches- wikipedia | unknown | Retail | Hacked |
| 330 | Friend Finder Network | 2016 | 412214295 | list of databreaaches- wikipedia | unknown | Web service | Hacked |
| 331 | Funimation | 2016 | 2500000 | list of databreaaches- wikipedia | unknown | Web service | Hacked |
| 332 | Global Payments | 2012 | 7000000 | list of databreaaches- wikipedia | unknown | Financial service | Hacked |
| 333 | Grozio Chirurgija | 2017 | 25000 | list of databreaaches- wikipedia | unknown | Healthcare | Hacked |
| 334 | HauteLook | 2018 | 28517244 | list of databreaaches- wikipedia | unknown | Retail | Hacked |
| 335 | HCA Healthcare | 2023 | 11270000 | list of databreaaches- wikipedia | unknown | Healthcare | Hacked |
| 336 | Hilton Hotels | 2014 | 363000 | list of databreaaches- wikipedia | unknown | Hospitality | Hacked |
| 337 | LexisNexis | 2014 | 1000000 | list of databreaaches- wikipedia | unknown | Technology | Hacked |
| 338 | LifeLabs | 2019 | 15000000 | list of databreaaches- wikipedia | unknown | Healthcare | Hacked |
| 339 | Lyca Mobile | 2023 | 16000000 | list of databreaaches- wikipedia | unknown | Telecommunication | Hacked |
| 340 | MacRumors.com | 2014 | 860000 | list of databreaaches- wikipedia | unknown | Web service | Hacked |
| 341 | Manipulated Caiman | 2023 | 40000000 | list of databreaaches- wikipedia | unknown | Financial service | Hacked |
| 342 | Marriott International | 2018 | 500000000 | list of databreaaches- wikipedia | unknown | Hospitality | Hacked |
| 343 | MediaWorks New Zealand | 2023 | 162710 | list of databreaaches- wikipedia | unknown | Media/Entertainment | Hacked |
| 344 | Medibank & AHM | 2022 | 9700000 | list of databreaaches-wikipedia | unknown | Healthcare | Hacked |
| 345 | MGM Resorts | 2019 | 10600000 | list of databreaaches-wikipedia | unknown | Hospitality | Hacked |
| 346 | Michaels | 2014 | 3000000 | list of databreaaches-wikipedia | unknown | Retail | Hacked |
| 347 | Morinaga Confectionery | 2022 | 1648922 | list of databreaaches-wikipedia | unknown | Retail | Hacked |
| 348 | NEC Networks, LLC | 2021 | 1600000 | list of databreaaches-wikipedia | unknown | Healthcare | Hacked |
| 349 | Nintendo (Club Nintendo) | 2013 | 240000 | list of databreaaches-wikipedia | unknown | Gaming | Hacked |
| 350 | Nintendo (Nintendo Account) | 2020 | 160000 | list of databreaaches-wikipedia | unknown | Gaming | Hacked |
| 351 | Nippon Television | 2016 | 430000 | list of databreaaches-wikipedia | unknown | Media/Entertainment | Hacked |
| 352 | NTT Business Solutions | 2023 | 9000000 | list of databreaaches-wikipedia | unknown | Telecommunication | Hacked |
| 353 | NTT Docomo | 2023 | 5960000 | list of databreaaches-wikipedia | unknown | Telecommunication | Hacked |
| 354 | OGUsers | 2022 | 529000 | list of databreaaches-wikipedia | unknown | Web service | Hacked |
| 355 | Optus | 2022 | 9800000 | list of databreaaches-wikipedia | unknown | Telecommunication | Hacked |
| 356 | Patreon | 2015 | 2300000 | list of databreaaches-wikipedia | unknown | Web service | Hacked |
| 357 | Popsugar | 2018 | 123857 | list of databreaaches-wikipedia | unknown | Media/Entertainment | Hacked |
| 358 | Rambler.ru | 2012 | 98167935 | list of databreaaches-wikipedia | unknown | Web service | Hacked |
| 359 | Scottrade | 2015 | 4600000 | list of databreaaches-wikipedia | unknown | Financial service | Hacked |
| 360 | Snapchat | 2013 | 4700000 | list of databreaaches-wikipedia | unknown | Social Media | Hacked |
| 361 | StockX | 2019 | 6800000 | list of databreaaches-wikipedia | unknown | Retail | Hacked |
| 362 | Taobao | 2016 | 20000000 | list of databreaaches-wikipedia | unknown | Retail | Hacked |
| 363 | Taringa! | 2017 | 28722877 | list of databreaaches-wikipedia | unknown | Web service | Hacked |
| 364 | Target Corporation | 2013 | 110000000 | list of databreaaches-wikipedia | unknown | Retail | Hacked |
| 365 | TaxSlayer.com | 2016 | 8800 | list of databreaaches-wikipedia | unknown | Web service | Hacked |
| 366 | TD Bank | 2012 | 260000 | list of databreaaches-wikipedia | unknown | Financial service | Hacked |
| 367 | Ticketfly (subsidiary of Eventbrite) | 2018 | 26151608 | list of databreaaches-wikipedia | unknown | Media/Entertainment | Hacked |
| 368 | Tianya Club | 2011 | 28000000 | list of databreaaches-wikipedia | unknown | Web service | Hacked |
| 369 | T-Mobile | 2021 | 45000000 | list of databreaaches-wikipedia | unknown | Telecommunication | Hacked |
| 370 | Tokopedia | 2020 | 91000000 | list of databreaaches-wikipedia | unknown | Retail | Hacked |
| 371 | Tumblr | 2013 | 65469298 | list of databreaaches-wikipedia | unknown | Web service | Hacked |
| 372 | Vastaamo | 2020 | 130000 | list of databreaaches-wikipedia | unknown | Healthcare | Hacked |
| 373 | VTech | 2015 | 5000000 | list of databreaaches-wikipedia | unknown | Retail | Hacked |
| 374 | Walmart | 2015 | 1300000 | list of databreaaches-wikipedia | unknown | Retail | Hacked |
| 375 | Wattpad | 2020 | 270000000 | list of databreaaches-wikipedia | unknown | Web service | Hacked |
| 376 | Wawa (company) | 2020 | 30000000 | list of databreaaches-wikipedia | unknown | Retail | Hacked |
| 377 | Weebly | 2016 | 43430316 | list of databreaaches-wikipedia | unknown | Web service | Hacked |
| 378 | Westpac | 2019 | 98000 | list of databreaaches-wikipedia | unknown | Financial service | Hacked |
| 379 | Yahoo | 2013 | 3000000000 | list of databreaaches-wikipedia | unknown | Web service | Hacked |
| 380 | Zynga | 2019 | 173000000 | list of databreaaches-wikipedia | unknown | Social Media | Hacked |
| 381 | Democratic National Committee | 2016 | 19252 | Wikipedia Region | unknown | political | Hacked |
| 382 | Directorate General of Immigration of Indonesia | 2023 | 34900867 | Wikipedia Region | unknown | Government | Hacked |
| 383 | Service | 2015 | 720000 | Wikipedia Region | unknown | Financial service | Hacked |
| 384 | Japan Pension Service | 2015 | 1250000 | Wikipedia Region | unknown | Government | Hacked |
| 385 | Puerto Rico Department of Health | 2010 | 515000 | Wikipedia Region | unknown | Healthcare | Hacked |
| 386 | Service NSW | 2020 | 104000 | Wikipedia Region | unknown | Government | Hacked |
| 387 | South Africa police | 2013 | 16000 | Wikipedia Region | unknown | Government | Hacked |
| 388 | Various | 2012 | 2434899 | Wikipedia Region | unknown | Government | Hacked |
| 389 | University of California, Berkeley | 2016 | 80000 | Wikipedia Region | unknown | Academic | Hacked |
| 390 | University of Maryland, College Park | 2014 | 300000 | Wikipedia Region | unknown | Academic | Hacked |
| 391 | University of Central Florida | 2016 | 63000 | Wikipedia Region | unknown | Academic | Hacked |
| 392 | MARA | 2019 | 1164540 | Wikipedia Region | unknown | Academic | Hacked |
| 393 | Virginia Prescription Monitoring Program | 2009 | 8257378 | Wikipedia Region | unknown | Healthcare | Hacked |
| 394 | Roblox | 2020 | 4000 | information is beautiful | PII | Gaming | Poor Security |
| 395 | Irish towing company | 2023 | 512000 | information is beautiful | Credit card details | Transportation | Poor Security |
| 396 | Toyota | 2022 | 296019 | information is beautiful | PII | Transportation | Poor Security |
| 397 | X (Twitter) | 2023 | 200000000 | information is beautiful | email IDs/online details | Web service | Poor Security |
| 398 | LINE Pay | 2021 | 133000 | information is beautiful | PII | Financial service | Poor Security |
| 399 | Brewdog | 2021 | 200000 | information is beautiful | email IDs/online details | Retail | Poor Security |
| 400 | Royal Enfield | 2020 | 420873 | information is beautiful | Credit card details | Transportation | Poor Security |
| 401 | Thailand visitors | 2021 | 100000000 | information is beautiful | PII | Government | Poor Security |
| 402 | Amazon Reviews | 2021 | 13124962 | information is beautiful | PII | Web service | Poor Security |
| 403 | Microsoft | 2020 | 250000000 | information is beautiful | email IDs/online details | Web service | Poor Security |
| 404 | Virgin Media | 2020 | 900000 | information is beautiful | email IDs/online details | Retail | Poor Security |
| 405 | Israeli government | 2020 | 6500000 | information is beautiful | PII | Government | Poor Security |
| 406 | Buchbinder Car Rentals | 2020 | 5000000 | information is beautiful | PII | Transportation | Poor Security |
| 407 | Quest Diagnostics | 2019 | 20000000 | information is beautiful | Health & other personal records | Healthcare | Poor Security |
| 408 | Chtrbox | 2019 | 49000000 | information is beautiful | email IDs/online details | misc | Poor Security |
| 409 | WiFi Finder | 2019 | 2000000 | information is beautiful | email IDs/online details | Technology | Poor Security |
| 410 | Unknown | 2019 | 1800000 | information is beautiful | Health & other personal records | Web service | Poor Security |
| 411 | Vårdguiden | 2019 | 2700000 | information is beautiful | Full details | Healthcare | Poor Security |
| 412 | Indian citizens | 2019 | 275265298 | information is beautiful | PII | Web service | Poor Security |
| 413 | Suprema | 2019 | 27800000 | information is beautiful | Full details | Technology | Poor Security |
| 414 | Facebook | 2019 | 419000000 | information is beautiful | PII | Web service | Poor Security |
| 415 | OxyData | 2019 | 380000000 | information is beautiful | PII | Technology | Poor Security |
| 416 | GovPayNow.com | 2018 | 14000000 | information is beautiful | PII | Financial service | Poor Security |
| 417 | Chinese resume leak | 2018 | 202000000 | information is beautiful | PII | Web service | Poor Security |
| 418 | Google+ | 2018 | 52500000 | information is beautiful | PII | Web service | Poor Security |
| 419 | Panerabread | 2018 | 37000000 | information is beautiful | PII | Retail | Poor Security |
| 420 | Texas voter records | 2018 | 14800000 | information is beautiful | PII | Web service | Poor Security |
| 421 | Nametests | 2018 | 120000000 | information is beautiful | email IDs/online details | Technology | Poor Security |
| 422 | Aadhaar | 2018 | 550000000 | information is beautiful | Health & other personal records | Government | Poor Security |
| 423 | MBM Company | 2018 | 1300000 | information is beautiful | Health & other personal records | Retail | Poor Security |
| 424 | LocalBlox | 2018 | 48000000 | information is beautiful | PII | Web service | Poor Security |
| 425 | Twitter | 2018 | 330000000 | information is beautiful | email IDs/online details | Technology | Poor Security |
| 426 | Urban Massage | 2018 | 309000 | information is beautiful | PII | Technology | Poor Security |
| 427 | SKY Brasil | 2018 | 32000000 | information is beautiful | email IDs/online details | Telecommunication | Poor Security |
| 428 | Apollo | 2018 | 200000000 | information is beautiful | email IDs/online details | Technology | Poor Security |
| 429 | RootsWeb | 2017 | 300000 | information is beautiful | Health & other personal records | Web service | Poor Security |
| 430 | Spambot | 2017 | 520000000 | information is beautiful | Health & other personal records | Web service | Poor Security |
| 431 | AI.type | 2017 | 31000000 | information is beautiful | Health & other personal records | Technology | Poor Security |
| 432 | Waterly | 2017 | 1000000 | information is beautiful | Credit card details | Technology | Poor Security |
| 433 | SVR Tracking | 2017 | 540000 | information is beautiful | Health & other personal records | Technology | Poor Security |
| 434 | World Check | 2016 | 2200000 | information is beautiful | Credit card details | misc | Poor Security |
| 435 | Uber | 2015 | 50000 | information is beautiful | email IDs/online details | Technology | Poor Security |
| 436 | Deep Root Analytics | 2015 | 198000000 | information is beautiful | PII | Web service | Poor Security |
| 437 | Sanrio | 2015 | 3300000 | information is beautiful | PII | Web service | Poor Security |
| 438 | Indiana University | 2014 | 146000 | information is beautiful | PII | Academic | Poor Security |
| 439 | New York Taxis | 2014 | 52000 | information is beautiful | email IDs/online details | Transportation | Poor Security |
| 440 | Mozilla | 2014 | 76000 | information is beautiful | PII | Web service | Poor Security |
| 441 | Accendo Insurance Co. | 2011 | 175350 | information is beautiful | PII | Healthcare | Poor Security |
| 442 | Automatic Data Processing | 2006 | 125000 | information is beautiful | PII | Financial service | Poor Security |
| 443 | 50 companies and government institutions | 2022 | 6400000 | list of databreaaches-wikipedia | unknown | various | Poor Security |
| 444 | Accendo Insurance Co. | 2020 | 175350 | list of databreaaches-wikipedia | unknown | Healthcare | Poor Security |
| 445 | Adobe Inc. | 2019 | 7500000 | list of databreaaches-wikipedia | unknown | Technology | Poor Security |
| 446 | Airtel | 2019 | 320000000 | list of databreaaches-wikipedia | unknown | Telecommunication | Poor Security |
| 447 | Ancestry.com | 2021 | 300000 | list of databreaaches-wikipedia | unknown | Healthcare | Poor Security |
| 448 | Apple Health Medicaid | 2021 | 91000 | list of databreaaches-wikipedia | unknown | Healthcare | Poor Security |
| 449 | BMO and Simplii | 2018 | 90000 | list of databreaaches-wikipedia | unknown | Financial service | Poor Security |
| 450 | Dedalus Biologie (a division of Dedalus Global) | 2021 | 500000 | list of databreaaches-wikipedia | unknown | Healthcare | Poor Security |
| 451 | DonorView | 2023 | 948029 | list of databreaaches-wikipedia | unknown | Non-Profit Organisation | Poor Security |
| 452 | ElasticSearch | 2019 | 108000000 | list of databreaaches-wikipedia | unknown | Technology | Poor Security |
| 453 | Exactis | 2018 | 340000000 | list of databreaaches-wikipedia | unknown | data broker | Poor Security |
| 454 | Facebook | 2019 | 540000000 | list of databreaaches-wikipedia | unknown | Social Media | Poor Security |
| 455 | Facebook | 2019 | 267000000 | list of databreaaches-wikipedia | unknown | Social Media | Poor Security |
| 456 | First American Corporation | 2019 | 885000000 | list of databreaaches-wikipedia | unknown | Financial service | Poor Security |
| 457 | Iberdrola | 2022 | 1300000 | list of databreaaches-wikipedia | unknown | energy | Poor Security |
| 458 | Instagram | 2020 | 200000000 | list of databreaaches-wikipedia | unknown | Social Media | Poor Security |
| 459 | MongoDB | 2019 | 202000000 | list of databreaaches-wikipedia | unknown | Technology | Poor Security |
| 460 | MongoDB | 2019 | 275000000 | list of databreaaches-wikipedia | unknown | Technology | Poor Security |
| 461 | Quest Diagnostics | 2019 | 11900000 | list of databreaaches-wikipedia | unknown | clinical laboratory | Poor Security |
| 462 | SlickWraps | 2020 | 377428 | list of databreaaches-wikipedia | unknown | Retail | Poor Security |
| 463 | Tetrad | 2020 | 120000000 | list of databreaaches-wikipedia | unknown | market analysis | Poor Security |
| 464 | TikTok | 2020 | 42000000 | list of databreaaches-wikipedia | unknown | Social Media | Poor Security |
| 465 | Verifications.io (first leak) | 2019 | 809000000 | list of databreaaches-wikipedia | unknown | online marketing | Poor Security |
| 466 | Verifications.io (total leaks) | 2019 | 2000000000 | list of databreaaches-wikipedia | unknown | online marketing | Poor Security |
| 467 | YouTube | 2020 | 4000000 | list of databreaaches-wikipedia | unknown | Social Media | Poor Security |
| 468 | Office of the Registrar General, Birth & Death Registration | 2023 | 50000000 | Wikipedia Region | unknown | Government | Poor Security |
| 469 | Consumer Financial Protection Bureau | 2023 | 256000 | Wikipedia Region | unknown | Government | Poor Security |
| 470 | Department of Homeland Security | 2016 | 30000 | Wikipedia Region | unknown | Government | Poor Security |
| 471 | Unknown | 2020 | 201000000 | Wikipedia Region | unknown | Government | Poor Security |
| 472 | Authority | 2019 | 808000 | Wikipedia Region | unknown | Healthcare | Poor Security |
| 473 | National Health Information Center (NCZI) of Slovakia | 2020 | 391250 | Wikipedia Region | unknown | Healthcare | Poor Security |
| 474 | Various law enforcement agencies (Philippine National Police, National Bureau of Investigation, Bureau of Internal Revenue) | 2023 | 1279437 | Wikipedia Region | unknown | Government | Poor Security |
| 475 | United States Postal Service | 2018 | 60000000 | Wikipedia Region | unknown | Government | Poor Security |
| 476 | Kaiser Permanente | 2024 | 13400000 | information is beautiful | Credit card details | Healthcare | Unknown |
| 477 | Dell | 2024 | 49000000 | information is beautiful | PII | Technology | Unknown |
| 478 | Atlassian | 2023 | 13200 | information is beautiful | email IDs/online details | Technology | Unknown |
| 479 | Japan | 2022 | 500000 | information is beautiful | Credit card details | Government | Unknown |
| 480 | Dubai Real Estate Leak | 2022 | 800000 | information is beautiful | email IDs/online details | Financial service | Insider Threat |
| 481 | Experian SA | 2020 | 24000000 | information is beautiful | Credit card details | Web service | Unknown |
| 482 | Spotify | 2020 | 500000 | information is beautiful | email IDs/online details | Technology | Unknown |
| 483 | Dutch Government | 2020 | 6900000 | information is beautiful | Health & other personal records | Government | Lost/Stolen Device |
| 484 | Marriott Hotels | 2020 | 5200000 | information is beautiful | PII | Retail | Insider Threat |
| 485 | Desjardins Group | 2019 | 4200000 | information is beautiful | PII | Financial service | Insider Threat |
| 486 | Blur | 2019 | 2400000 | information is beautiful | email IDs/online details | Technology | Unknown |
| 487 | NMBS | 2018 | 700000 | information is beautiful | PII | Transportation | Unknown |
| 488 | ViewFines | 2018 | 934000 | information is beautiful | Health & other personal records | Transportation | Unknown |
| 489 | Mount Olympus | 2016 | 1100 | information is beautiful | Full details | Financial service | Insider Threat |
| 490 | CEX | 2017 | 2000000 | information is beautiful | Credit card details | Retail | Unknown |
| 491 | Hong Kong Registration & Electoral Office | 2017 | 3700000 | information is beautiful | PII | Government | Lost/Stolen Device |
| 492 | River City Media | 2017 | 340000000 | information is beautiful | PII | Web service | Unknown |
| 493 | Mutuelle Generale de la Police | 2016 | 112000 | information is beautiful | Full details | Healthcare | Insider Threat |
| 494 | Service | 2016 | 550000 | information is beautiful | Health & other personal records | Healthcare | Unknown |
| 495 | Australian Immigration Department | 2015 | 30 | information is beautiful | Health & other personal records | Government | Unknown |
| 496 | Privatization Agency of the Republic of Serbia | 2014 | 5190396 | information is beautiful | PII | Government | Unknown |
| 497 | Korea Credit Bureau | 2014 | 20000000 | information is beautiful | Full details | Financial service | Insider Threat |
| 498 | Affinity Health Plan, Inc. | 2013 | 344579 | information is beautiful | Health & other personal records | Healthcare | Lost/Stolen Device |
| 499 | Citigroup | 2013 | 150000 | information is beautiful | PII | Financial service | Unknown |
| 500 | Crescent Health Inc., Walgreens | 2013 | 100000 | information is beautiful | Health & other personal records | Healthcare | Lost/Stolen Device |
| 501 | Group | 2013 | 4000000 | information is beautiful | PII | Healthcare | Lost/Stolen Device |
| 502 | Agency | 2013 | 1500000 | information is beautiful | Full details | Government | Insider Threat |
| 503 | Facebook | 2013 | 6000000 | information is beautiful | email IDs/online details | Web service | Unknown |
| 504 | TerraCom & YourTel | 2013 | 170000 | information is beautiful | PII | Telecommunication | Unknown |
| 505 | Court Ventures | 2013 | 200000000 | information is beautiful | PII | Financial service | Insider Threat |
| 506 | South Carolina Government | 2012 | 228000 | information is beautiful | Health & other personal records | Healthcare | Insider Threat |
| 507 | California Department of Child Support Services | 2012 | 800000 | information is beautiful | PII | Government | Lost/Stolen Device |
| 508 | Emory Healthcare | 2012 | 315000 | information is beautiful | Health & other personal records | Healthcare | Lost/Stolen Device |
| 509 | Attorney General | 2012 | 6500000 | information is beautiful | PII | Government | Unknown |
| 510 | New York State Electric & Gas | 2012 | 1800000 | information is beautiful | PII | misc | Insider Threat |
| 511 | Memorial Healthcare System | 2012 | 102153 | information is beautiful | PII | Healthcare | Lost/Stolen Device |
| 512 | "Apple" | 2012 | 12367232 | information is beautiful | PII | Retail | Unknown |
| 513 | New York City Health & Hospitals Corp. | 2011 | 1700000 | information is beautiful | Health & other personal records | Healthcare | Lost/Stolen Device |
| 514 | South Shore Hospital, Massachusetts | 2011 | 800000 | information is beautiful | Full details | Healthcare | Lost/Stolen Device |
| 515 | Yale University | 2011 | 43000 | information is beautiful | PII | Academic | Unknown |
| 516 | Morgan Stanley Smith Barney | 2011 | 34000 | information is beautiful | Credit card details | Financial service | Lost/Stolen Device |
| 517 | State of Texas | 2011 | 3500000 | information is beautiful | PII | Government | Unknown |
| 518 | Health Net - IBM | 2011 | 1900000 | information is beautiful | Credit card details | Healthcare | Lost/Stolen Device |
| 519 | Eisenhower Medical Center | 2011 | 514330 | information is beautiful | Health & other personal records | Healthcare | Lost/Stolen Device |
| 520 | Spartanburg Regional Healthcare System | 2011 | 400000 | information is beautiful | Health & other personal records | Healthcare | Lost/Stolen Device |
| 521 | NHS | 2011 | 8600000 | information is beautiful | Health & other personal records | Healthcare | Lost/Stolen Device |
| 522 | Nemours Foundation | 2011 | 1600000 | information is beautiful | Health & other personal records | Healthcare | Lost/Stolen Device |
| 523 | Sutter Medical Foundation | 2011 | 4243434 | information is beautiful | PII | Healthcare | Lost/Stolen Device |
| 524 | Tricare | 2011 | 4901432 | information is beautiful | Health & other personal records | Healthcare | Lost/Stolen Device |
| 525 | AvMed, Inc. | 2010 | 1220000 | information is beautiful | PII | Healthcare | Lost/Stolen Device |
| 526 | Blue Cross Blue Shield of Tennessee | 2010 | 1023209 | information is beautiful | PII | Healthcare | Lost/Stolen Device |
| 527 | US Military | 2010 | 260000 | information is beautiful | Full details | Government | Insider Threat |
| 528 | Triple-S Salud, Inc. | 2010 | 398000 | information is beautiful | Health & other personal records | Healthcare | Lost/Stolen Device |
| 529 | Emergency Healthcare Physicians, Ltd. | 2010 | 180111 | information is beautiful | Health & other personal records | Healthcare | Lost/Stolen Device |
| 530 | Colorado government | 2010 | 105470 | information is beautiful | PII | Healthcare | Lost/Stolen Device |
| 531 | Lincoln Medical & Mental Health Center | 2010 | 130495 | information is beautiful | Health & other personal records | Healthcare | Lost/Stolen Device |
| 532 | Educational Credit Management Corp | 2010 | 3300000 | information is beautiful | PII | Financial service | Lost/Stolen Device |
| 533 | Classified Iraq War documents | 2010 | 392000 | information is beautiful | PII | Government | Insider Threat |
| 534 | US National Guard | 2009 | 131000 | information is beautiful | PII | Government | Lost/Stolen Device |
| 535 | Health Net | 2009 | 1500000 | information is beautiful | Health & other personal records | Healthcare | Lost/Stolen Device |
| 536 | US Military | 2009 | 76000000 | information is beautiful | PII | Government | Lost/Stolen Device |
| 537 | Compass Bank | 2008 | 1000000 | information is beautiful | Credit card details | Financial service | Insider Threat |
| 538 | University of Miami | 2008 | 2100000 | information is beautiful | Credit card details | Academic | Lost/Stolen Device |
| 539 | BNY Mellon Shareowner Services | 2008 | 4500000 | information is beautiful | email IDs/online details | Financial service | Lost/Stolen Device |
| 540 | Countrywide Financial Corp | 2008 | 2500000 | information is beautiful | PII | Financial service | Insider Threat |
| 541 | UK Home Office | 2008 | 84000 | information is beautiful | PII | Government | Lost/Stolen Device |
| 542 | GS Caltex | 2008 | 11100000 | information is beautiful | PII | misc | Insider Threat |
| 543 | AT&T | 2008 | 113000 | information is beautiful | email IDs/online details | Telecommunication | Lost/Stolen Device |
| 544 | Stanford University | 2008 | 72000 | information is beautiful | PII | Academic | Lost/Stolen Device |
| 545 | University of Utah Hospitals & Clinics | 2008 | 2200000 | information is beautiful | Health & other personal records | Academic | Lost/Stolen Device |
| 546 | Texas Lottery | 2008 | 89000 | information is beautiful | PII | Government | Insider Threat |
| 547 | Starbucks | 2008 | 97000 | information is beautiful | PII | Retail | Lost/Stolen Device |
| 548 | UK Ministry of Defence | 2008 | 1700000 | information is beautiful | Full details | Government | Lost/Stolen Device |
| 549 | T-Mobile, Deutsche Telecom | 2008 | 17000000 | information is beautiful | email IDs/online details | Telecommunication | Lost/Stolen Device |
| 550 | Authorities | 2008 | 3950000 | information is beautiful | PII | Government | Unknown |
| 551 | Service Personnel and Veterans Agency (UK) | 2008 | 50500 | information is beautiful | PII | Government | Lost/Stolen Device |
| 552 | Agency | 2007 | 3000000 | information is beautiful | PII | Government | Lost/Stolen Device |
| 553 | Fidelity National Information Services | 2007 | 8500000 | information is beautiful | Credit card details | Financial service | Insider Threat |
| 554 | City and Hackney Teaching Primary Care Trust | 2007 | 160000 | information is beautiful | PII | Government | Lost/Stolen Device |
| 555 | Gap Inc | 2007 | 800000 | information is beautiful | PII | Retail | Lost/Stolen Device |
| 556 | Dai Nippon Printing | 2007 | 8637405 | information is beautiful | email IDs/online details | Retail | Insider Threat |
| 557 | JP Morgan Chase | 2007 | 2600000 | information is beautiful | Credit card details | Financial service | Lost/Stolen Device |
| 558 | UK Revenue & Customs | 2007 | 25000000 | information is beautiful | email IDs/online details | Government | Lost/Stolen Device |
| 559 | AOL | 2006 | 20000000 | information is beautiful | email IDs/online details | Web service | Unknown |
| 560 | US Dept of Vet Affairs | 2006 | 26500000 | information is beautiful | PII | Government | Lost/Stolen Device |
| 561 | Hewlett Packard | 2006 | 200000 | information is beautiful | PII | Retail | Lost/Stolen Device |
| 562 | Ameritrade Inc. | 2005 | 200000 | information is beautiful | PII | Financial service | Lost/Stolen Device |
| 563 | Citigroup | 2005 | 3900000 | information is beautiful | Credit card details | Financial service | Lost/Stolen Device |
| 564 | AOL | 2004 | 92000000 | information is beautiful | email IDs/online details | Web service | Insider Threat |
| 565 | Apple, Inc./BlueToad | 2021 | 12367232 | list of databreaaches-wikipedia | unknown | Retail | Accidental Exposure |
| 566 | AT&T | 2021 | 72000000 | list of databreaaches-wikipedia | unknown | Telecommunication | Unknown |
| 567 | Bank of America | 2005 | 1200000 | list of databreaaches-wikipedia | unknown | Financial service | Lost/Stolen Device |
| 568 | BlueCross BlueShield of Tennessee | 2009 | 1023039 | list of databreaaches-wikipedia | unknown | Healthcare | Lost/Stolen Device |
| 569 | Capital One | 2019 | 106000000 | list of databreaaches-wikipedia | unknown | Financial service | Misconfiguration |
| 570 | CheckPeople | 2020 | 56000000 | list of databreaaches-wikipedia | unknown | Government | Unknown |
| 571 | Collection No. 1 | 2019 | 773000000 | list of databreaaches-wikipedia | unknown | various | Credential Stuffing |
| 572 | DC Health Link | 2023 | 56000 | list of databreaaches-wikipedia | unknown | Healthcare | Misconfiguration |
| 573 | Desjardins | 2019 | 9700000 | list of databreaaches-wikipedia | unknown | Financial service | Insider Threat |
| 574 | Duolingo | 2023 | 2676696 | list of databreaaches-wikipedia | unknown | Academic | Web Scraping |
| 575 | Facebook | 2019 | 1500000 | list of databreaaches-wikipedia | unknown | Social Media | Accidental Exposure |
| 576 | Facebook Marketplace | 2023 | 200000 | list of databreaaches-wikipedia | unknown | Social Media | Unknown |
| 577 | IKEA | 2022 | 95000 | list of databreaaches-wikipedia | unknown | Retail | Accidental Exposure |
| 578 | International Committee of the Red Cross | 2022 | 515000 | list of databreaaches-wikipedia | unknown | Non-Profit Organisation | Unknown |
| 579 | Inuvik hospital | 2016 | 6700 | list of databreaaches-wikipedia | unknown | Healthcare | Insider Threat |
| 580 | Japanet Takata | 2004 | 510000 | list of databreaaches-wikipedia | unknown | Retail | Insider Threat |
| 581 | Les Editions Protègez-vous | 2020 | 380000 | list of databreaaches-wikipedia | unknown | Media/Entertainment | Unknown |
| 582 | Mitsubishi Tokyo UFJ Bank | 2006 | 960000 | list of databreaaches-wikipedia | unknown | Financial service | Lost/Stolen Device |
| 583 | Nemours Foundation | 2011 | 1055489 | list of databreaaches-wikipedia | unknown | Healthcare | Lost/Stolen Device |
| 584 | NHS | 2011 | 8300000 | list of databreaaches-wikipedia | unknown | Healthcare | Lost/Stolen Device |
| 585 | Now:Pensions | 2020 | 30000 | list of databreaaches-wikipedia | unknown | Financial service | Insider Threat |
| 586 | PayPay | 2020 | 20076016 | list of databreaaches-wikipedia | unknown | Financial service | Misconfiguration |
| 587 | Rakuten | 2020 | 1381735 | list of databreaaches-wikipedia | unknown | Retail | Misconfiguration |
| 588 | Spoutible | 2024 | 207114 | list of databreaaches-wikipedia | unknown | Social Media | Misconfiguration |
| 589 | Tangerine Telecom | 2024 | 243462 | list of databreaaches-wikipedia | unknown | Telecommunication | Credential Stuffing |
| 590 | Tesla | 2023 | 75000 | list of databreaaches-wikipedia | unknown | Transportation | Insider Threat |
| 591 | Tic Hosting Solutions (known as Torchbyte) | 2023 | 46 | list of databreaaches-wikipedia | unknown | Web service | Misconfiguration |
| 592 | Trello | 2024 | 15111945 | list of databreaaches-wikipedia | unknown | Technology | Misconfiguration |
| 593 | Truecaller | 2019 | 299055000 | list of databreaaches-wikipedia | unknown | Telecommunication | Unknown |
| 594 | View Media | 2020 | 38000000 | list of databreaaches-wikipedia | unknown | online marketing | Misconfiguration |
| 595 | Xat.com | 2015 | 6054459 | list of databreaaches-wikipedia | unknown | Web service | Social Engineering |
| 596 | Central Intelligence Agency | 2017 | 91 | Wikipedia Region | unknown | Government | Insider Threat |
| 597 | Embassy Cables | 2010 | 251000 | Wikipedia Region | unknown | Government | Insider Threat |
| 598 | England and Wales Cricket Board | 2024 | 43299 | Wikipedia Region | unknown | Government | Unknown |
| 599 | Ministry of Health | 2019 | 14200 | Wikipedia Region | unknown | Healthcare | Insider Threat |
| 600 | South Carolina Department of Revenue | 2012 | 6400000 | Wikipedia Region | unknown | Healthcare | Insider Threat |
| 601 | U.S. Army | 2011 | 50000 | Wikipedia Region | unknown | Government | Exposure |